\documentclass[doctor,11pt]{iscs-thesis}

\usepackage{url}
\usepackage[fleqn]{amsmath}
\usepackage[psamsfonts]{amssymb}
\usepackage[final,dvipdfm]{graphicx}
\usepackage{amsthm}
\usepackage{braket}
\usepackage{algorithmicx}
\usepackage{algpseudocode}
\usepackage{multirow,bigbrace}
\usepackage{subfigure}

\newfont{\bg}{cmr9 scaled \magstep4}
\theoremstyle{plain}
\newtheorem{theorem}{Theorem}[chapter]
\newtheorem{corollary}{Corollary}[chapter]
\newtheorem{lemma}{Lemma}[chapter]
\newtheorem{claim}{Claim}[chapter]
\theoremstyle{definition}
\newtheorem{definition}{Definition}[chapter]
\newtheorem{axiom}{Axiom}[chapter]
\newtheorem{example}{Example}[chapter]
\newtheorem{algorithm}{Algorithm}[chapter]
\newtheorem{conjecture}{Conjecture}[chapter]
\theoremstyle{remark}

\etitle{Voronoi Diagrams for Quantum States and Its Application to
a Numerical Estimation of a Quantum Channel Capacity}
\jtitle{}
\jauthor{}
\date{December 17, 2007}
\eauthor{Kimikazu Kato
\thanks{This version is slightly chaged from the original thesis. 
In the original thesis, the title and the abstract are also written in 
Japanese. They are omitted to comply with arXiv's policy, and so that
this can be complied by a usual latex without Japanese capability.}
}
\esupervisor{Hiroshi Imai}
\jsupervisor{}
\supervisortitle{Professor}

\begin{eabstract}

 In quantum information theory, a geometric approach, known as ``quantum
 information geometry,'' has been considered as a powerful method. In
 this thesis, we give a computational geometric interpretation to the
 geometric structure of a quantum system. Especially we introduce the
 concept of the Voronoi diagram and the smallest enclosing ball problem
 to the space of quantum states. With those tools in computational
 geometry, we analyze the adjacency structure of a point set in the
 quantum state space. Additionally, as an application, we show an
 effective method to compute the capacity of a quantum channel.
 
 In the first part of this thesis, we show some coincidences of Voronoi
 diagrams in a quantum state space with respect to some distances. That
 helps us to reinterpret the structure of the space of quantum pure
 states as a subspace of the whole space. More properly, we investigate
 the Voronoi diagrams with respect to the divergence, Fubini-Study
 distance, Bures distance, geodesic distance and Euclidean distance.

For one qubit (or two level) states, the whole space is expressed as the
 Bloch ball. In the Bloch ball, we analyze the coincidence of the
 Voronoi diagrams in two different settings: 1)the diagram in pure
 states when Voronoi sites are taken as pure states and 2)the diagram
 in mixed states when sites are taken as pure states. We show that in
 both cases, all the diagrams coincide. This clear result is because of
 the symmetry specific for one qubit states.

 For three or higher level systems, we investigate the diagrams in pure
 states. The natural embedding of the quantum state space into a
 Euclidean space is no longer symmetric as in one qubit
 case. Consequently the coincidence of the Euclidean Voronoi diagram and
 the divergence-Voronoi diagram does not hold in a higher level
 system. However the coincidence of the divergence-, Fubini-Study-
 and Bures-Voronoi diagrams still holds.

 In the second part, we propose a method to compute a capacity of a
 quantum communication channel and show the result of the actual
 computation. We show that our method is sufficiently effective not
 only for one qubit states but for three level states. It is a
 practical application of the theoretical result shown in the first
 part; the theorems in the first part guarantee the correctness of the
 algorithm used in the second part. The algorithm uses Welzl's
 algorithm to compute a smallest enclosing ball. Although the original
 algorithm introduced by Welzl is only for the Euclidean space, we
 show that the same method is useful for non-Euclidean space. We also
 implement the algorithm and experiment it to prove it is practical.
\end{eabstract}

\begin{jabstract}
\end{jabstract}

\def\Tr{{\rm Tr}\;}

\allowdisplaybreaks[1]
\begin{document}
\sloppy
\maketitle
\begin{acknowledge}
First of all, my Ph.D work is totally financially supported by Nihon
 Unisys, Ltd., which I belong to as an employee. I am very grateful to
 the company for providing me such an opportunity. I am also grateful to
 my adviser Hiroshi Imai for his advice and continuous support, and
 above all, for accepting me as a new member of his lab.

I thank coauthors of the papers which consists most of this
 dissertation. In particular, the very start of my research project is
 the discussion with  Mayumi Oto. I owe her very much. I
 am also grateful to Keiko Imai and Jiro Nishitoba, who are also
 coauthors of the papers.

Discussions with my colleagues and related researchers are of course,
 essential to progress my work. Although I cannot list up all of them,
 an incomplete list includes Fran\c{c}ois Le Gall, Jun Hasegawa,
 Masahito Hayashi, Tsuyosi Ito, Masaki Owari, Toshiyuki Shimono. I am
 grateful to them for their advice and fruitful discussions.

I am also thankful to Kokichi Sugihara, who advised me about figural
 presentation of some Voronoi diagrams; and Hidetoshi Muta, who provided some
 figures of Voronoi diagrams.

Last but not least, I would like to thank my family. I am specially
 thankful to my wife, Masumi Kato who supported my decision to go back
 to the university, and also grateful to my two daughters; Koko and
 Koto, who supported me by just being there.
\end{acknowledge}

\frontmatter
\tableofcontents
\listoffigures
\listoftables
\mainmatter
\chapter{Introduction}
In this chapter, we explain the background for our research and the
summary of our contribution. Our contribution has mainly two aspects:
computational geometry and quantum information theory. 

In computational geometry, our contribution is shortly described as
introduction of another Voronoi diagram with a distortion measure and an
algorithm to solve the smallest enclosing ball problem in that measure.

In quantum information theory, we contribute to re-interpret the
structure of a quantum state space to some extent, and proposed a
practical algorithm to compute the capacity of a quantum channel.

We explain summarized background for topics
related to our contribution. The explanation of the background is
divided into two parts: one is about computational geometry and the
other is about quantum information theory. Then, we explain the outline
of the contribution of this dissertation, and show how this dissertation
is organized.

\section{Computational geometry}
\subsection{Voronoi diagrams}
\label{sec:vd-computational-geometry}
Voronoi diagrams and Delaunay triangulations have had an important role
in computational geometry. Voronoi diagrams are not only useful for
such applications as numerical calculation and
visualization, but also useful for theoretical interpretation of a geometric
object.

Practically, Voronoi diagrams are used in many fields such as
geophysics, meteorology, astrology, geometric information system, city
planning, and so on. Additionally, nowadays one of the most important
applications of Voronoi diagrams is computer vision. In particular, because
of the recent rapid development of computer
graphics in entertainment media such as movies and games, Voronoi
diagrams are getting more and more important. Other practical examples
of applications are explained in \cite{okabe00}.

\begin{figure}
 \begin{center}
  \includegraphics[scale=.3]{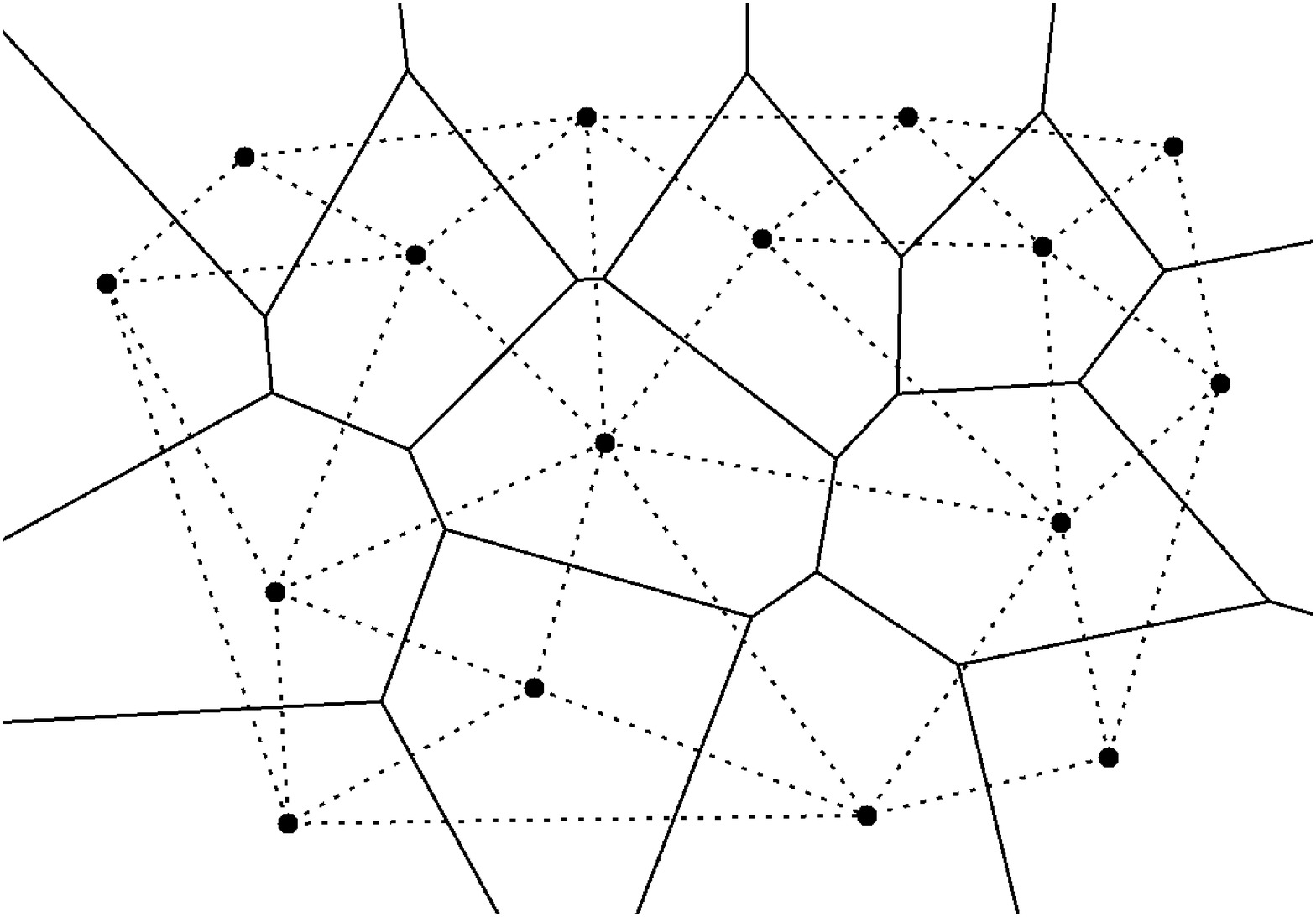}
 \end{center}
\caption[An example of Voronoi diagram and Delaunay triangulation]
{An example of Voronoi diagram and Delaunay triangulation: Solid
 lines are a Voronoi diagram, and dotted lines are a Delaunay triangulation}
 \label{fig:example-vd-dt}
\end{figure}

A Voronoi diagram is a division of a space. The applications listed
above all deal with a geometric space and need to divide it so that it
can be computed in reasonable time. The essence of Voronoi diagrams can
be explained as follows. Suppose that some points (which we call {\em
sites}) are given and you want to divide the space into some regions so
that each region expresses dominance of a point. If for any point in a
region, the nearest site is the site included in the region in problem,
then the division is called a Voronoi diagram. A Delaunay
triangulation is a dual of a Voronoi diagram; for any two sites, draw a
line between them if there is a Voronoi edge between
them, and you obtain a Delaunay triangulation. Fig.~\ref{fig:example-vd-dt} shows a example of Voronoi diagram
and Delaunay triangulation.

\begin{figure}
\begin{center}
 \includegraphics{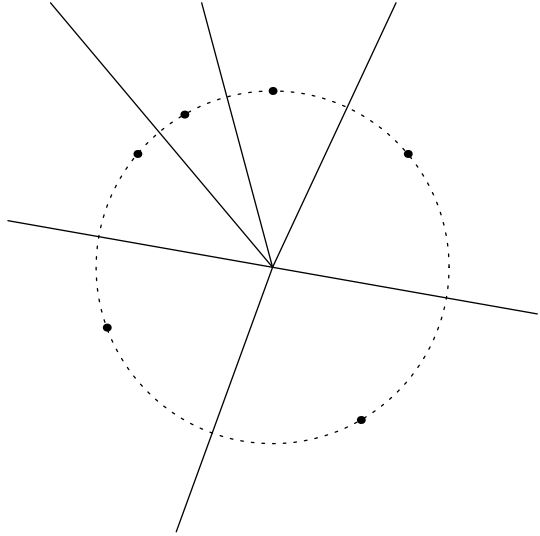}
\end{center}
 \caption[An example of degenerate Voronoi diagram]
 {An example of degenerate Voronoi diagram: Sites are in a cocircular
 position and Voronoi edges meet at one point.}
 \label{fig:degenerate}
\end{figure}

As Voronoi diagrams used in the wider area, the more complicated
diagrams with many Voronoi cells became needed and consequently, the
robustness of the computation became emphasized. Actually, the
general real world problems which needs Voronoi diagrams are badly
positioned so that the resulting diagrams are degenerating. Degeneracy
of Voronoi diagrams happens when the sites are cocircular. The evilness
of degeneracy lies on the weakness for perturbation; only slight move of
a site will change the topological position (See
Fig.~\ref{fig:degenerate}). In calculation in a computer where numbers are
not expressed rigidly, this kind of evilness may cause inconsistent
situation such as ``A point is geometrically in a certain area, but
topologically out of the area.''  One off the researches to overcome
this problem is by Sugihara and Iri \cite{sugihara92, sugihara94}. They
achieved the robustness by focusing only on topologies and ignoring the
geometries when computing about the relative position of edges. Thanks
to those researches about the robustness of actual computation, now
Voronoi diagram with billions of sites is shown to be
computed. Isenburg et al.~\cite{isenburg06} showed an algorithm to
calculate a tessellation of terrain data with a billion of triangles
using some heuristics.

The significant application which emerged recently is fine art. Fritzsche
et al.~\cite{fritzsche05} proposed a algorithm to synthesize an
authentic look mosaic structure from a picture. Sugihara
\cite{sugihara07} synthesize a artistic pattern. Since there is often
a fractal structure behind a beauty in the nature, he proposed an
algorithm to create a nature-like shape by combining Voronoi diagrams
with fractal.

\subsection{Generalized Voronoi diagram}
\label{sec:generalized-vd}
In the original ``normal'' Voronoi diagrams, the sites are given as
points and Euclidean distance is used to decide the dominance of each
region. The generalization of Voronoi diagrams mainly goes into two
ways: a) define a site as a set of point instead of one point, or b) use
a general distance instead of the Euclidean distance.

In the direction to consider a general site, Voronoi diagrams for line
segments were intensively researched from the late 70's, by Drysdale and
Lee \cite{drysdale78}, Drysdale~\cite{drysdale79},
Kirkpatrick~\cite{kirkpatrick79}, Lee and Drysdale~\cite{lee81}, Imai
et al.~\cite{imai85}, Sharir~\cite{sharir85}, Fortune~\cite{fortune86},
Yap~\cite{yap87}, Clarkson and Shor~\cite{clarkson89}, Goodrich et
al.~\cite{goodrich93}, Burnikel et al.~\cite{burnikel94}, Rajasekaran
and Ramaswami~\cite{rajasekaran94, rajasekaran95}, and Deng and
Zhu~\cite{deng96}. The rather recent ones are Voronoi diagrams for
circles or balls. Concerning the Voronoi diagrams whose sites are given
as circles, Kim et al.~\cite{kim01a,kim01b} proposed a algorithm which
is robust even for the degenerating case.

Here, note that when we say ``distance'' in the context of Voronoi
diagrams, it does not necessarily satisfy the axioms of a distance. In
this dissertation, we use the word ``measure'', or ``pseudo-distance''
in a confusing context.

\begin{figure}
\begin{center}
  \includegraphics{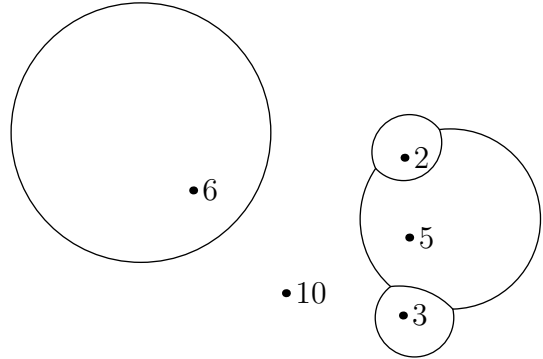}
\end{center} 
\caption[An example of weighted Voronoi diagram]{An example of weighted
 Voronoi diagram: Each number assosiated to the sites means its weight}
 \label{fig:weighted-voronoi}
\end{figure}

Another direction is to use a general distance in a general space. One
of the simplest in this direction is Voronoi diagrams for the weighted
distance \cite{okabe00}. It means that each site has a weight and the
distance is measured according to the weight. If $w_i$ is the weight for
the site $s_i$, then the multiplicatively weighted distance to the point
$x$ is defined as $d_\text{weighted}(s_i,x)= \left|x-s_i\right| /
w_i$. Fig.~\ref{fig:weighted-voronoi} shows an example of
multiplicatively weighted Voronoi diagrams. The Voronoi edges for a
multiplicatively weighted Voronoi diagram becomes a part of a circle, so
called {\em Apollonius circle}.

The weighted distance is distorted to some extent, but it still based on
Euclidean distance. Using more general distance, more distorted or
sometimes pathological Voronoi diagrams can be obtained. For example,
Onishi and Itoh investigated Riemannian Voronoi diagram
\cite{onishi03}. In classical information theory, Onishi and
Imai~\cite{onishi97,onishi97a} and Nielsen et al.~\cite{nielsen07a} are
for divergences. The detail of divergence Voronoi diagrams are explained
in the next section.

\begin{figure}
\begin{center}
 \includegraphics[scale=0.2]{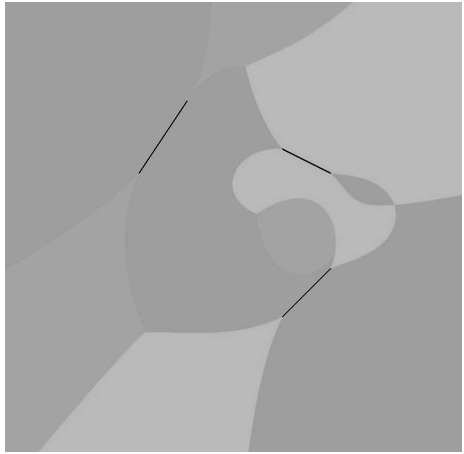}
\end{center}
\caption[An example of an angular Voronoi diagram]{An example of an
 angular Voronoi diagram (drawn by H. Muta): Although it can be
 drawn in two dimensional space, its measure is distorted and the
 Voronoi edges are generally cubic curves}
\label{fig:angular-voronoi}
\end{figure}

The combination of the two directions described above can also be
considered. Generalizing the way to decide the polygon mesh used in
computer vision, Asano introduced an aspect-ratio Voronoi diagram and
analyzed its computational complexity \cite{asano07b}, and Asano et al.\
also introduced an angular Voronoi diagram \cite{asano06}. In those
diagrams, Voronoi sites are given as line segments and the distance is
given as a visual angle or the aspect ratio of the triangle composed by
a point and a line segment respectively. In both cases, the Voronoi
edges are curves of degree three and the regions can be complicated; the
region dominated by the same line segment can separated by points. Fig.\
\ref{fig:angular-voronoi} shows the example of angular Voronoi diagrams.
Asano et al.\ also generalized those Voronoi diagrams to obtain an
abstract notation of Voronoi diagrams and analyzed the characteristics
of those diagrams \cite{asano07a}.

When we think about actual computation of those generalized distortion
Voronoi diagrams, the arising problem is how we can achieve its
robustness. Muta and the author analyzed the extended the notion of
degeneracy of Voronoi diagrams from the viewpoint of the computational
robustness \cite{muta07}. The main discussion in the paper is about the
conditions to make the number of crossing points of Voronoi edges jump
with perturbation. It is only about angular Voronoi diagrams, but the
concept of degeneracy in the paper is also applicable to general Voronoi
diagrams, although there is only few attempt to actually compute the
diagrams explicitly.

There are still other Voronoi diagrams with some ``strange'' measure
spaces. In the book by Okabe et al.~\cite{okabe00}, many of them are
introduced with analysis about their computational complexity.

\subsection{Voronoi diagrams in classical and quantum information}
\label{sec:vd-information}
Information theory is considered to have been founded by
Shannon~\cite{shannon48}. He showed the bound for the capacity of a
channel by coding the source message of the channel. Its coding strategy
is decided by the probabilistic distribution of the source
message. Thus, information theory is mainly based on probability theory
and statistics. The most important quantities used in this field are
entropy, and Kullback Leibler divergence ({\em or} relative
entroby). Kullback-Leibler divergence is defined as some kind of
``distance'' of two probabilistic distribution. Hence, geometry in
information space can be considered but its structure is very distorted
and far from intuition. Kullback-Leibler divergence does not satisfy the
axioms of distance (so it is not a distance in a rigid sense); for
example it does not satisfy the law of triangle inequity, or is not
commutative either. However, its distorted and strange properties give a
rich field of a research for computational geometry.

A computational geometric analysis was done by Onishi and
Imai~\cite{onishi97,onishi97a}, Onishi~\cite{onishi98}, and Sadakane et
al.~\cite{sadakane98}.  A Voronoi diagram and Delaunay triangulation are
defined with respect to the Kullback-Leibler divergence, and are shown
to be the extensions of the Euclidean counterparts. The Voronoi diagram
is computed from an associated potential function instead of a
paraboloid which is used in a Euclidean Voronoi diagram.

In the same line, Nielsen et al.~\cite{nielsen07a} showed some
properties of Voronoi diagram with respect to Bregman divergence, which
is generalization of Kullback-Leibler divergence. Using the Voronoi
diagram, Nielsen et al.~\cite{nielsen07b} also showed that Welzl's
algorithm to solve the smallest enclosing ball problem is also
applicable to Bregman divergence.

We extend the Voronoi diagram in classical information to the quantum
world. In quantum information theory, there is a natural extension of
the Kullback-Leibler divergence, and it is called a quantum
divergence. We introduce a Voronoi diagram with respect to the quantum
divergence, and analyze its structure. Additionally we consider other
diagrams with respect to some distances. Comparing the diagrams, we can
compare the structures of some distance spaces and consequently some
problem concerning a certain distance can be replaced by another problem
of another distance.

\subsection{Smallest enclosing ball problem}
\label{sec:seb}
The smallest enclosing ball is namely a problem to compute the smallest
ball which contains given points. It has variety of applications;
collision detection, facility location, automated manufacturing, and so
on.  It is a geometric problem but has some aspect of combinatorial
optimization.

The first theoretically effective algorithm was given by Megiddo
\cite{meggido84}. In spite of its astonishing idea of {\em
pruning-and-search}, Megiddo's algorithm was impractical because there
is a big constant hidden behind a big-O notation. Welzl \cite{welzl91}
gave the first practical algorithm based on Seidel's randomized linear
programming algorithm \cite{seidel90}. In these algorithms, however, the
complexity is the exponential of the dimension. Eventually Matou\v{s}ek
et al.~\cite{matousek96} discovered subexponential time algorithm. The
most efficient algorithm known so far is Fischer and G\"artner's
algorithm \cite{fischer04}.  They gave an $O(d^3(1.438)^d)$-time
algorithm. Fischer also implemented a program to compute the smallest
enclosing ball problem as a part of CGAL~\cite{cgal}.

Nishitoba et al.~\cite{nishitoba06} connected this line of the research
to the computation of the Holevo capacity in quantum information
theory.  He also shed light on
combinatorial aspect of this problem. This direction is followed by
Nishitoba~\cite{nishitoba07} to analyze the combinatorial
structure. Although Hayashi et al.\ had already showed the method using the
smallest enclosing ball problem, they first mentioned the necessity of
the fast algorithm with respect to the dimension to extend the existing method
to the higher level system. Actually the dimension of the space when we
think of the smallest enclosing ball problem is $d^2-1$ for $d$-level
system; this grows too rapidly from the viewpoint of
practical computation.

\section{Quantum information theory}
\subsection{Quantum computation and quantum information}
\label{sec:q-compuatation-information}
Feynman \cite{feynman82} is considered to be one of the earliest to show an idea to apply
quantum mechanics to computation. His idea comes from the fact that in
quantum mechanics, huge amount of computation is needed to compute the
behavior of particles; he considered that particles which behave
according to the law of quantum mechanics can be used to compute a quantum
behavior itself. 

Deutsch \cite{deutsch85} is the first to show this idea is really useful
in some problem. He showed a problem which can be solved by a quantum
computer exponentially faster than a classical computer. Although the
problem proposed by him is rather artificial and it is only to
show the computational gap between a classical computer and a quantum computer, it
has a significant meaning as a first example to show the power of
superposition of quantum states in computation. The original algorithm 
proposed by
Deutsch is only for one bit but it was shown to be extended to $n$-bit
by Deutsch and Jozsa~\cite{deutsch92}, and Cleve et al.~\cite{cleve98}
gave another improvement .

One of the greatest works in quantum computation was by
Shor~\cite{shor98}. He showed that using a quantum computer, factoring of
integer and discrete logarithms can be solved in a polynomial time. It
is the first example in which quantum computer is exponentially faster
than classical computer. It was sensational because the difficulty of
factoring is a guarantee for the security of the existing public key
cryptosystem.

Another famous algorithm for quantum computer is database search
algorithm by Grover~\cite{grover96}. His algorithm is quadratically
faster than classical one. His original algorithm is improved and
generalized by Grover himself \cite{grover98, grover98b} and Biham et
al.~\cite{biham99}. Another algorithm is about integration. Abrams and
Williams~\cite{abrams99} showed an algorithm for multi-dimensional
integration. Analysis on some classes of functions is done by
Novak~\cite{novak01} and Heinrich~\cite{heinrich03}. For quantum
algorithms for numerical integration, surveys are written by
Heinrich~\cite{heinrich01, heinrich06}.

Miyake and Wadati~\cite{MW01} showed that the Fubini-Study distance in a
quantum state space has
a special meaning for quantum search algorithm. In the continuous
version of Grover algorithm, a quantum state follows a path which is
geodesic in the Fubini-Study distance.

Quantum information theory has been considered as a primitive backbone
for the quantum computation. The typical theme of quantum information is
``What can be possible using a quantum channel?'' Its difficulty
is based on the characteristics of the measuring of quantum states. Even
for the two completely same quantum states, the result of the
measurement may be different and may distribute probabilistically. 
Consequently, the typical objective of quantum
information theory is to distinguish some different quantum states by
measuring. 

For quantum information theory, the invention of the quantum
cryptosystem \cite{bb84} is an important epoch-making event; it became a
trigger to make active the research in that field although it was not
the very start of quantum information theory. From the practical point
of view, quantum cryptosystem is believed to be a very near to
utilization in the real world. Tajima et al.~\cite{tajima07} reported a
success of an experiment toward realization of quantum cryptosystem with
a realistic settings. Additionally, some venture companies, such as
MagiQ Technologies~\cite{magiq} and id Quantique~\cite{idquantique}, are
emerging in this field.

Some aspect of quantum
information theory is to investigate a kind of distance between two
different quantum states. Depending on the situation, several distances
are defined in quantum states. In quantum information geometry, the
structure of those distances is researched \cite{amari00,petz96}.

The quantum divergence have been used as an informational distance from
a quantum state to another. In particular, it played an important role in
quantum hypothesis testing \cite{hiai91, ogawa00} and an estimation of a
capacity of a quantum channel \cite{holevo73, holevo79}. This
informational measure is the main of our interest. It is not symmetric
and so distorted that its structure is difficult to understand. We have
been motivated to understand its structure and clarify its geometric
properties.

\subsection{Power of entanglement and additivity conjecture}
\label{sec:entanglement-additivity} Entanglement is considered to be one
of the most important and interesting objects in quantum information
theory, and actually provides a hot field of research. As is described above,
the result of the measurement of a quantum state may distribute
probabilistically; let us compare it to the coin tossing. Then entangled
states are like correlated coins; the probability whether one will show
the top or tail is related to the result of the toss of the others.

This correlation, a strange behavior of particles, were pointed out by
Einstein et al.~\cite{einstein35} and the claim in this paper is known as
Einstein--Podolsky--Rosen (EPR) paradox. Its original intention was to
show the paradox of quantum mechanics. If there were such correlation,
the locationally separated particles can provide a mean to convey
information more rapidly than the speed of the light; they claimed it is
a contradiction. In spite of its original intention, their result is now
known to be a fundamental principle of quantum mechanics. Since the
correlation given by the EPR paradox was known to be a break of the
inequality shown by Bell which shows the necessary condition for a given
artificially made probabilistic distribution to be really
realizable. Some of the recent research concerning Bell's inequity is by
Tsirelson~\cite{tsirelson80, tsirelson87, tsirelson93}, Avis et
al.~\cite{avis07}, and Ito~\cite{ito07}

One of the lines of the research concerning entanglement is based on
the rather philosophical question: ``What is communication?''  It is
shown that sharing entangled states helps two parties to win some sort
of games. The researches in quantum games are based on the idea that
whether ones are communicating or not is judged by whether they can do
something they could not do without any share of information. Avis et
al.~\cite{avis06} showed that two parties sharing entangled states can
make a good performance in the graph coloring game compared to the case that
there is no communication. This result indicates that they are surely
communicating something, although it is much weaker than classical
communication. It is called a {\em pseudo-telepathy} because it looks
like a telepathy but can win in only limited games.

Another direction of the research is to evaluate numerically how much the
states are entangled. One direction is to measure some kind of distance
from the maximally entangled state to the state in problem. Some entanglement
measures were proposed by Bennett et al.~\cite{bennett96}. The
generalization of the measures which means investigation for the
condition which the entanglement measures must satisfy is done by
Vidral et al.~\cite{vedral97a,vedral97b,vedral98}, and Rains \cite{rains97}.

``How much do entangled states contribute to the capacity of a quantum
communication channel?'' has been considered as an important
problem. The problem can be described more precisely as follows: does
the Holevo capacity of a given channel make any difference depending on
whether its domain is restricted to separable states (i.e.\ not
entangled states) or not? In a mathematical sense, such a problem is
stated as an ``additivity problem.'' It is conjectured that the
additivity holds for any quantum state space. In other words, it is
believed that entangled states give no power to the quantum channel with
respect to some measures.

Concerning the problem of sending a classical message via a quantum
channel, Holevo showed the upper bound for its
capacity \cite{holevo73, holevo79}. Holevo~\cite{holevo98} and
Schumacher--Westmoreland~\cite{schumacher97}
independently showed theoretically that the upper bound can be attained.

Shor~\cite{shor04a} proved that some open problems concerning the
additivity with respect to some measures are all equivalent. In particular,
they proved the equivalence of the additivity of the Holevo capacity and
the additivity of the minimum entropy output. The additivity of the
minimum entropy output is equivalent to the limit of the multiplicativity of
the $p$-norm as $p\to 1$.

Although the conjectures are not solved completely, it is confirmed to
hold for some classes of channels. About unital channels,
King~\cite{king02a} proved for unital qubit channels; Fujiwara and
Hashizum\'e~\cite{fujiwara02}, King~\cite{king03}, and
Amosov~\cite{amosov04} for depolarizing channel;
Matsumoto~\cite{matsumoto04}, Datta et al.~\cite{datta04a}, and
Alicki~\cite{alicki04} for Werner-Holevo channels; Fannes et
al.~\cite{fannes05} and Datta et al.~\cite{datta04b} for the transpose
depolarizing channel; Datta and Ruskai~\cite{datta05} for some
asymmetric unital channel. About non-unital channels,
Shor~\cite{shor04b} and King~\cite{king02b} proved for
entanglement-breaking channels; Wolf~\cite{wolf05} for a modification of
the Werner-Holevo channel; and King~\cite{king04b} for diagonal
channels.

Although it is also conjectured that the multiplicativity of the
$p$-norm holds, a counterexample for $p> 4.79$ is discovered by Werner
and Holevo \cite{werner02}.  However, it is still believed it holds for
a sufficiently small $p$. King and Ruskai~\cite{king04a} showed a
condition under which the multiplicativity holds if $p=2$. The condition
shown there holds for typical examples, and accordingly, it is a strong
support for the multiplicativity conjecture especially when $p=2$. On
the other hand, almost nothing is known around $p=1$, and thus, the
additivity problem is considered to be extremely difficult.

Those problems related to entanglement also motivated us. Our
computational geometric approach clarifies the structure of an
entanglement measure. Our numerical computation for the Holevo capacity
is also related. If there were a time effective and numerically robust
algorithm to compute the Holevo capacity even for a high level system,
it would be a good tool to check the additivity conjecture. Our
algorithm is a step toward it.

\subsection{Numerical estimation of a quantum channel}
\label{sec:numerical-q-channel} 
Generally, a space of quantum states
has a complicated structure. For $d$-level system, the whole space is
known to be closed convex object in a Euclidean space of real dimension
$d^2-1$. Kimura~\cite{kimura03} showed the list of the inequalities
which formalize the conditions for embedding of a quantum state space
into Euclidean space and it is convincing the complicatedness of the space.
Because of the complicatedness, the capacity of a quantum channel is
difficult to compute although it is important for quantum related
engineering. Whether there is a practically efficient algorithm to
compute the capacity is one of our interests.

\begin{figure}
\subfigure[]{
 \includegraphics[scale=0.5]{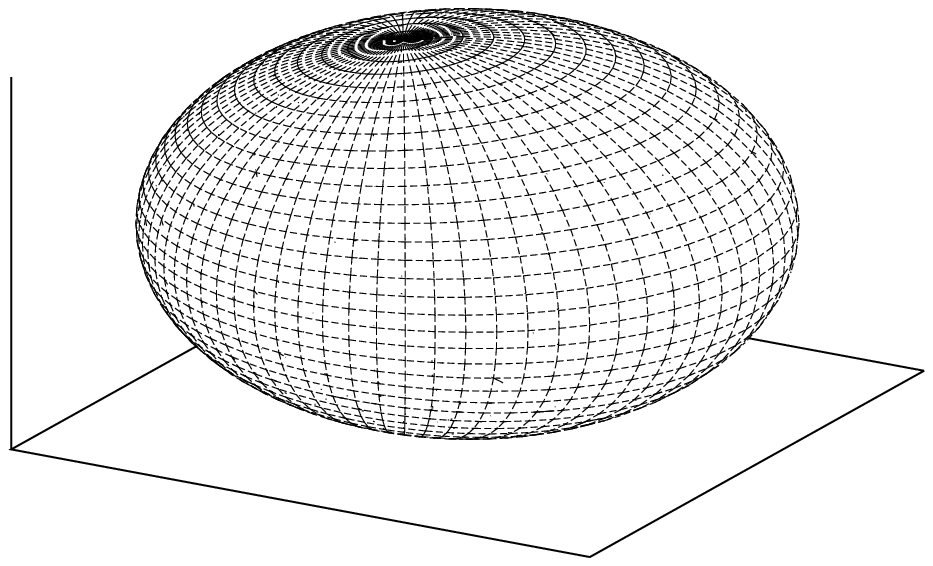} 
\label{subfig:div-sphere1}
}
\subfigure[]{
 \includegraphics[scale=0.5]{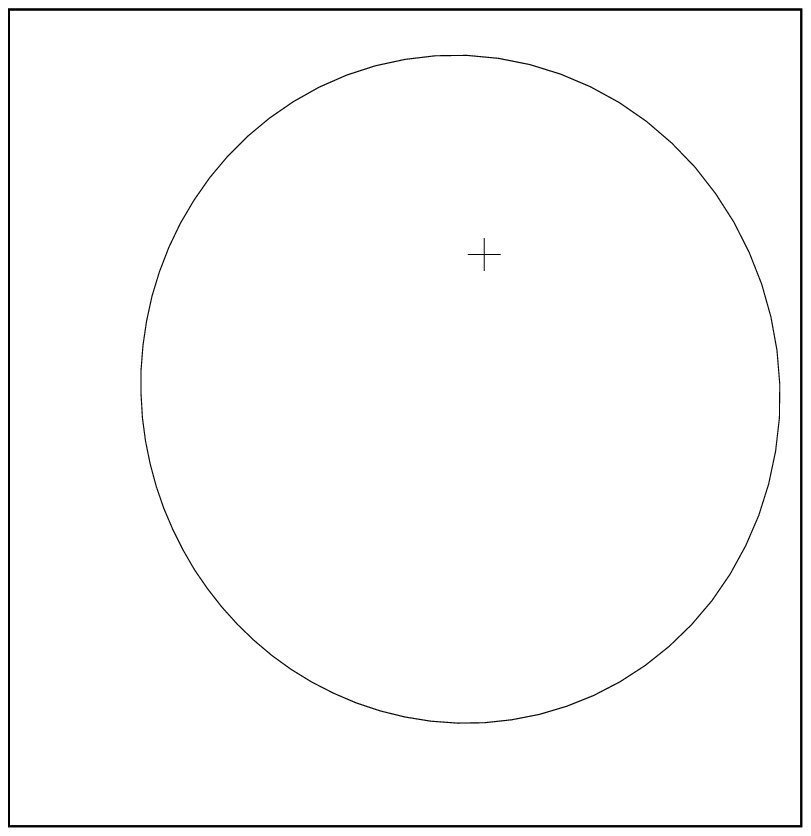} 
\label{subfig:div-sphere2}
}
\caption[An example of a divergence-sphere]{An example of a
 divergence-sphere : 3D view of a divergence-sphere (a) and its section
 by a plane passing through its center (b); the center appears as a
 plus sign.} \label{fig:div-sphere}
\end{figure}

Hayashi et al. \cite{hayashi05} and Oto et al.~\cite{oto04a, oto04b}
showed an effective method to numerically compute the Holevo capacity of
one-qubit quantum channel. With an actual numerical computation, Hayashi
et al.~\cite{hayashi05} showed that there is a case that needs maximal
number of points to determine the smallest enclosing ball; this means
the quantum state space with respect to the divergence is distorted
compared to Euclidean space. Actually Fig.~\ref{fig:div-sphere} shows an
example of a divergence-sphere and we can observe it is really
distorted. About the three level system, Osawa and
Nagaoka~\cite{osawa01} were the first to show an example of numerical
computation. They proposed a quantum version of Arimoto-Blahut algorithm
\cite{arimoto72, blahut72} to numerically compute a Holevo capacity, and
confirmed that the additivity holds for some three-level examples. Our
motivation partially comes from their work. If there is a faster
algorithm to compute the Holevo capacity, although this direction will
never help us to prove it, we can be more convinced with the additivity
conjecture or otherwise, can find a counterexample.

\begin{figure}
\begin{center}
  \includegraphics[angle=-90, scale=0.5]{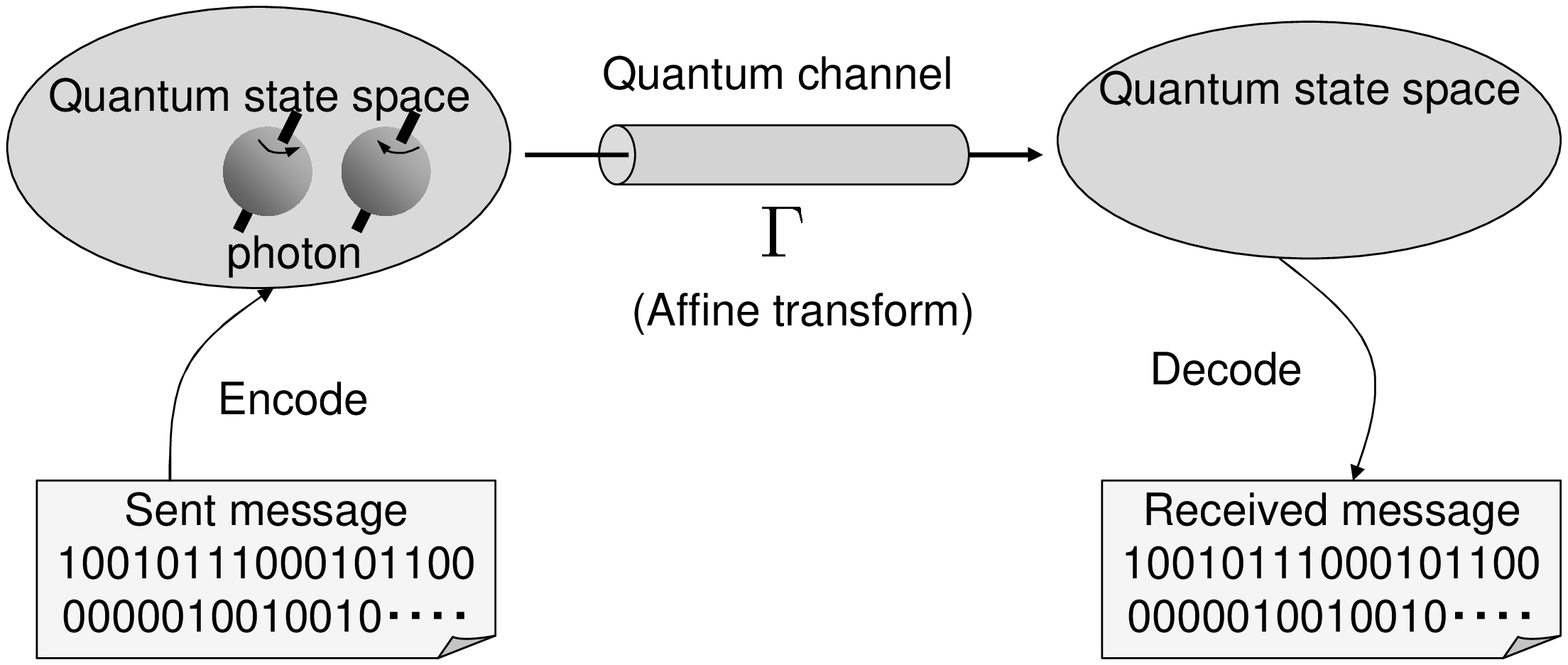}
 \caption{An explanation of the setting for the Holevo capacity}
 \label{fig:holevo-capacity}
\end{center}
\end{figure}

The Holevo capacity is defined as the capacity of a quantum channel when
it sends classical message. Consider the setting that you send a
classical message via a quantum channel and suppose that a probabilistic
distribution of source messages and the way of encoding a message are
varying parameter. The Holevo capacity is the maximum of conveyed
information with those varying parameters. Its concept is illustrated in
Fig.~\ref{fig:holevo-capacity}.  An upper bound for the capacity is
proved by Holevo~\cite{holevo73, holevo79}, and it is proved to be
attained \cite{holevo98,schumacher97}.

The method we introduce in this dissertation is based on the algorithm to
compute the smallest enclosing ball. This is an extension of the method
used by Hayashi et al.~\cite{hayashi05} and Oto et al.~\cite{oto04a,oto04b}. The smallest enclosing ball
problem itself is also important in computational geometry and is still to
be solved from some aspect; even in the Euclidean distance space, a
polynomial time algorithm in the dimension of the space is not known. 

\section{Contribution of this dissertation}\label{sec:contribution} 

\begin{figure}
\begin{center}
  \includegraphics[angle=-90, scale=0.5]{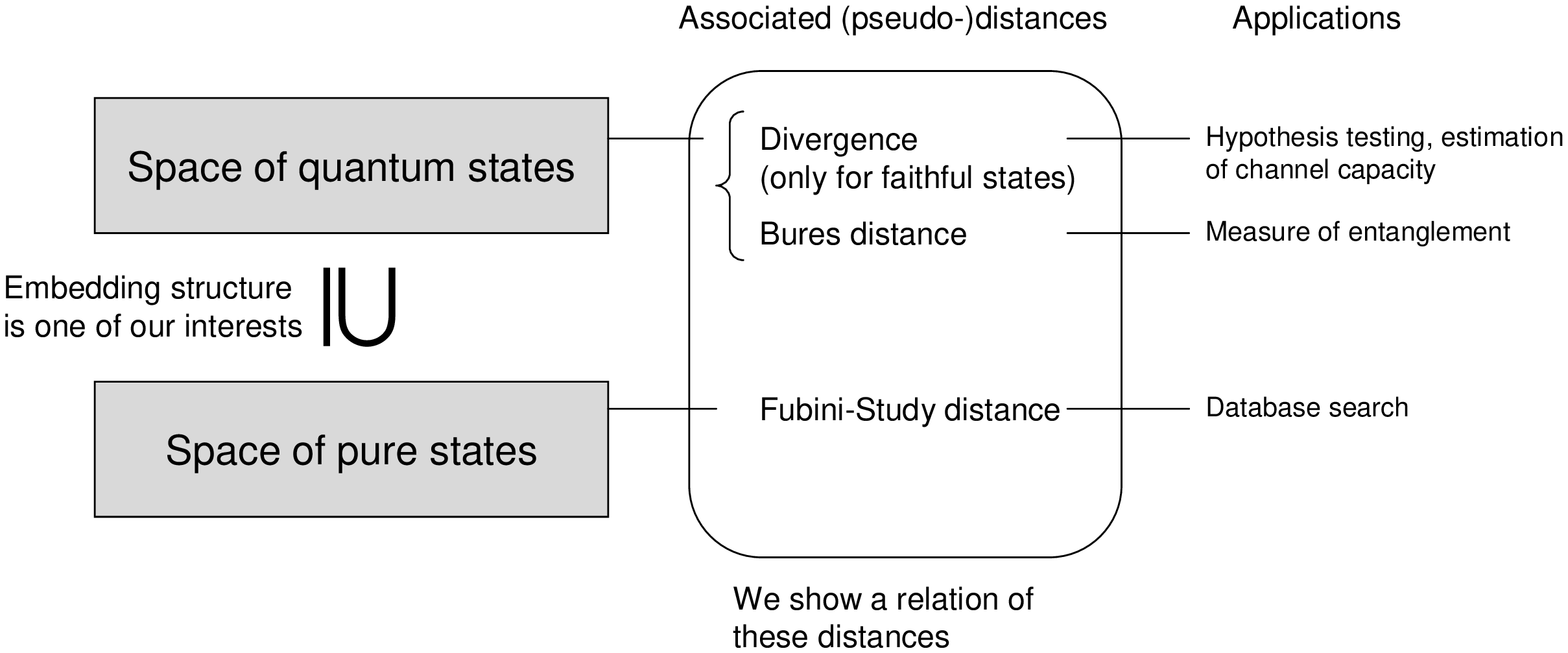}
 \caption{A description of a part of our contribution}
 \label{fig:spaces-distances}
\end{center}
\end{figure}

Our main contribution is a computational geometric interpretation of a
 quantum state space. In particular, we introduce a concept of Voronoi
 diagrams and the smallest enclosing ball problem in a quantum state
 space. Those standard tools in computational geometry help to clarify
 the adjacency structure of a point set in a quantum state space.
 Another aspect of our interpretation is that we
 show some relation between quantum information and quantum
 computation. Actually we show some different distances used in quantum
 information and quantum computation give a same Voronoi diagram. The
 setting for the problems described above and the variety of the
 (pseudo-)distances in a quantum state space are illustrated in
 Fig.~\ref{fig:spaces-distances}. Moreover as an application of such a
 geometric interpretation, we propose an algorithm to compute a capacity
 of a quantum channel, and show, by an experiment, that it is really
 practical.

By proving coincidences of Voronoi diagrams, we show a connection
between some distances which were considered differently. An especially
significant point is that we showed a coincidence of Voronoi diagrams
with respect to the divergence and the Fubini-Study distance. The
divergence is an important measure in quantum information theory, and is
used for quantum hypothesis testing \cite{hiai91, ogawa00} and
estimation of the capacity of a quantum channel \cite{holevo73,
holevo79}, while the Fubini-Study distance gives a convergence path in
Grover's search algorithm \cite{MW01}. We bridge those topics which had
not seemed to be related, but had been considered to be both important
for quantum-related researches. Moreover, we also show a connection
between the Bures distance and the divergence. The Bures distance is
used as a measure of entanglement \cite{schumacher97, hayashi06}.
Although for pure states, the Bures distance is fundamentally the same
thing as the Fubini-Study distance, it is meaningful to know a
connection between the Bures distance and the divergence which are used
in different contexts.

Another interesting point is that the Fubini-Study distance is only for pure
states while the divergence is not defined for pure states. To find the
relation between those exclusive measures, we introduce a natural
definition of Voronoi diagram with respect to the divergence in a space of
pure states. Here, note that although the divergence for pure states is
not defined, the divergence-Voronoi diagram is naturally extended to
a space of pure states by taking a topological closure. A space of pure
states has a simple structure and the Fubini-Study distance is defined
as a very natural distance in it. It is meaningful because the whole
space including pure and mixed states has a complicated structure and
the quantum divergence is a distortion measure in it. Thus we give a
connection between a simple natural structure and a distorted structure.

From the viewpoint of computational geometry, our contribution is that
we introduce Voronoi diagrams in a distortion space and characterize
it. Our main interest is the quantum divergence --- the most distorted one
among the measures defined in quantum state space. A Voronoi diagram
with respect to the quantum divergence is a natural extension of a
diagram with respect to Kullback-Leibler divergence in classical
information theory, and we reveal its geometric properties.

 For pure states in the space of one-qubit
quantum states, we show the coincidence of Voronoi diagrams with respect
to some distances --- the divergence, the Fubini-Study distance, the
Bures distance, the geodesic distance and the Euclidean distance
\cite{kato05,kato07b}. As an
application of this fact, we introduce a method to compute numerically
the Holevo capacity of a quantum channel \cite{oto04a,oto04b,hayashi05}. The
effectiveness of this method is partially based on the coincidence of
the diagrams. Moreover, also as to the diagrams in mixed states, we
found the coincidence of some of them. The diagrams with respect to the
three distances --- the divergence, the Fubini-Study distance, and the
Bures distance --- coincide \cite{kato06b,kato07b}.

A natural question that arises after this story is ``What happens in a
higher level system?'' For a higher level system, the diagrams with
respect to the divergence and the Euclidean distance do not coincide
\cite{kato06a,kato07b}. On the other hand, the diagrams with respect to the
divergence, the Bures distance and the Fubini-Study distance still
coincide for a higher level.

We also show that Welzl's algorithm is also applicable to quantum state
space. Most of the idea of its proof is by Nielsen et
al.~\cite{nielsen07a}, but they did not mention about quantum
divergence. Following their idea, we show that the smallest
enclosing problem in quantum state space with respect to quantum
divergence obey to the axioms of LP-type problem which is essential
condition to show the Welzl's algorithm is effective.

As an application of the theoretical result we proved, we propose an
algorithm to compute the Holevo capacity of a quantum channel, It is a
natural but non-trivial extension of the existing algorithm for one
qubit states. The merit
of our algorithm is robustness of computation as a global
optimization. The algorithm by Osawa and Nagaoka \cite{osawa01} can
converges to a local optimum and might need some iterations of
optimization process. Our algorithm overcomes that
problem. Approximating a continuous object by a point mesh, our
algorithm can compute a global optimum although it only yields an
approximation and its preciseness depends on the fineness of the mesh.

\section{Organization of this dissertation}
\label{sec:organization} 
This dissertation is organized as follows. In
Chapters \ref{chap:comp-geom} and \ref{chap:quant-info}, we explain some
preliminary facts about computational geometry and quantum information
theory respectively. Our contribution is described in Chapters
\ref{chap:one-qubit}, \ref{chap:higher-level}, and
\ref{chap:numerical}. 

In Chapter \ref{chap:one-qubit}, we show explicit correspondence between
primal and dual quantum state space. We show some coincidences of Voronoi
diagram in the space of one-qubit space. The main result of this chapter
is divided into two parts: about the space of pure states and the whole
space including mixed states. This chapter is based on the papers \cite{kato05,kato06b}.

The similar problem in a higher level system is described in Chapter
\ref{chap:higher-level}. Here, we also show the correspondence between
primal and dual quantum state space. The correspondence is less explicit
than one-qubit case but mathematically proven as for one-qubit. About
the coincidence of Voronoi diagrams, because of its complicatedness, the
space of pure states is only analyzed here. This chapter is based on the
papers \cite{kato06a, kato06b}.

In Chapter \ref{chap:numerical}, we propose an
algorithm to compute the Holevo capacity of a quantum channel and
experiment it to show it is really useful. It is also proved that
Welzl's algorithm to compute the smallest enclosing ball problem is
applicable to a quantum state space. The idea of those algorithms
was mentioned in \cite{kato06b}.

The summary of all our
contribution is described in Chapter \ref{chap:conclusion}, and we also
explain a perspective of the future research.


\chapter{Computational Geometry}
\label{chap:comp-geom}

\section{Voronoi diagrams and Delaunay triangulations}
First we start with the abstract notation of Voronoi diagrams.

\begin{definition}[Voronoi diagram]
 For a given tuple $(X,d,P)$ where $X$
 is a metric space, $d$ is a distance attached to $X$, and
 $P=\{p_i\}_{i=1}^N$ is a set of points of $X$, the {\em Voronoi
 diagram} $V$ is defined as
 \begin{align}
  V&=\Bigl\{ V^{(i)} \Bigr\}_{i=1}^N\\
 V^{(i)}&=\Bigl\{ x \in X \Bigm| d(x,p_i)\leq d(x,p_j) \text{ for any }
  j \Bigr\}, 
 \end{align}
and each $p_i$ is called a {\em site} ({\em generator}, {\em Voronoi vertex}).

When it is necessary to make a distance associated to the diagram clear,
 we denote $V$ by $V_d$.
\end{definition}

In the definition above, the space $S$ and the distance $d$ can be
arbitrary. The Voronoi diagram most commonly used is a Euclidean Voronoi
diagram. It is defined as follows.
\begin{definition}[Euclidean Voronoi diagram]
For a set of sites $P\{p_i\}_{i=1}^N\subset \mathbb{R}^n$, the Euclidean
 Voronoi diagram is defined as
 \begin{align}
  V&=\Bigl\{ V^{(i)} \Bigr\}_{i=1}^N\\
 V^{(i)}&=\Bigl\{ x \in \mathbb{R}^n \Bigm| |x-p_i|\leq |x-p_j| \text{ for any }
  j \Bigr\}, 
 \end{align}
\end{definition}

Just by saying ``Voronoi diagram,'' we usually mean the Euclidean Voronoi
diagram. Intuitively a Voronoi diagram is a diagram of dominance
of sites in the terms of the distance. In other words, it is the
coloring of the space according to which site is the nearest. The
following example is the application actually wanted by the author. 
\begin{example}[Nearest station problem]
 There are four subway stations near Hongo campus of the University of
 Tokyo. The campus is so large that if you choose a wrong station to
 access, it takes unnecessarily a long time to walk. 
If the map of campus were colored as a Voronoi diagram regarding
 stations as sites, you can find the nearest station easily by just telling
 the color of your current location.
\end{example}

The region dominated by each site is called a {\em Voronoi region} ({\em
Voronoi polygon}). The
edge appears in a boundary of a Voronoi region is called a {\em Voronoi
edge}. As a practical implementation of a Voronoi diagram, just knowing
the boundary of each region is enough. Each Voronoi edge is a part of
{\em bisector line} (or in general case {\em bisector curve}) of a
certain pair of sites. Thus, computing a Voronoi diagram can be just
described as deciding which part of bisector curve appears in a
diagram. Here, note that some bisector curves do not appear at all.

The dual diagram for a Voronoi diagram is called a {\em Delaunay
pretriangulation} ({\em Delaunay tessellation}). The Delaunay
pretriangulation and triangulation of a Euclidean space is defined as
follows.
\begin{definition}[Delaunay pretriangulation and triangulation]
For a given set of sites $P=\{p_i \in \mathbb{R}^n \}_{i=1}^N$, the Delaunay
 triangulation $D$ is defined as
\begin{multline}
  D=\Bigl\{ e \Bigm| e \text{ is a line segment between } p_i \text{ and
  } p_j \text{, where the bisector of } (p_i,p_j) \\
 \text{ appears as
  an edge of Voronoi diagram of } P\Bigr\}
\end{multline}

If Delaunay pretriangulation is a triangulation, i.e.\ if for any $x\in X$,
 there is a triple of line segments $(s_1,s_2,s_3)\;(s_i \in D)$ which
 formulate a triangle and it has $x$ as a inner point, then $D$ is
 called {\em Delaunay triangulation.}

If $D$ is not a triangulation, the triangulation
 made by joining the vertices of non-triangle polygon of $D$ is called
 {\em Delaunay triangulation}

\end{definition}

In other words, Delaunay pretriangulation is a diagram obtained by
connecting by a line each pair of sites whose Voronoi region is
adjacent. Although definition above is only for a Euclidean space, but
it can be extended for a general space; for a general space, just take a
segment of a geodesic curve instead of a line segment. 

\section{Computation of Voronoi diagrams}
\label{sec:computation-vd}
There are some algorithms known for construction of Voronoi
diagrams. Here, we introduce the {\em incremental method,} which is an
intuitive algorithm for two-dimensinal Euclidean Voronoi diagram. We only
explain the outline of the algorithm to show the main idea. For the
detail of the algorithm, refer to \cite{okabe00}. Note that the internal
data structure to represent the Voronoi diagram is not trivial at all,
but we skip it.

\begin{figure}
 \label{fig:alg-incremental}
 \includegraphics{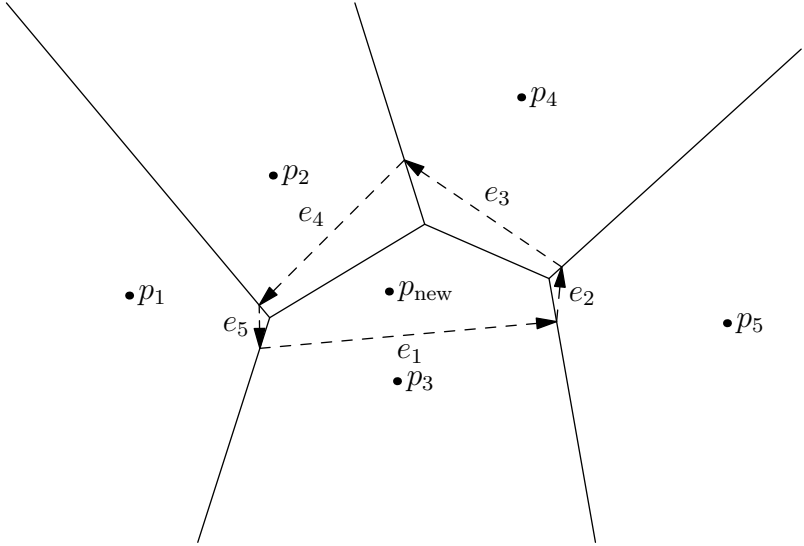}
 \caption[An explanation of the incremental algorithm to draw a Voronoi diagram]
 {An explanation of the incremental algorithm to draw a Voronoi diagram:
 This shows the situation when $p_\mathrm{new}$ is added to an existing
 digram formed by the sites $p_1,\ldots,p_5$}
\end{figure}

The incremental method to construct a Voronoi diagram is described as
follows. Suppose that a set of point $P=\left\{ p_1,\ldots,p_N\right\}$
is given, we construct a Voronoi diagram $V_i$ of the point set $\left\{
p_1,\ldots,p_i \right\}$ step by step and finally obtain the required
diagram $V=V_N$. Suppose that we are going to add a point $p$ as in
Fig.~\ref{fig:alg-incremental}. The rough sketch of the process to add
this point is described as follows:
\begin{enumerate}
 \item Find a region which the new point $p_\mathrm{new}$ belongs to
 \item \label{enu:loop}Draw a bisector between the new point $p$ and the site which
       dominate the region found
 \item Find neighboring region to the current region
 \item Go back to \ref{enu:loop} until it comes back to the original region
\end{enumerate}

In Fig.~\ref{fig:alg-incremental}, the process above is explained as
follow. First the region by the $p_3$ is found and draw a bisector
between $p_1$ and $p_\mathrm{new}$. The intersection of the bisector and
the $p_3$'s region appears as a new Voronoi edge $e_1$. Then find a point the
bisector and the edge of the region meet, and go into the neighboring
region, the $p_5$'s region. Now $e_2$ is drawn as the intersection of
the bisector of $p_\mathrm{new}$, $p_5$ and the $p_5$'s region. The same
process is iterated until it get to the original region, and $e_3$,
$e_4$ and $e_5$ are drawn.

In Fig.~\ref{fig:alg-incremental}, mathematical rigid computation will
guarantee that the start point of $e_1$ and the end point of $e_5$ is
exactly the same. However, in a computer, coordinates are usually
expressed in floating point number, and have some numerical error. It
means that in Fig.~\ref{fig:alg-incremental}, the start point of $e_1$
and the end point of $e_5$ might be slightly different. In usual case as
Fig~\ref{fig:alg-incremental}, it is not a problem at all because the
algorithm only have to maintain the neighboring structure of regions.  

\begin{figure}
 \begin{center}
  \includegraphics[clip,scale=0.8]{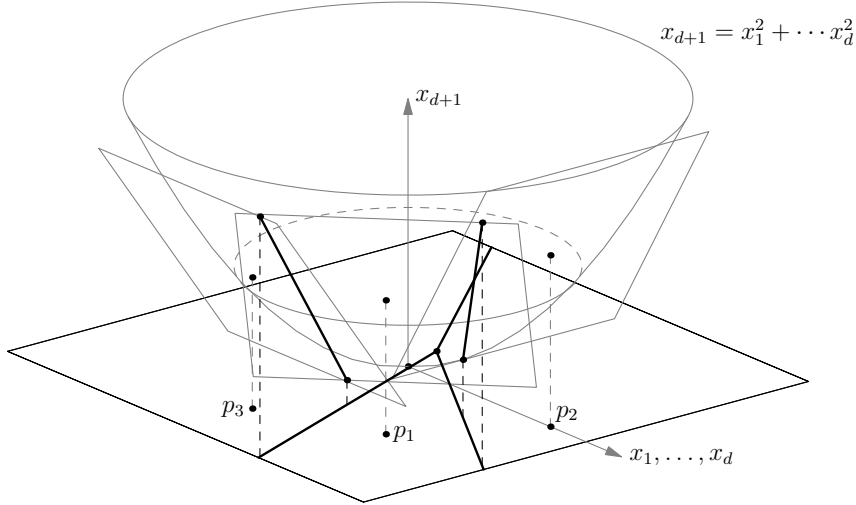}
 \end{center}
 \caption{An exapmle of a Euclidean Voronoi diagram determined by a lower envelope}
 \label{fig:lower-envelope}
\end{figure}

For a general dimensional space, a Voronoi diagram can be computed via a
lower envelope. Actually, a Voronoi diagram in $d$-dimensional space can
obtained as follows. Consider a paraboloid in $d+1$-dimensional space
expressed by $x_{d+1}=x_1^2+\cdots +x_d^2$ and tangent planes at the
points which are obtained by lift-up of sites. Then, the lower envelope
of the tangent planes is a Voronoi diagram
(Fig.~\ref{fig:lower-envelope}). Here, a lower envelope means the lowest
part of a given set of surfaces, and its computation is that of convex
hull of a polytope. Thus, the computational complexity of a Voronoi
diagram in $d$-dimensional space is the same as that of a convex hull in
$d+1$-dimensional space.

In a $d$-dimensional space, the complexity for computation of a convex
hull has been proven to be $O(n\log n + n^{\lfloor d/2 \rfloor})$ by
several different algorithms
\cite{preparata85,seidel86,edelsbrunner87,buckley88}. Consequently, the
complexity of $d$-dimensional Voronoi diagram is $O(n \log n + n^{\lfloor
(d+1)/2 \rfloor})$. Note that although this is a polynomial for a fixed
dimension $d$, it is exponential about $d$.

For a general distance function, a Voronoi diagram can also be
considered as a projection of a lower envelope of some potential
function. Halperin and Sharir \cite{halperin93,halperin94} showed that
in three dimensional space, a lower envelope of algebraic surfaces can
be computed in $O(n^{2+\varepsilon})$-time for any small
$\varepsilon$. Sharir \cite{sharir94} extend its result to a
$d$-dimensional space and showed it can be computed in
$O(n^{d-1+\varepsilon})$.  This means a Voronoi diagram in $d$
dimensional space whose edges are expressed by algebraic equation can be
computed in $O(n^{d+\varepsilon})$. Another non-Euclidean specific
distance is analyzed by Icking and Ma \cite{icking01}.
\section{Smallest enclosing ball problem}

The smallest enclosing ball problem is described as follows: for a
given set of points $P$, compute the smallest ball which includes all
the points of $P$.  The first practical algorithm for the smallest
enclosing ball problem is
given by Welzl \cite{welzl91}. The extension of it to the {\it smallest
enclosing ball of balls} (SEBB) problem is shown by Fischer and
G\"artner \cite{fischer04}, and its implementation is freely available
as a part of CGAL package \cite{cgal}. However, SEBB solver in CGAL only
works for Euclidean distance, and it does not fit our objective.

The following is Welzl's algorithm to compute the smallest enclosing
ball.
\begin{algorithm} (Welzl \cite{welzl91})
\label{alg:welzl}
 \begin{algorithmic}
  \Procedure{minball}{$P$ : set of points}\Comment{Compute the smallest enclosing ball}
  \State {\sc b\_minball($P$, $\emptyset$)}
  \EndProcedure
  \Procedure{b\_minball}{$P$,$R$}\Comment{Compute the ball which includes $P$
  and has $R$ in its boundary}
  \If{$P=\emptyset$ or $R=d+1$ (where $d$ is a dimension of the space)}
   \State \textbf{return} the ball which has $R$ on its boundary
  \Else
   \State Choose $p\in P$
   \State $B \gets \text{\sc b\_minball}(P-\left\{p\right\}, R)$
   \If{$p\not\in B$}
    \State $B \gets \text{\sc b\_minball}{P-\left\{p\right\}, R \cup
  \{p\}}$
   \EndIf
   \State \textbf{return} $B$
  \EndIf
  \EndProcedure
 \end{algorithmic}
\end{algorithm}

The function {\sc b\_minball} is the main part of this
algorithm. $\text{{\sc b\_minball}}(P,R)$) computes the smallest ball
that includes $P$ under the constraint that all the points or $R$ must be
on its boundary. This algorithm is based on the idea that more
constraints on the boundary make the computation easier. So, the chosen
point $p$ is not included in the current optimal ball, it tries to a new
bigger optimal ball which has $p$ on its boundary. It works because of
the following lemma.
\begin{lemma}
 For point sets $P$ and $R$, let $\mathrm{SEB}(P,R)$ be the smallest
 ball which includes $P$ and has $R$ on its boundary. Then, for any
 $p\in P$
\begin{equation}
 \mathrm{SEB}(P,R)=\mathrm{SEB}(P-\left\{p\right\}, R\cup \left\{p\right\}),
\end{equation}
\end{lemma}
\begin{proof}
 See \cite{welzl91}.
\end{proof}

Another important point of this algorithm is the first ``if'' part. Note
that if $\#R = d+1$, the optimal ball is uniquely determined. If the
function is called when $P\neq \emptyset$ and $\#R = d+1$, it simply
returns the ball which has $R$ on its boundary. It returns a wrong
answer if some point of $P$ is not included in the ball constructed by
$R$. In that case, to keep the consistency of the specification of this
function, it is correct to return ``undefined'' because there is no ball
which includes $P$ and has $R$ on its boundary. However, our main
objective is implement {\sc minball} and this exceptional behavior of
{\sc b\_minball} makes it easier. 

Suppose that ``or $|R|=d+1$'' of the first ``if'' condition of Algorithm
\ref{alg:welzl} is omitted, and the second ``if'' condition is replaced
by ``$B$ is defined and $p\not\in B$.'' Denote this different version of
function as {\sc b\_minball}$'$. {\sc b\_minball}$'$ has a
consistent specification as itself, but returns the same value if called
by {\sc minball}. If {\sc b\_minball} is called from {\sc minball} with
$\#R=d+1$ and $P=\left\{p\right\},\;p\not\in R$, then the condition for
second ``if'' becomes ``true'' and {\sc b\_minball} is called with
$\#R\cup \left\{p\right\}$. $\#R$ is already maximum and so, the
returned value becomes ``undefined.'' If once a returned value is
``undefined'' in the depth of the call of {\sc b\_minball}, the returned
value of the top level also becomes ``undefined.'' It is easily checked
by induction. However, the value of {\sc minball} is certainly defined,
so it is a contradiction.

By the observation above, we can say it is the same thing whether it
checks if $\#R=d+1$ in the first ``if'' or it checks if $D$ is defined
in the second ``if.'' However, for a practical performance, the original
algorithm is better although it doesn't affect the order of
computational complexity.

Welzl showed this algorithm ends in an expected $O(n)$ time if the
dimension is fixed, where $n$ is the number of the given points (i.e.\
$n=\#P$). Its effectiveness is shortly based on the low probability of
the recomputation of an optimum; the probability for the two
if-conditions to be true is very low. For the detail of the
probabilistic analysis, see \cite{welzl91}.

Although Welzl's original algorithm is only for the Euclidean distance,
Nielsen \cite{nielsen07a} showed this algorithm is also applicable for
Bregman divergence, which is a pseudo-distance used in classical
information theory. Whether it is also applicable for the quantum
divergence is important for numerical computation of Holevo capacity,
and is proved to be true as a mostly straightforward corollary of
Nielsen's result. The detail of its proof is given in Section
\ref{sec:seb-qdiv}.

\chapter{Quantum Infomation Theory}
\label{chap:quant-info}

\section{Quantum states and their parameterization}
A pure state can be expressed as a state vector. A state vector in a
$d$-level system is defined as
\begin{equation}
 \bra{\phi}=\sum_{i=1}^{d} \alpha_i \bra{i}, \quad \sum_{i=1}^{d} |\alpha_i|^2=1
\end{equation}
where $\bra{\cdot}$ is Dirac's braket notation and means
mathematically a complex vector in $\mathbb{C}^{d}$. The vector
$\bra{i}$ means the $i$-th element of the orthogonal basis of $d$
dimensional complex vector space. Additionally the definition of a state
vector is up to a scalar multiplication, i.e.\ $\bra{\phi}=\sum_{i}\alpha_i$
and $\bra{\psi}=\sum_{i}\beta_i$ are equivalent when there exists a
scalar $\gamma\in \mathbb{C}^{d}$ such that $\alpha_i=\gamma \beta_i$
for all $i$.

A density matrix is representation of some probabilistic distribution of
states of particles. Mathematically it is defined as follows.

\begin{definition}
\label{def:density-matrix}
 A {\em density matrix} $\rho$ is a complex square matrix which
 satisfies the following three conditions:
\begin{itemize}
 \item[a)] Hermitian, i.e.\ $\rho=\rho^*$,
 \item[b)] The trace is one,
 \item[c)] It is positive semi-definite.
\end{itemize}
Moreover, we denote the space of
all density matrices of size $d\times d$ by $\mathcal{S}(\mathbb{C}^d)$,
and we called it a {\em $d$-level system.} Especially when $d=2^n$, a $d$-level
 system is also called an {\em $n$-qubit system.} When $d$ is obvious, we denote
$\mathcal{S}(\mathbb{C}^d)$ by $\mathcal{S}$.
\end{definition}

A density matrix can express both a pure and mixed states. The state
vector $\bra{\phi}$ correspond to $\ket{\phi}\bra{\phi}$ as a density
matrix. Here $\ket{\phi}$ means a Hermitian conjugate of $\bra{\phi}$
as in the convention of Dirac's braket notation. A mixed state is a
state which is not pure. Namely a mixed state corresponds to a state
which is mixture of multiple states. Actually a density matrix is
expressed as:
\begin{equation}
 \rho=\sum_{i=1}^d a_i\ket{i}\bra{i}\quad a_i\geq 0,a_i\in \mathbb{R},
\end{equation}
and the condition for $\rho$ to be mixed is equivalent to that at least
two of $a_i$ are non-zero.

We give another mathematically simple definition for pureness and
mixedness and also for faithfulness.
\begin{definition}
 A density matrix $\rho$ is called {\em pure} if
 $\mathrm{rank}\;\rho=1$, {\em mixed} if it is not pure, i.e.\
 $\mathrm{rank}\; \rho>1$, and {\em faithful} if $\mathrm{rank}\; \rho
 =\mathrm{dim}\; \mathcal{S}$.
\end{definition}
We also use a notation for subspace of $\mathcal{S}$ as follows.
\begin{definition}
 For a given quantum state space $\mathcal{S}$, denote
 $\mathcal{S}^\mathrm{pure}$ by
\begin{equation}
 \mathcal{S}^\mathrm{pure} = \left\{ \rho \mid \rho\in \mathcal{S},\;
			     \rho \text{ is pure} \right\}.
\end{equation}
 $\mathcal{S}^\mathrm{faithful}$
and  $\mathcal{S}^\mathrm{nonfaithful}$ are defined similarly.
(Note that especially for one-qubit system,
$\mathcal{S}^\mathrm{nonfaithful}=\mathcal{S}^\mathrm{pure}$)
\end{definition}

\begin{figure}
\begin{center}
  \includegraphics[scale=1.2]{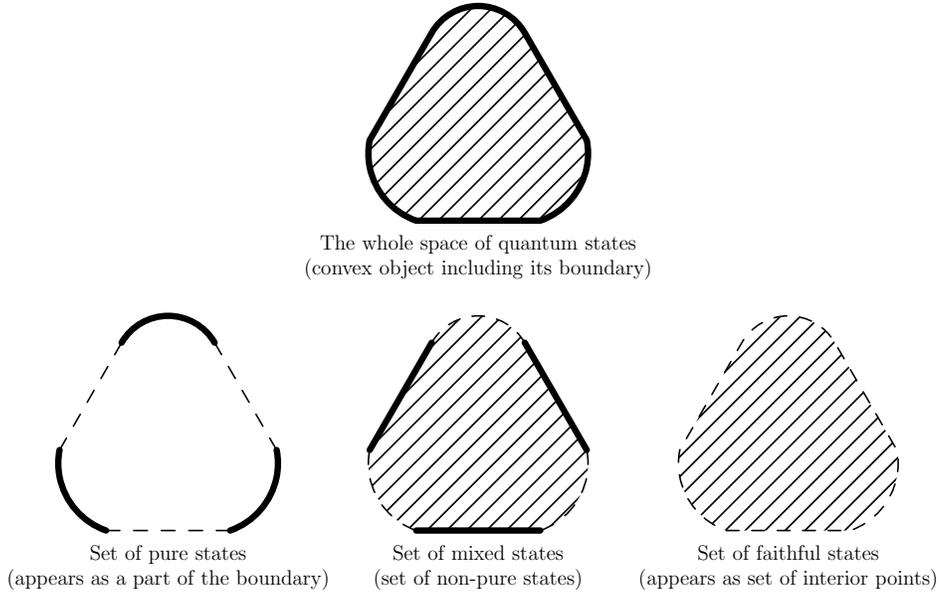}
\end{center}
 \caption{An explanation of pureness, mixedness, and faithfulness}
\label{fig:pure-mixed}
\end{figure}

Note that while a quantum state is either pure or mixed, any faithful
state is mixed. The geometric image of pureness, mixedness, and
faithfulness is illustrated in Fig.~\ref{fig:pure-mixed}.

Especially in two-level system, which is often called {\em one-qubit
system}, the conditions a), b) and c) in Definition
\ref{def:density-matrix} are equivalently expressed as 
\begin{align}
 \label{eq:one-qubit-rho}
 \lefteqn{ \rho= \left(
 \begin{array}{cc}
 \displaystyle \frac{1+z}{2}& \displaystyle \frac{x-iy}{2} \bigskip\\
 \displaystyle \frac{x+iy}{2}& \displaystyle \frac{1-z}{2}
 \end{array}
 \right),}\nonumber\\
 &x^2+y^2+z^2 \leq 1,\quad x,y,z\in \mathbb{R}.
\end{align}
This is called a Bloch ball because it is a ball in the $xyz$ coordinate system.
The parameterized matrix corresponds to the conditions a) and b), and the
inequality corresponds to the condition c).

There have been some attempts to extend this Bloch ball expression to a
higher level system. A matrix which satisfies only first two conditions,
Hermitianness and unity of its trace, is expressed as:
\begin{align} \label{multi-level-rho}
\lefteqn{\rho=}\nonumber\\ 
&
\begin{pmatrix}
 \displaystyle\frac{\xi_1+1}{d} &
 \!\!\!\displaystyle\frac{\xi_d-i\xi_{d+1}}{2} &
 \cdots & &   
 \displaystyle\frac{\xi_{3d-4}-i\xi_{3d-3}}{2} 
 \smallskip\\
 \displaystyle\frac{\xi_d+i\xi_{d+1}}{2} & 
 \!\!\!\displaystyle\frac{\xi_2+1}{d} &
 \cdots & & 
 \displaystyle\frac{\xi_{5d-8}-i\xi_{5d-7}}{2} 
 \smallskip\\
 \vdots & & \!\!\!\!\ddots & & \vdots \\
 \displaystyle\frac{\xi_{3d-6}+i\xi_{3d-5}}{2} & \cdots& &
 \!\!\!\!\!\!\!\displaystyle\frac{\xi_{d-1}+1}{d}& 
 \displaystyle\frac{\xi_{d^2-2}-i\xi_{d^2-1}}{2}\smallskip\\
 \displaystyle\frac{\xi_{3d-4}+i\xi_{3d-3}}{2} & \cdots& &
 \!\!\!\!\!\!\!\displaystyle\frac{\xi_{d^2-2}+i\xi_{d^2-1}}{2} &
 \displaystyle
 \frac{-\sum_{i=1}^{d-1}\xi_i+1}{d}
\end{pmatrix}
,\nonumber\\
&\quad \xi_i\in \mathbb{R}.
\end{align}
Actually, any matrix which is Hermitian and whose trace is one is
expressed this way with some adequate $\{\xi_i\}$.  This condition
doesn't contain a consideration for a semi-positivity. To add the
condition for a semi-positivity, it is not simple as in one-qubit case,
and we have to consider complicated
inequalities \cite{byrd03,kimura03}. Note that this is not the only way
to parameterize all the density matrices, but it is reasonably natural
way because it is natural extension of one-qubit case and has a special
symmetry.

Additionally our interest is a pure state. A pure state is expressed by
a density matrix whose rank is one. A density matrix which is not pure
is called a mixed state. A pure state has a special meaning in quantum
information theory and also has a geometrically special meaning because
it is on the boundary of the convex object. In one-qubit case, the
condition for $\rho$ to be pure is
\begin{equation}
  x^2+y^2+z^2=1.
\end{equation}
This is a surface of a Bloch ball. On the other hand, in general case,
the condition for pureness is again expressed by complicated
inequalities.  

\section{Distances}

Before explaining about distances for quantum states, we prepare a
mathematical notation.
\begin{definition}
  Suppose that the matrix $\rho$ is diagonalized as
 \begin{equation}
  \rho= U
 \begin{pmatrix}
  \lambda_1& & & \\
 &\lambda_2&&\\
 &&\ddots&\\
 &&&\lambda_d
 \end{pmatrix}
 U^*.
 \end{equation}
 with a unitary matrix $U$.
 For a given function $f:\;\mathbb{R}\to \mathbb{R}$, taking $f$ of the
 matrix $\rho$ is defined as
 \begin{equation}
 f(\rho)= U
 \begin{pmatrix}
  f(\lambda_1)& & & \\
 &f(\lambda_2)&&\\
 &&\ddots&\\
 &&&f(\lambda_d)
 \end{pmatrix}
 U^*.
 \end{equation}
 Especially we define
 \begin{equation}
  \sqrt{\rho}= U
 \begin{pmatrix}
  \sqrt{\lambda_1}& & & \\
 &\sqrt{\lambda_2}&&\\
 &&\ddots&\\
 &&&\sqrt{\lambda_d}
 \end{pmatrix}
 U^*, 
 \end{equation}
 and
 \begin{equation}
  \log\rho= U
 \begin{pmatrix}
  \log\lambda_1& & & \\
 &\log\lambda_2&&\\
 &&\ddots&\\
 &&&\log\lambda_d
 \end{pmatrix}
 U^*. 
 \end{equation}
\end{definition}

Now we define two distances.

\begin{definition}[See \cite{hayashi98}]
 For two pure states $\rho$ and $\sigma$, the Fubini-Study distance 
 $d_{\rm FS}(\rho,\sigma)$ is defined as
 \begin{equation}
 \cos d_{\rm FS}(\rho,\sigma)=\sqrt{\Tr(\rho\sigma)},
 \qquad 0\leq d_{\rm FS}(\rho,\sigma)\leq\frac{\pi}{2}.
 \end{equation}
\end{definition}

\begin{definition}
 For two arbitrary quantum states (i.e.\ mixed or pure states) $\rho$ and $\sigma$, the Bures distance $d_{\rm B}(\rho,\sigma)$
 \cite{bures69} is defined as
\begin{equation}
  d_{\rm B}(\rho,\sigma)=\sqrt{1-\Tr\sqrt{\sqrt{\sigma}\rho\sqrt{\sigma}}}.
\end{equation} 
\end{definition}

Especially if $\rho$ and $\sigma$ are pure states and expressed as
$\ket{\varphi}$ and $\ket{\psi}$ respectively, the Bures distance is as follows:
\begin{align}
   d_{\rm B}(\rho,\sigma)&=\sqrt{1-\Tr(\rho\sigma)}\\
 &=\sqrt{1-\left|\braket{\rho | \sigma}\right|}
\end{align}

This means the Bures distance and the Fubini-Study distance are
fundamentally the same thing for pure states. Moreover, since a space of
pure states is defined as a unit ball divided by a multiplication, those
distances are both natural and have a Euclidean-like property.

\section{Divergence}

In a classical context, Kullback-Leibler divergence means, in a sense, a
``distance'' from a probabilistic distributions to another. When two probabilistic
distributions $p_i$ and $q_i$ are given, the Kullback-Leibler divergence is defined as:
\begin{equation}
 D_{\mathrm{KL}}(p||q)=\sum_{i} p_i \log \frac{p_i}{q_i}
\end{equation}

The quantum divergence is the quantum version of Kullback-Leibler divergence. Just
like Kullback-Leibler divergence has an important role in classical information theory,
the quantum divergence is essential in quantum information theory. It is
defined by a similar formula.
\begin{definition}
 Suppose that two quantum states $\rho$ and $\sigma$ are given and
 $\sigma$ is faithful. The
 quantum divergence is defined as
\begin{equation}
  D(\sigma||\rho) = \Tr \sigma (\log \sigma - \log \rho) .
\end{equation}
\end{definition}
Note that though this has some distance-like properties, it is not
commutative, i.e.~$D(\sigma||\rho)\neq D(\rho||\sigma)$. Also note that
$\rho$ does not necessarily need to be faithful because $0 \log 0$ can
be naturally defined as $0$.

\section{Quantum channel and its capacity}
A quantum channel is the linear transform that maps quantum states to
quantum states. In other words, a linear transform
$\Gamma:M(\mathbb{C};d)\to M(\mathbb{C};d)$ is a quantum channel if
$\Gamma(\mathcal{S}(\mathbb{C}^d))\subset \mathcal{S}(\mathbb{C}^d)$. To
preserve the condition for density matrix, there is a natural
restriction for a quantum channel. A quantum channel $\Gamma$ satisfies
following condition:
\begin{enumerate}
 \item it must be trace-preserving, i.e.\ $\Tr\Gamma(\rho)=\Tr\rho$, and
 \item it must be completely positive, i.e.\ For any identity map $I$,
       the map $\Gamma\otimes I$ maps a semi-positive Hermitian matrix
       into a semi-positive Hermitian matrix.
\end{enumerate}
Such a map can be shortly denoted by a ``TPCP map.'' In other words, the
condition for a linear transform to be a quantum channel is to be a TPCP
map.

The Holevo capacity is considered as a classical information capacity
of a given quantum channel under the consumption that the input state is
not entangled and the output particle is properly measured. 
\begin{definition}[Holevo capacity \cite{holevo98}]
The Holevo capacity of a given channel $\Gamma$ is defined as follows:
\begin{equation}
 C(\Gamma) = \max_{p_1,\ldots,p_n,\rho_1,\ldots,\rho_n} 
S(\sum_{i=1}^n p_i \Gamma(\rho_i) )+
\sum_{i=1}^n p_i S(\Gamma(\rho_i) ),
\end{equation}
where $(\cdot)$ means von Neumann entropy, i.e.\ $S(\rho)=-\rho \log \rho$.
\end{definition}

Another formulation of the Holevo capacity is given by the following
theorem.
\begin{theorem}[Ohya, Petz, Watanabe \cite{ohya97}]
 \begin{equation}
   C(\Gamma)= \min_{\sigma\in \mathcal{S}(\mathbb{C}^d)} 
 \max_{\rho\in \mathcal{S}(\mathbb{C}^d)} D(\Gamma(\rho)||\Gamma(\sigma)).
 \end{equation}
\end{theorem}
This theorem means that the Holevo capacity is equal to the radius of
the smallest enclosing ball with respect to the quantum divergence. From
now on, we mainly use this smallest-enclosing-ball formulation.

\section{Calculation of Holevo capacity}\label{sec:holevo}
Our first motivation to investigate a Voronoi diagram in quantum states
is the numerical calculation of the Holevo capacity for one-qubit
quantum states \cite{oto04a}. We explain its method in this section.  In
order to calculate the Holevo capacity, some points are plotted in the
source of channel, and it is assumed that just thinking of the images of
plotted points is enough for approximation. Actually, the Holevo
capacity is reasonably approximated taking the smallest enclosing ball
of the images of the points.  More precisely, the procedure for the
approximation is the following:
\begin{enumerate}
 \item Plot equally distributed points on the Bloch ball which is the
       source of the channel in problem.
 \item Map all the plotted points by the channel.
 \item Compute the smallest enclosing ball of the image with respect to
       the divergence. Its radius is the Holevo capacity.
\end{enumerate}
In this procedure, Step 3 uses a farthest Voronoi diagram. That is the
essential part to make this algorithm effective because Voronoi diagram
is the known fastest tool to seek a center of a smallest enclosing ball
of points.

However, when you think about the effectiveness of this algorithm, there might
arise a question about its reasonableness. Since the Euclidean distance and the
divergence are completely different, Euclideanly uniform points are
not necessarily uniform with respect to the divergence. We gave partial
answer to that problem by Theorem \ref{th:coincidence}. At least, on the
surface of the Bloch ball, the coincidence of Voronoi diagrams implies
that the uniformness of points with respect to Euclidean distance is
equivalent to the uniformness with respect to the divergence.

\section{Entanglement and additivity problem}
The additivity of Quantum channel is simply stated as follows.

\begin{conjecture}\label{conj:additivity-holevo}
  For any two channels $\Gamma_1:\mathcal{S}_1 \to \mathcal{S}_1$ and
 $\Gamma_2:\mathcal{S}_2 \to \mathcal{S}_2$, it is conjectured that the
  following equation holds.
\begin{equation}\label{eq:additivity-holevo}
 C(\Gamma_1 \otimes \Gamma_2) = C(\Gamma_1) + C(\Gamma_2)
\end{equation}
\end{conjecture}

The right hand side of Equation \ref{eq:additivity-holevo} means the
total capacity when two channels $\Gamma_1$ and $\Gamma_2$ are used
separatedly. The difference between $\mathcal{S}_1 \otimes
\mathcal{S}_2$ and $\mathcal{S}_1 \times \mathcal{S}_2$ is entangled
states; more precisely, for an entangled state is an element
of the set $E$ defined as:
\begin{equation}
 E=\mathcal{S}_1 \otimes \mathcal{S}_2 - \Bigl\{ \rho_1 \otimes \rho_2
  \Bigm| \rho_1 \in \mathcal{S}_1, \rho_2 \in \mathcal{S}_2 \Bigr\}.
\end{equation}
So, the additivity conjecture of the Holevo capacity is simply stated as
``Do entangled states not contribute to the capacity of a product
channel?'' or ``Are entangled states powerless in the terms of Holevo
capacity?''

The conjecture above is stated in the terms of capacity, but there are some
direct measures of how much a quantum state is entangled, and similar
properties for them are conjectured. Now we define some measures.
\begin{definition}[Entanglement of formation]
 For a state $\rho$ in a bipartite system $ \mathcal{S}_A
 \otimes \mathcal{S}_B$, the {\em entanglement
 of formation} $E_F$ is defined as
\begin{equation}
 E_F(\rho) = \min_{\rho=\sum_{i} p_i\ket{v_i}\bra{v_i},\;\sum_{i}p_i=1}
  \sum_{i} p_i  S(\Tr_B \ket{v_i}\bra{v_i})
\end{equation}
where the minimization is over all possible expressions such that $\rho=\sum_{i}
 p_i\ket{v_i}\bra{v_i}$ and $\sum_{i} p_i=1$.
\end{definition}

\begin{definition}[Minimum output entropy]
 For a given channel $\Gamma: \mathcal{S}\to \mathcal{S}$, the {\em minimal
 output entropy} $S$ of $\Gamma$ is defined as
\begin{equation}\label{eq:moe}
 S(\Gamma) = \min_{\rho \in \mathcal{S}} S(\Gamma(\rho)).
\end{equation}
\end{definition}

The very important fact we have to mention here is that globally, some
conjectured properties about the measures defined above are
equivalent. It is stated as follows.

\begin{theorem}[Shor \cite{shor04a}]
 The following four propositions are equivalent.
\begin{enumerate}
 \item The additivity of the minimum entropy output of a quantum
       channel. Suppose that two channels $\Gamma_1$ and $\Gamma_2$ are
       given. Then 
\begin{equation}
 S(\Gamma_1 \otimes \Gamma_2) = S(\Gamma_1) + S(\Gamma_2).
\end{equation}
 \item The additivity of the Holevo capacity (See
       Conjecture \ref{conj:additivity-holevo})
 \item The additivity of the entanglement of formation. Suppose that two
       states $\rho_1 \in \mathcal{S}_{A1} \otimes \mathcal{S}_{B1}$ and
       $\rho_2 \in \mathcal{S}_{A2} \otimes \mathcal{S}_{B2}$. Then
\begin{equation}
 E_F(\rho_1\otimes\rho_2) = E_F(\rho_1) + E_F(\rho_2),
\end{equation}
       where $E_F$ is calculated over the bipartite $A$-$B$ partition.
 \item The strong superadditivity of the entanglement of
       formation. Suppose a state $\rho \in
       \mathcal{S}_{A1}\otimes\mathcal{S}_{A2}\otimes\mathcal{S}_{B1}\otimes\mathcal{S}_{B2}$
       is given. Then 
\begin{equation}
 E_F(\rho) \ge E_F(\mathrm{Tr}_1 \;\rho) + E_F(\mathrm{Tr}_2 \;\rho),
\end{equation}
 where $E_F$'s are over the bipartite $A$-$B$ system, and $\mathrm{Tr}_i$ means
       a trace out for the space $\mathcal{S}_{Ai}\otimes\mathcal{S}_{Bi}$.
\end{enumerate}
\end{theorem}

This means that some conjectures about the measures of entanglement are
equivalent. All the conjectures are about the power of entangled
states. 


\chapter{Voronoi diagrams for one-qubit quantum
 states and Its Application}\label{chap:one-qubit}

\section{Overview}
In this chapter, we investigate the one-qubit quantum state space. In
this case, the theoretic analysis is much simpler than general case
because the whole space is a three dimensional ball ({\em Bloch
ball}). In spite of its simple structure, the divergence defined here is
still distorted and far from the intuition which we have for a normal
``distance.''

In this chapter, we show some theorems concerning the coincidences of
Voronoi diagrams with respect to some pseudo-distances. Although some of
the theorems are the just special cases of the theorems for a higher level
case which we explain later, we introduce them here because historically
they were proved earlier and the proofs are much simpler than the general
cases.

Generally, the importance of investigation for one-qubit system as a
start-up for a new stream of a research has been emphasized in quantum
information theory. The same thing can be said for Voronoi
diagrams. More can be known for the one-qubit system than a higher-level
system because of its simplicity. Although the Voronoi sites are
restricted to be pure states, some coincidences of diagrams in a set of
mixed states can be proved for the one-qubit system, while in a
higher-level system, mixed states are much more complicated and similar
thing is still not known.

Our motivation originally started from the algorithm by Hayashi et
al.~\cite{hayashi05} to compute the Holevo capacity of a quantum
channel. Since it uses Voronoi diagrams with respect to the quantum
divergence, it is considered to be important to investigate the
structure of the Voronoi diagram with respect to the quantum divergence.

In Hayashi et al.'s algorithm, the key factor to attain the accuracy is an
approximation of a set of pure states by discrete points. Especially,
how well-distributed points you can obtain determines the accuracy of
the computation. The wellness here is in terms of the divergence. Our
result about coincidences of Voronoi diagrams tells that well
distributed points in Euclidean space is also well distributed in terms
of the divergence. Hence, although the divergence is difficult to deal
with, we can generate well distributed points appropriate for the
computation.

The refinement of the point set is another considerable
application. Suppose that you have computed the capacity with some
generated approximating points, and wish to refine the precision of the
result. Then, where should the new points be plotted? Voronoi diagrams
give the answer to that question. Since some diagrams with respect to
some pseudo-distances are the same, all you have to think about is the
Euclidean Voronoi diagram. Just plotting new points on the Voronoi edges
would be a reasonable refinement.

Although the algorithm by Hayashi et al.\ is the only existing application
of our result so far, it can be considered in a general context. The
difficulty of the numerical computation in quantum information theory is
due to a computation of a continuous geometric object. It is essentially
different from classical information theory. Using Voronoi diagrams can
be one of the options to overcome the difficulty.

\section{Voronoi diagrams in a quantum state space}
We consider Voronoi diagrams with respect to the divergence. Since
the quantum divergence $D(\cdot||\cdot)$ is not symmetric, we can
consider two different Voronoi diagrams about this measure. 

\begin{definition}
 For a given
 quantum state space $\mathcal{S}$, 
 Voronoi diagrams for a given set of sites
 $\Sigma=\left\{\sigma_i\right\}$ are defined as:
 \begin{align}
 V_D(\Sigma)&=\Bigl\{
  V_D^{(i)}=\bigl\{ \rho \in \mathcal{S} \bigm| D(\rho ||
 \sigma_i)\leq D(\rho ||
 \sigma_j) \text{ for any } j \bigr\} \Bigr\}\\
 V_D^*(\Sigma)&=\Bigl\{
 V_D^{*(i)}=\bigl\{ \rho \in \mathcal{S} \bigm| D(\sigma_i || \rho)\leq
 D(\sigma_j || \rho) \text{ for any } j \bigr\}\Bigr\}
 \end{align}
\end{definition}

The definition of the coincidence of Voronoi diagrams is natural.
Suppose that a space X and distances $d_1$ and $d_2$ on $X$ are
given. For a given set $\left\{p_i\in X \right\}$, the two Voronoi diagrams $V_{d_1}$
and $V_{d_2}$ are said to coincide when they are equal as a set. This
property is equivalent to the coincidence of bisector curves.

To show that given Voronoi diagrams coincide, it is sufficient to check
their bisector curves. This fact is stated as follows:
\begin{theorem}
Suppose that a space X and distances $d_1$ and $d_2$ on $X$ are
given and they satisfy $d_i(x,x)=0$ for any $x\in X$.
Then, the following two conditions are equivalent.
\begin{enumerate}
 \item For any set of points $P=\{p_1,\ldots,p_n\}$, Voronoi
       diagrams $V_{d_1}(P)$ and $V_{d_2}(P)$ are equivalent.
 \item For any given pair of points $(p_1,p_2)$, the bisector curve of
       $p_1,p_2$ with respect to $d_1$ and $d_2$ are equivalent.
\end{enumerate}
\end{theorem}
\begin{proof}
 We show this by induction. For two sites, the only edge is the bisector,
 and because $d_i(x,x)=0$, the region dominated by a site is the same
 side as the site.

Suppose that $V_{d_1}(P)$ and $V_{d_2}(P)$ are the same for first $n-1$
 points of $P$. If we add an $n$-th point as a new site, newly appearing
 edges are the bisectors of the new site and other site. The region
 dominated by the new site is the same side as the new site. Then, the
 diagrams for $n$ points are also the same.
\end{proof}

\section{Primal and dual Voronoi diagrams}
In this section, we characterize the Bloch ball by means of
computational geometry. We define primal and dual space of quantum
state. We show one of the two divergence Voronoi diagrams is linear in
the primal space and the other is linear in the dual space.  All the
theorems shown in this section can be extended to an arbitrary quantum
state space. However, more explicit computation is possible for
one-qubit states, and it helps us to understand the things more
deeper. It is a common merit for an investigation of one-qubit states.

\begin{theorem}\label{th:qubit-primal-linear}
 For one-qubit states, $V_D$ is linear.
\end{theorem}

The following is the essential for the proof of this theorem.

\begin{lemma}\label{lem:1-qubit}
For a one-qubit mixed state $\sigma=\sigma(\tilde{x},\tilde{y},\tilde{z})$
and a general one-qubit state $\rho=\rho(x,y,z)$,
\begin{equation}
 D(\rho\|\sigma)
  =
\left\{
\begin{array}{lr}
\displaystyle \frac{1}{2}\log\frac{1-r^2}{4}
 +\frac{r}{2}\log\frac{1+r}{1-r}
 -\frac{1}{2}\log\frac{1-\tilde{r}^2}{4}
&\\
\displaystyle
\qquad
  -\frac{1}{2\tilde{r}}
  \log\frac{1+\tilde{r}}{1-\tilde{r}}
  \left(x\tilde{x}+y\tilde{y}+z\tilde{z}\right)
 & ((\tilde{x},\tilde{y},\tilde{z})\not=(0,0,0))\bigskip\\
\displaystyle
\displaystyle \frac{1}{2}\log\frac{1-r^2}{4}
 +\frac{r}{2}\log\frac{1+r}{1-r} -\frac{1}{2}\log \frac{1}{4}
 & ((\tilde{x},\tilde{y},\tilde{z})=(0,0,0))\\
\end{array}
\right.,
\end{equation}
where the parameterizations are given as Formula
 (\ref{eq:one-qubit-rho}) and $r=\sqrt{x^2+y^2+z^2}$.
(Note that the formula for $(\tilde{x},\tilde{y},\tilde{z})=(0,0,0)$ is
the limit of the formula for
$(\tilde{x},\tilde{y},\tilde{z})\neq(0,0,0)$ as
$(\tilde{x},\tilde{y},\tilde{z})\to (0,0,0)$.)
\end{lemma}

\paragraph{Note for notation:}
By $\sigma=\sigma(\tilde{x},\tilde{y},\tilde{z})$, we mean $\sigma$ is
parameterized by $(\tilde{x},\tilde{y},\tilde{z})$ as in Formula
(\ref{eq:one-qubit-rho}) and implicitly define $\tilde{r}$ by
$\tilde{r}=\sqrt{\tilde{x}^2+\tilde{y}^2+\tilde{z}^2}$

\begin{proof}
 The eigenvalues of $\rho$ are
\begin{equation}
 \frac{1+r}{2},\quad \frac{1-r}{2}.
\end{equation}
When $(x,y)\neq (0,0)$, defining a unitary matrix $U$ as
\begin{equation}
 U=\frac{1}{\sqrt{2}}
\begin{pmatrix}
 \displaystyle\frac{x-iy}{\sqrt{x^2+y^2}}\sqrt{\frac{r+z}{r}} &
 \displaystyle\frac{x-iy}{\sqrt{x^2+y^2}}\sqrt{\frac{r-z}{r}}
 \bigskip \\
 \displaystyle\sqrt{\frac{r-z}{r}} &
 -\displaystyle\sqrt{\frac{r+z}{r}} \\
\end{pmatrix}
,
\end{equation}
$\rho$ is expressed as
\begin{equation}
 \sigma=U
 \begin{pmatrix}
  \displaystyle\frac{1+r}{2} & 0\\
  0 & \displaystyle\frac{1-r}{2}\\
 \end{pmatrix}
 U^*.
\end{equation}
Then,
\begin{align}\label{eq:rho-r}
 \Tr\rho\log\rho&=
\Tr U
 \begin{pmatrix}
  \displaystyle\frac{1+r}{2} & 0\\
  0 & \displaystyle\frac{1-r}{2}\\
 \end{pmatrix}
U^*
\cdot
U
 \begin{pmatrix}
  \displaystyle\log\frac{1+r}{2} & 0\\
  0 & \displaystyle\log\frac{1-r}{2}\\
 \end{pmatrix}
U^*\notag\\
&=\Tr
U
 \begin{pmatrix}
  \displaystyle\frac{1+r}{2}\log\frac{1+r}{2} & 0\\
  0 & \displaystyle\frac{1-r}{2}\log\frac{1-r}{2}\\
 \end{pmatrix}
U^*\notag\\
&=\Tr
 \begin{pmatrix}
  \displaystyle\frac{1+r}{2}\log\frac{1+r}{2} & 0\\
  0 & \displaystyle\frac{1-r}{2}\log\frac{1-r}{2}\\
 \end{pmatrix}
U^*U\notag\\
&=  \frac{1+r}{2}\log\frac{1+r}{2}
  +\frac{1-r}{2}\log\frac{1-r}{2}\notag\\
&=  \frac{1}{2}\log\frac{1-r^2}{4}
  +\frac{r}{2}\log\frac{1+r}{1-r}.
\end{align}
If $(x,y)=(0,0)$, $z=r$ and
\begin{align}
 \Tr\rho\log\rho&=
\begin{pmatrix}
 \displaystyle\frac{1+z}{2}& 0\\
 0 & \displaystyle\frac{1-z}{2}
\end{pmatrix}
\begin{pmatrix}
 \displaystyle\log\frac{1+z}{2}& 0\\
 0 & \displaystyle\log\frac{1-z}{2}
\end{pmatrix}\notag\\
&=  \frac{1+z}{2}\log\frac{1+z}{2}
  +\frac{1-z}{2}\log\frac{1-z}{2}\notag\\
&=  \frac{1+r}{2}\log\frac{1+r}{2}
  +\frac{1-r}{2}\log\frac{1-r}{2}.
\end{align}
This means Equation (\ref{eq:rho-r}) also holds for $(x,y)=(0,0)$.

Now, we will consider three cases:
\begin{itemize}
 \item $(\tilde{x},\tilde{y})\neq(0,0)$
 \item $(\tilde{x},\tilde{y})=(0,0)$ and $\tilde{z}\neq 0$
 \item $(\tilde{x},\tilde{y},\tilde{z})=0$
\end{itemize}

For $(\tilde{x},\tilde{y})=(0,0)$,
denoting similarly as $\rho$ by
\begin{equation}
\sigma=
\tilde{U}
 \begin{pmatrix}
  \displaystyle\frac{1+r}{2} & 0\\
  0 & \displaystyle\frac{1-r}{2}\\
 \end{pmatrix}
\tilde{U}^*,
\end{equation}
we obtain
\begin{align}
&\Tr(\rho \log\sigma)\notag\\
&=
\Tr \left[
 \rho\times
 \tilde{U}
  \begin{pmatrix}
   \displaystyle\log\frac{1+\tilde{r}}{2} & 0\\
   0 & \displaystyle\log\frac{1-\tilde{r}}{2}\\
  \end{pmatrix}
 \tilde{U}^*
\right]\notag\\
&=
\Tr \left[
 \rho
 \tilde{U}
 \frac{1}{\sqrt{2}}
  \begin{pmatrix}
   \displaystyle \frac{\tilde{x}+i\tilde{y}}
   {\sqrt{\tilde{x}^2+\tilde{y}^2}}
   \sqrt{\frac{\tilde{r}+\tilde{z}}{\tilde{r}}}
   \log\frac{1+\tilde{r}}{2} 
   & \displaystyle 
   \sqrt{\frac{\tilde{r}-\tilde{z}}{\tilde{r}}} 
   \log\frac{1+\tilde{r}}{2}\bigskip\\
   \displaystyle 
   \frac{\tilde{x}+i\tilde{y}}
   {\sqrt{\tilde{x}^2+\tilde{y}^2}}
   \sqrt{\frac{\tilde{r}-\tilde{z}}{\tilde{r}}}
   \log\frac{1-\tilde{r}}{2} 
   & \displaystyle 
   -\sqrt{\frac{\tilde{r}+\tilde{z}}{\tilde{r}}}
   \log\frac{1-\tilde{r}}{2}
  \end{pmatrix}
 \right]\notag\\
&=
\Tr \left[
 \rho
 \frac{1}{2}
  \begin{pmatrix}
   \frac{\tilde{r}+\tilde{z}}{\tilde{r}}
   \log\frac{1+\tilde{r}}{2} +
   \frac{\tilde{r}-\tilde{z}}{\tilde{r}}
   \log\frac{1-\tilde{r}}{2} 
   & 
   \frac{\tilde{x}-i\tilde{y}}{\tilde{r}}
   \log\frac{1+\tilde{r}}{1-\tilde{r}}\bigskip\\
   \frac{\tilde{x}+i\tilde{y}}{\tilde{r}}
   \log\frac{1+\tilde{r}}{1-\tilde{r}}
   &
   \frac{\tilde{r}-\tilde{z}}{\tilde{r}}
   \log\frac{1+\tilde{r}}{2} +
   \frac{\tilde{r}+\tilde{z}}{\tilde{r}}
   \log\frac{1-\tilde{r}}{2} 
  \end{pmatrix}
 \right]\notag\\
&=
\Tr\!
 \frac{1}{4}\!
  \begin{pmatrix}
   (1+z)\left(
   \frac{\tilde{r}+\tilde{z}}{\tilde{r}}
   \log\frac{1+\tilde{r}}{2} +
   \frac{\tilde{r}-\tilde{z}}{\tilde{r}}
   \log\frac{1-\tilde{r}}{2} \right)
   +
   \frac{(x-iy)(\tilde{x}+i\tilde{y})}{\tilde{r}}
   \log\frac{1+\tilde{r}}{1-\tilde{r}}
   & 
   *\bigskip\\
   \hspace{-85mm}
   *
   &\hspace{-90mm}
   (1-z)\left(
   \frac{\tilde{r}-\tilde{z}}{\tilde{r}}
   \log\frac{1+\tilde{r}}{2} +
   \frac{\tilde{r}+\tilde{z}}{\tilde{r}}
   \log\frac{1-\tilde{r}}{2} \right)
   +
   \frac{(x+iy)(\tilde{x}-i\tilde{y})}{\tilde{r}}
   \log\frac{1+\tilde{r}}{1-\tilde{r}}
  \end{pmatrix}
 \notag\\
 &\phantom{=}
\text{(``}*\text{'' means an unnecessary element for the computation below)}
\notag\\
&=\!\!\frac{1}{4}
   (1+z)\left(
   \frac{\tilde{r}+\tilde{z}}{\tilde{r}}
   \log\frac{1+\tilde{r}}{2} +
   \frac{\tilde{r}-\tilde{z}}{\tilde{r}}
   \log\frac{1-\tilde{r}}{2} \right)\!
   +
   \frac{(x-iy)(\tilde{x}+i\tilde{y})}{4\tilde{r}}
   \log\frac{1+\tilde{r}}{1-\tilde{r}}\notag\\
&\ \ +\!
 \frac{1}{4}
   (1-z)\left(
   \frac{\tilde{r}-\tilde{z}}{\tilde{r}}
   \log\frac{1+\tilde{r}}{2} +
   \frac{\tilde{r}+\tilde{z}}{\tilde{r}}
   \log\frac{1-\tilde{r}}{2} \right)\!
   +
   \frac{(x+iy)(\tilde{x}-i\tilde{y})}{4\tilde{r}}
   \log\frac{1+\tilde{r}}{1-\tilde{r}}\notag\\
&=
\frac{1}{2}\log\frac{1-\tilde{r}^2}{4}+
\frac{1}{2\tilde{r}}\log\frac{1+\tilde{r}}{1-\tilde{r}}
\left(x\tilde{x}+y\tilde{y}+z\tilde{z}\right).
\end{align}
Thus,
\begin{align}
 D(\rho||\sigma)&=
\Tr\left(\rho\log\rho\right)
-
\Tr\left(\rho\log\sigma\right)\notag\\
&=
 \frac{1}{2}\log\frac{1-r^2}{4}
 +\frac{r}{2}\log\frac{1+r}{1-r}
 -\frac{1}{2}\log\frac{1-\tilde{r}^2}{4}\notag\\
&\qquad
  -\frac{1}{2\tilde{r}}
  \log\frac{1+\tilde{r}}{1-\tilde{r}}
  \left(x\tilde{x}+y\tilde{y}+z\tilde{z}\right)
\end{align}
and this completes the proof for $(x,y)\neq(0,0)$.

If $\tilde{x}=\tilde{y}=0$, $\tilde{z}=\tilde{r}$ and
\begin{align}\label{eq:rho-sigma-0-0}
&\Tr(\rho \log\sigma)\notag\\
&=
\Tr \left[
 \rho
  \begin{pmatrix}
   \displaystyle\log\frac{1+\tilde{r}}{2} & 0\\
   0 & \displaystyle\log\frac{1-\tilde{r}}{2}\\
  \end{pmatrix}
\right]\notag\\
 &=
 \frac{1+z}{2}\log\frac{1+\tilde{r}}{2}
 +\frac{1-z}{2}\log\frac{1-\tilde{r}}{2}\notag\\
 &=
 \frac{1}{2}\log\frac{1-\tilde{r}^2}{4}
 +\frac{z}{2}\log\frac{1+\tilde{r}}{1-\tilde{r}}.
\end{align}
Here,
\begin{align}
 &\frac{1}{2}\log\frac{1-\tilde{r}^2}{4}+
 \frac{1}{2\tilde{r}}\log\frac{1+\tilde{r}}{1-\tilde{r}}
 \left(x\tilde{x}+y\tilde{y}+z\tilde{z}\right)\notag\\
&=
 \frac{1}{2}\log\frac{1-\tilde{r}^2}{4}+
 \frac{1}{2\tilde{r}}\log\frac{1+\tilde{r}}{1-\tilde{r}}
 \left(z\tilde{z}\right)\notag\\
&=
 \frac{1}{2}\log\frac{1-\tilde{r}^2}{4}+
 \frac{1}{2}\log\frac{1+\tilde{r}}{1-\tilde{r}}
 \cdot z
\end{align}
This completes the proof for $x=y=0, z\neq 0$.

When $x=y=z=0$, (\ref{eq:rho-sigma-0-0}) becomes
\begin{equation}
 \frac{1}{2}\log\frac{1}{4},
\end{equation}
and this completes the proof for $x=y=z=0$.
\end{proof}

\begin{proof}[Proof of Theorem~\ref{th:qubit-primal-linear}]
Suppose that $\rho$, $\sigma_1$ and $\sigma_2$ are parameterized by
 $(x,y,z)$, $(\tilde{x}_1,\tilde{y}_1,\tilde{z}_1)$,
 $(\tilde{x}_2,\tilde{y}_2,\tilde{z}_2)$ respectively.
Using Lemma~\ref{lem:1-qubit}, we obtain
\begin{align}
& D(\rho||\sigma_1)-D(\rho||\sigma_2)\notag\\
&=
-\frac{1}{2}\left( \log\frac{1-\tilde{r}_1}{4}
 -\log\frac{1-\tilde{r}_2}{4}\right)\notag\\
&\qquad
-x\left(
\tilde{x}_1\frac{1}{2\tilde{r}_1}\log\frac{1+\tilde{r}_1}{1-\tilde{r}_1}
-\tilde{x}_2\frac{1}{2\tilde{r}_2}\log\frac{1+\tilde{r}_2}{1-\tilde{r}_2}
\right)\notag\\
&\qquad\quad
-y\left(
\tilde{y}_1\frac{1}{2\tilde{r}_1}\log\frac{1+\tilde{r}_1}{1-\tilde{r}_1}
-\tilde{y}_2\frac{1}{2\tilde{r}_2}\log\frac{1+\tilde{r}_2}{1-\tilde{r}_2}
\right)\notag\\
 &\qquad\qquad
-z\left(
\tilde{z}_1\frac{1}{2\tilde{r}_1}\log\frac{1+\tilde{r}_1}{1-\tilde{r}_1}
-\tilde{z}_2\frac{1}{2\tilde{r}_2}\log\frac{1+\tilde{r}_2}{1-\tilde{r}_2}
\right).
\end{align}
This is linear in $x,y,z$.
\end{proof}

\begin{theorem}
Define a transformation from $\rho=\rho(x,y,z)$
 to $\hat{\rho}$ by
\begin{equation}
 \hat{\rho}=-\log \rho + \frac{1}{2} \log \frac{1-r^2}{4}\cdot I
\end{equation}
and
\begin{align}
 u&=-\frac{\partial}{\partial x} \Tr \rho \log \rho\notag\\
 v&=-\frac{\partial}{\partial y} \Tr \rho \log \rho\notag\\
 w&=-\frac{\partial}{\partial z} \Tr \rho \log \rho
\end{align}
Then, $\hat{\rho}$ can be expressed as
\begin{equation}
 \hat{\rho}(u,v,w)
=
\begin{pmatrix}
 w& u-iv\\
u+iv & -w
\end{pmatrix}.
\end{equation}
\end{theorem}
\begin{proof}
 $\rho$ can be diagonalized as
\begin{equation}
 \rho=U
\begin{pmatrix}
 \displaystyle \frac{1+r}{2}&0 \\
 0 & \displaystyle \frac{1-r}{2}
\end{pmatrix}
U^*
\end{equation}
with
\begin{equation}
U= \begin{pmatrix}
  \displaystyle\frac{x-iy}{\sqrt{x^2+y^2}}\sqrt{\frac{r+z}{r}}&
  \displaystyle\frac{x-iy}{\sqrt{x^2+y^2}}\sqrt{\frac{r-z}{r}}\\
  \displaystyle\sqrt{\frac{r-z}{r}}&
 -\displaystyle\sqrt{\frac{r+z}{r}}
 \end{pmatrix}.
\end{equation}
Then,
\begin{align}
u&= \frac{\partial}{\partial x}\left(- \Tr \rho \log \rho\right)\notag\\
&=\frac{\partial r}{\partial x}\cdot\frac{\partial}{\partial r}\left(-\Tr
 \rho \log \rho\right).
\end{align}
Here,
\begin{align}
\frac{\partial}{\partial r}\left(\Tr
 \rho \log \rho\right)&=
\frac{\partial}{\partial r}\left(\Tr
 U
\begin{pmatrix}
 \frac{1+r}{2}&0\\
 0&\frac{1-r}{2}
\end{pmatrix}
U^* \cdot U
\begin{pmatrix}
 \log\frac{1+r}{2}&0\\
 0&\log\frac{1-r}{2}
\end{pmatrix}
U^*\right)\notag\\
&=
\frac{\partial}{\partial r}\left(
\Tr
 U
\begin{pmatrix}
 \frac{1+r}{2}\log\frac{1+r}{2}&0\\
 0&\frac{1-r}{2}\log\frac{1-r}{2}
\end{pmatrix}
U^*\right)
\notag\\
&=
\frac{\partial}{\partial r}\left(
\Tr
\begin{pmatrix}
 \frac{1+r}{2}\log\frac{1+r}{2}&0\\
 0&\frac{1-r}{2}\log\frac{1-r}{2}
\end{pmatrix}
U^*U\right)\notag\\
&=
\frac{\partial}{\partial r}\left(
 \frac{1+r}{2}\log\frac{1+r}{2} +
 \frac{1-r}{2}\log\frac{1-r}{2}
\right)\notag\\
&=
\frac{1}{2}\log\frac{1+r}{1-r},
\end{align}
and
\begin{equation}
 \frac{\partial r}{\partial x} = \frac{x}{r}.
\end{equation}
Thus,
\begin{equation}
u=
-\frac{x}{2r}\log\frac{1+r}{1-r}.\label{eq:u-r}
\end{equation}
In the similar way, we can obtain
\begin{align}
v&=
-\frac{y}{2r}\log\frac{1+r}{1-r},\label{eq:v-r}\\
w&=
-\frac{z}{2r}\log\frac{1+r}{1-r}.\label{eq:w-r}
\end{align}
On the other hand,
\begin{align}
 &\hat{\rho}=-\log \rho + \frac{1}{2}\log\frac{1-r^2}{4}\cdot I\notag\\
&=-U
\begin{pmatrix}
 \displaystyle \log\frac{1+r}{2}&0\\
 0 & \displaystyle \log\frac{1-r}{2}
\end{pmatrix}
\frac{1}{\sqrt{2}}
\begin{pmatrix}
 \displaystyle \frac{x+iy}{\sqrt{x^2+y^2}}\sqrt{\frac{r+z}{r}}&
 \displaystyle\sqrt{\frac{r-z}{r}}\\
 \displaystyle \frac{x+iy}{\sqrt{x^2+y^2}}\sqrt{\frac{r-z}{r}}&
 \displaystyle-\sqrt{\frac{r+z}{r}}\\
\end{pmatrix}\notag\\
&\qquad +\frac{1}{2}\log\frac{1-r^2}{4}\cdot I\notag\\
&=
-\frac{1}{2}
\begin{pmatrix}
  \frac{x-iy}{\sqrt{x^2+y^2}}\!\sqrt{\frac{r+z}{r}}&\!\!\!
  \frac{x-iy}{\sqrt{x^2+y^2}}\!\sqrt{\frac{r-z}{r}}\\
  \sqrt{\frac{r-z}{r}}&\!\!\!
 -\sqrt{\frac{r+z}{r}}
 \end{pmatrix}\!\!
\begin{pmatrix}
 \frac{x+iy}{\sqrt{x^2+y^2}}\!\sqrt{\frac{r+z}{r}} \log\frac{1+r}{2}
&\!\!
 \sqrt{\frac{r-z}{r}}\log\frac{1-r}{2}\\
 \frac{x+iy}{\sqrt{x^2+y^2}}\!\sqrt{\frac{r-z}{r}} \log\frac{1+r}{2}&\!\!
 -\sqrt{\frac{r+z}{r}}\log\frac{1-r}{2}
\end{pmatrix}\notag\\
&\qquad +\frac{1}{2}\log\frac{1-r^2}{4}\cdot I\notag\\
&=
-\frac{1}{2}
\begin{pmatrix}
\displaystyle\frac{r+z}{r}\log\frac{1+r}{2}+\frac{r-z}{r}\log\frac{1-r}{2}&\!\!
\displaystyle\frac{x-iy}{r}\log\frac{1+r}{1-r}\bigskip\\
\displaystyle\frac{x+iy}{r}\log\frac{1+r}{1-r}&\!\!
\displaystyle\frac{r-z}{r}\log\frac{1+r}{2}+\frac{r+z}{r}\log\frac{1-r}{2}
\end{pmatrix}\notag\\
&\qquad
+\frac{1}{2}\log\frac{1-r^2}{4}\cdot I\notag\\
&=
\begin{pmatrix}
\displaystyle-\frac{z}{2r}\log\frac{1+r}{1-r}&
\displaystyle-\frac{x-iy}{2r}\log\frac{1+r}{1-r}\bigskip\\
\displaystyle-\frac{x+iy}{2r}\log\frac{1+r}{1-r}&
\displaystyle\frac{z}{2r}\log\frac{1+r}{1-r}\\
\end{pmatrix}\notag\\
&=
\begin{pmatrix}
 w& u-iv\\
u+iv & -w
\end{pmatrix},
\end{align}
and this completes the proof.
\end{proof}
\begin{theorem}
For a one-qubit state $\rho$ and $\sigma$, we use plain notation for
 $\rho$ and tilde notation for $\sigma$, i.e.\ $\sigma$ and
 $\hat\sigma$ are parameterized by $\tilde{x},\tilde{y},\tilde{z}$ and
 $\tilde{u},\tilde{v},\tilde{w}$ respectively.
 Define $\hat{D}$ by 
\begin{equation}
\label{eq:hatD-psi}
 \hat{D}(\hat\rho||\hat\sigma)=\psi(\hat{\rho})-\psi(\hat{\sigma})-
\left\langle 
\begin{pmatrix}
 u\\
 v\\
 w
\end{pmatrix}
-
\begin{pmatrix}
 \tilde{u}\\
 \tilde{v}\\
 \tilde{w}
\end{pmatrix}
 , \nabla_\psi
\begin{pmatrix}
 \tilde{u}\\
 \tilde{v}\\
 \tilde{w}
\end{pmatrix}
\right\rangle,
\end{equation}
where $\langle\cdot,\cdot\rangle$ means the inner product of vectors and
\begin{equation}
 \label{eq:dual-psi}
 \psi(\hat\rho) = \log \left( \Tr (\exp \hat\rho) \right).
\end{equation}
Then, 
$D(\rho||\sigma)=\hat{D}(\hat\sigma||\hat\rho)$ and the
 Voronoi diagram with respect to $\hat{D}$ with sites as second argument
 of $\hat{D}$ (denoted by $V_{\hat{D}}$) is linear.
\end{theorem}
\begin{proof}
 \begin{align}
  \exp \hat\rho&=\exp \left[ 
-U
\begin{pmatrix}
\log \frac{1+r}{2}& 9\\
0 & \log \frac{1-r}{2}
\end{pmatrix}
U^*
+\frac{1}{2}\log \frac{1-r^2}{4}\cdot I
\right]\notag\\
&=\exp U\left[ 
-\begin{pmatrix}
\log \frac{1+r}{2}& 9\\
0 & \log \frac{1-r}{2}
\end{pmatrix}
+\frac{1}{2}\log \frac{1-r^2}{4}\cdot I
\right]U^*\notag\\
&=U
\begin{pmatrix}
\exp \left(-\log \frac{1+r}{2}+\frac{1}{2}\log
 \frac{1-r^2}{4}\right)& 0\\
0 & \exp \left(-\log \frac{1-r}{2}+\frac{1}{2}\log
 \frac{1-r^2}{4}\right)
\end{pmatrix}
U^*\notag\\
&=U
\begin{pmatrix}
\displaystyle \sqrt{\frac{1-r}{1+r}}& 0\\
0 & \displaystyle \sqrt{\frac{1+r}{1-r}}
\end{pmatrix}
U^*.
\end{align}
Thus,
\begin{align}\label{eq:psi-r}
\psi(\hat{\rho})&=\log \Tr U
\begin{pmatrix}
\displaystyle \sqrt{\frac{1-r}{1+r}}& 0\\
0 & \displaystyle \sqrt{\frac{1+r}{1-r}}
\end{pmatrix}
U^*\notag\\
 &=\log \Tr
\begin{pmatrix}
\displaystyle \sqrt{\frac{1-r}{1+r}}& 0\\
0 & \displaystyle \sqrt{\frac{1+r}{1-r}}
\end{pmatrix}
U^* U\notag\\
&=
\log\left(\sqrt{\frac{1-r}{1+r}}+\sqrt{\frac{1+r}{1-r}}\right)\notag\\
&=-\frac{1}{2}\log\frac{1-r^2}{4}.
\end{align}
By taking square of (\ref{eq:u-r}), (\ref{eq:v-r}) and (\ref{eq:w-r});
 and adding them, we obtain
\begin{equation}
 u^2+v^2+w^2 =
\frac{1}{4}\left( \log\frac{1+r}{1-r} \right)^2.
\end{equation}
Taking $\frac{\partial}{\partial u}$ of this formula, 
\begin{align}
 2u&=\frac{\partial r}{\partial u}\frac{\partial}{\partial r}
\left[\frac{1}{4}\left( \log\frac{1+r}{1-r} \right)^2\right]\notag\\
&=\frac{\partial r}{\partial
 u}\frac{1}{1-r^2}\log\frac{1+r}{1-r},\notag\\
u&=\frac{\partial r}{\partial
 u}\frac{1}{2(1-r^2)}\log\frac{1+r}{1-r},
\end{align}
and similarly,
\begin{align}
 v&=\frac{\partial r}{\partial v}\frac{1}{2(1-r^2)}\log\frac{1+r}{1-r},\\
 w&=\frac{\partial r}{\partial w}\frac{1}{2(1-r^2)}\log\frac{1+r}{1-r}.
\end{align}
Thus, we obtain
\begin{align}\label{eq:nabla-psi}
 \nabla_\psi&=
\begin{pmatrix}
 \frac{\partial r}{\partial u}\\
 \frac{\partial r}{\partial v}\\
 \frac{\partial r}{\partial w}
\end{pmatrix}
\cdot
\frac{\partial}{\partial r}
\left(
-\frac{1}{2}\log \frac{1-r^2}{4}
\right)\notag\\
&=\frac{r}{1-r^2}
\begin{pmatrix}
 \frac{\partial r}{\partial u}\\
 \frac{\partial r}{\partial v}\\
 \frac{\partial r}{\partial w}
\end{pmatrix}\notag\\
&=2r\left(\log \frac{1+r}{1-r}\right)^{-1}
\begin{pmatrix}
 u\\
 v\\
 w
\end{pmatrix}
\end{align}
Using (\ref{eq:psi-r}) and (\ref{eq:nabla-psi}),
 $\hat{D}(\hat\rho||\hat\sigma)$ can be expanded as
\begin{align}
 \hat{D}(\hat\rho||\hat\sigma)&=
-\frac{1}{2}\log \frac{1-r^2}{4}
+\frac{1}{2}\log \frac{1-\tilde{r}^2}{4}
\notag\\
&\qquad
-
\left\langle
\begin{pmatrix}
 u-\tilde{u}\\
 v-\tilde{v}\\
 w-\tilde{w}
\end{pmatrix}
,\;
2r\left(\log \frac{1+r}{1-r}\right)^{-1}
\begin{pmatrix}
 \tilde{u}\\
 \tilde{v}\\
 \tilde{w} 
\end{pmatrix}
\right\rangle\notag\\
 &=
-\frac{1}{2}\log \frac{1-r^2}{4}
+\frac{1}{2}\log \frac{1-\tilde{r}^2}{4}
\notag\\
&\qquad
-
2\tilde{r}\left(\log \frac{1+\tilde{r}}{1-\tilde{r}}\right)^{-1}
\left[
\left(u\tilde{u}+v\tilde{v}+w\tilde{w}\right)
-
\left(\tilde{u}^2+\tilde{v}^2+\tilde{w}^2\right)
\right]\label{eq:qubit-dual-linear}\\
&=
-\frac{1}{2}\log \frac{1-r^2}{4}
+\frac{1}{2}\log \frac{1-\tilde{r}^2}{4}
\notag\\
&\qquad
-
\frac{1}{2r}\log\frac{1+r}{1-r}\cdot(x\tilde{x}+y\tilde{y}+z\tilde{z})
+
\frac{1}{2\tilde{r}}\log\frac{1+\tilde{r}}{1-\tilde{r}}
\cdot(\tilde{x}^2+\tilde{y}^2+\tilde{z}^2)\notag\\
&=
-\frac{1}{2}\log \frac{1-r^2}{4}
+\frac{1}{2}\log \frac{1-\tilde{r}^2}{4}
\notag\\
&\qquad
-
\frac{1}{2r}\log\frac{1+r}{1-r}\cdot(x\tilde{x}+y\tilde{y}+z\tilde{z})
+
\frac{\tilde{r}}{2}\log\frac{1+\tilde{r}}{1-\tilde{r}}\notag\\
&=D(\sigma||\rho)
\end{align}
Note here we used Theorem~\ref{lem:1-qubit}.
Additionally, Formula~(\ref{eq:qubit-dual-linear}) is linear for
 $u,v,w$, and consequently the equation for the boundary
\begin{equation}
 \hat{D}(\hat\rho||\hat\sigma_1)-
 \hat{D}(\hat\rho||\hat\sigma_2)=0
\end{equation}
is also linear.
\end{proof}

\section{Voronoi diagrams for one-qubit pure states}
 The definition for the Voronoi diagrams with respect to the divergence
 in the space of pure states is not obvious because the divergence
 $D(\rho||\sigma)$ is not defined for pure $\sigma$.  Actually, while
 $D(\rho||\sigma)=\Tr\rho(\log\rho - \log\sigma)$ can be defined when an
 eigenvalue of $\rho$ equals $0$ because $0\log 0$ can be naturally
 defined as $0$, it is not defined when an eigenvalue of $\sigma$ is
 $0$. Here we show that this Voronoi diagram of mixed states can be
 extended to pure states. We shall prove that even though the
 divergence $D(\rho||\sigma)$ can not be defined when $\sigma$ is a pure
 state, the Voronoi edges are naturally extended to pure states. In
 other words, we can define a Voronoi diagram for pure states by taking
 a natural limit of the diagram for mixed states.  When we say ``a
 Voronoi diagram with respect to divergence for pure states'', it means
 a diagram obtained by taking a limit of a diagram for mixed states.

Note that when we Voronoi sites are pure states, we assume a certain
kind of a parameterization for the convergence. Actually the Voronoi
diagram depends on how the sites converge. We assume that when sites are
given as $\rho_1(s_1), \ldots,\rho_n(s_n)$ and they are all pure, the
diagram is considered as the limit of the diagram with the sites $rs_1,
\ldots,rs_n$ in the Euclidean coordinate system, where $0<r<1$ and the
limit is taken for $r\to 1$. This definition might seem to be unnatural
because it trying to fix the way of uncertain convergence but we believe it
is natural because it is symmetric. Note that this definition is only
possible for the one-qubit system because it has special symmetry, and
in general, for a higher-level system, no longer exists.

To summarize the facts explained above, we give the following
definition.
\begin{definition}
When a set of Voronoi sites $\Sigma=\left\{\sigma_i\right\}$ is given,
 the {\em Voronoi diagram with respect to the divergence in the space of pure
 states} is defined as
 \begin{align}
 V_D^{\mathrm{pure}}(\Sigma)&=
  \mathrm{Closure} (V_D^{(i)}) \cap
  \mathcal{S}^\mathrm{pure},\notag\\
 V_D^{*\mathrm{pure}}(\Sigma)&=
  \mathrm{Closure} (V_D^{*(i)}) \cap
  \mathcal{S}^\mathrm{pure},\\
 \end{align}
where
 \begin{align}
 V_D^{(i)}&=
  \lim_{a\to 1}\bigl\{ \rho \in \mathcal{S}^\mathrm{faithful} \bigm| D(\sigma_i(a) || \rho)\leq
 D(\sigma_i(a) || \rho) \text{ for any } j \bigr\},\notag\\
  V_D^{*(i)}&=\bigl\{ \rho \in \mathcal{S}^\mathrm{faithful} \bigm| D(\rho ||
 \sigma_i)\leq 
  D(\rho ||
 \sigma_j) \text{ for any } j \bigr\}.
 \end{align}
 Here $\mathrm{Closure}(\cdot)$ means a topological closure and
 $\sigma_i(a)$ is defined by
\begin{equation}
\sigma_i(a) =
 \begin{pmatrix}
 \displaystyle\frac{1+a\tilde{z}_i}{2}& \displaystyle\frac{a\tilde{x}_i-ia\tilde{y}_i}{2}\bigskip\\
\displaystyle\frac{a\tilde{x}_i+ia\tilde{y}_i}{2} &
 \displaystyle\frac{1-a\tilde{z}_i}{2}
\end{pmatrix}
\end{equation}
when $\sigma_i$ is parameterized by
 $\sigma_i=\sigma_i(\tilde{x}_i,\tilde{y}_i,\tilde{z}_i)$, i.e.
\begin{equation}
 \sigma_i(a)=\frac{1}{2}I+a(\sigma_i-\frac{1}{2}I).
\end{equation}
\end{definition}

\begin{theorem}
\label{th:coincidence}
For given one-qubit pure states as sites, the following four Voronoi
diagrams are equivalent in the space of pure states:
\begin{enumerate}
\item
the Voronoi diagram with respect to the Fubini-Study distance
\item
the Voronoi diagram with respect to the Bures distance
\item
the Voronoi diagram on the sphere with respect to the ordinary
geodetic distance
\item
the section of the three-dimensional Euclidean Voronoi diagram with the
sphere and
\item
the two Voronoi diagram with respect to the divergences,
     i.e.\ $V_D^{\mathrm{pure}}$ and $V_D^{*\mathrm{pure}}$.
\end{enumerate}
\end{theorem} 

In Theorem~\ref{th:coincidence}, the coincidence of 1--4 is easy to
prove. 
Actually, for $\rho=\rho(x,y,z)$ and
$\sigma=\sigma(\tilde{x},\tilde{y},\tilde{z})$,
\begin{align}
  \Tr(\rho\sigma)
 &=
 \frac{1+z}{2} \frac{x-iy}{2}\frac{\tilde{x}+i\tilde{y}}{2}
 + \frac{1+\tilde{z}}{2}
 \frac{x+iy}{2}\frac{\tilde{x}-i\tilde{y}}{2}\notag\\
&=\frac{1+x\tilde{x}+y\tilde{y}+z\tilde{z}}{2}.
\end{align}
Thus,
\begin{align}
 d_\mathrm{FS}(\rho,\sigma)&=\cos^{-1}(\Tr(\rho\sigma))\notag\\
&=\cos^{-1}\sqrt{\frac{1+x\tilde{x}+y\tilde{y}+z\tilde{z}}{2}},\\
 d_\mathrm{B}(\rho,\sigma)&=\sqrt{1-\Tr\rho\sigma}\notag\\
&=\sqrt{\frac{1-x\tilde{x}-y\tilde{y}-z\tilde{z}}{2}}.
\end{align}
On the other hand, the Euclidean distance $d_\mathrm{E}$ is computed as
\begin{align}
 d_\mathrm{E}(\rho,\sigma)&=\sqrt{(x-\tilde{x})^2+(y-\tilde{y})^2+(z-\tilde{z})^2}\notag\\
&=\sqrt{(r^2+\tilde{r}^2)-2(x\tilde{x}+y\tilde{y}+z\tilde{z})}.
\end{align}
Especially when $\rho$ and $\sigma$ are pure, since $r=1$ and
$\tilde{r}=1$,
\begin{equation}
 d_\mathrm{E}(\rho,\sigma)=\sqrt{2(1-x\tilde{x}-y\tilde{y}-z\tilde{z})}.
\end{equation}
Then, the equivalence is
proved by simple workout as
\begin{align}
&d_\mathrm{FS}(\rho_1,\sigma)=d_\mathrm{FS}(\rho_2,\sigma)\notag\\
 &\quad \Leftrightarrow 
\cos^{-1}\sqrt{\frac{1+x_1\tilde{x}+y_1\tilde{y}+z_1\tilde{z}}{2}}
=\cos^{-1}\sqrt{\frac{1+x_2\tilde{x}+y_2\tilde{y}+z_2\tilde{z}}{2}}\notag\\
 &\quad \Leftrightarrow
x_1\tilde{x}+y_1\tilde{y}+z_1\tilde{z}
=x_2\tilde{x}+y_2\tilde{y}+z_2\tilde{z},
\end{align}
and similarly we can easily show
\begin{align}
 &d_\mathrm{B}(\rho_1,\sigma) = d_\mathrm{B}(\rho_2,\sigma)\notag\\
 &\quad\Leftrightarrow
x_1\tilde{x}+y_1\tilde{y}+z_1\tilde{z}
=x_2\tilde{x}+y_2\tilde{y}+z_2\tilde{z},\notag\\
 &d_\mathrm{E}(\rho_1,\sigma) = d_\mathrm{B}(\rho_2,\sigma)\notag\\
 &\quad\Leftrightarrow
x_1\tilde{x}+y_1\tilde{y}+z_1\tilde{z}
=x_2\tilde{x}+y_2\tilde{y}+z_2\tilde{z}.
\end{align}
This this complete the proof of 1--4.

The rest of Theorem~\ref{th:coincidence} is proved using
Lemma~\ref{lem:1-qubit} and the following lemma.

\begin{lemma}
\label{lem:1-qubit-2}
 For a mixed state $\sigma$ and an arbitrary $\rho$,
\begin{align}
 &\lefteqn{D(\rho_1||\sigma) =D(\rho_2||\sigma)}\nonumber\\
 &\quad \Leftrightarrow
 x_1\tilde{x}+y_1\tilde{y}+z_1\tilde{z}
 =x_2\tilde{x}+y_2\tilde{y}+z_2\tilde{z}.\label{eq:1-qubit-2a}
\end{align}
Moreover, under the condition $\tilde{r}_1=\tilde{r}_2$, 
\begin{align}
 &\lefteqn{D(\rho||\sigma_1) =D(\rho||\sigma_2)}\nonumber\\
 &\quad \Leftrightarrow
x\tilde{x}_1+y\tilde{y}_1+z\tilde{z}_1
 =x\tilde{x}_2+y\tilde{y}_2+z\tilde{z}_2.\label{eq:1-qubit-2b}
 \end{align}
\end{lemma}
\begin{proof}
 Because $\tilde{r}_1=\tilde{r}_2$, by Lemma \ref{lem:1-qubit}, 
\begin{align}
 &D(\rho||\sigma_1) -D(\rho||\sigma_2)\notag\\
=& -\log\frac{1+\tilde{r}_1}{1-\tilde{r}_1}
\left[
(x\tilde{x}_1+y\tilde{y}_1+z\tilde{z}_1)-
 (x\tilde{x}_2+y\tilde{y}_2+z\tilde{z}_2)
\right].
\end{align}
Thus,
\begin{align}
 &\lefteqn{D(\rho||\sigma_1) =D(\rho||\sigma_2)}\nonumber\\
 &\quad \Leftrightarrow
x\tilde{x}_1+y\tilde{y}_1+z\tilde{z}_1
 =x\tilde{x}_2+y\tilde{y}_2+z\tilde{z}_2.
 \end{align}
\end{proof}

Note that the setting for Lemma~\ref{lem:1-qubit-2} is more general than
 needed for the proof of Theorem~\ref{th:coincidence} because $\rho$ is not
 restricted to a pure state. Actually, the coincidence of Euclidean
 Voronoi and divergence-Voronoi also holds in the set of mixed
 states. It is proved in the next section.

Theorem~\ref{th:coincidence} means that all the diagrams are the same as
the ordinal Euclidean one, which has been researched enough. It tells us
the computational complexity of the diagrams stated as follows:
\begin{corollary}
 In the space of pure one-qubit states, the following Voronoi diagrams
 can be computed in $O(n\log n)$-time for $n$ sites.
 \begin{enumerate}
 \item
 the Voronoi diagram with respect to the Fubini-Study distance
 \item
 the Voronoi diagram with respect to the Bures distance
 \item
 the Voronoi diagram on the sphere with respect to the ordinary
 geodetic distance
 \item
 the section of the three-dimensional Euclidean Voronoi diagram with the
 sphere and
 \item
 the Voronoi diagram with respect to the divergences, i.e.~$V_D$ and $V_{D^*}$.
 \end{enumerate}
\end{corollary}
\begin{proof}
 The geodesic Voronoi diagram on a sphere is computed in $O(n\log
 n)$-time \cite{renka97}. Then, apply Theorem~\ref{th:coincidence}.
\end{proof}

\section{Voronoi diagrams for one-qubit mixed states}
In Theorem~\ref{th:coincidence}, we showed some Voronoi diagrams are the
same in the set of pure states. Some part of the theorem can be extended
to mixed states. The coincidence of Voronoi diagrams
in a set of mixed states is stated as follows.

\begin{theorem}\label{th:coincidence-one-qubit-whole-space}
For given one-qubit pure states as sites, the following four Voronoi
diagrams are equivalent in the whole space $\mathcal{S}$:
\begin{enumerate}
\item
the Voronoi diagram with respect to the Bures distance
\item
the Voronoi diagram with respect to the Euclidean distance
\item
the Voronoi diagram with respect to the divergences, i.e.~$V_D$ and
     $V_{D^*}$. 
\end{enumerate}
(Note that here, the sites are restricted to be pure states while the diagram
 is considered in the whole space.)
\end{theorem}
\begin{proof}
 Suppose that $\sigma_j$ $(j=1,2)$ are given as sites. Because they are
 pure, $\tilde{r}_1=\tilde{r}_2=1$. Under that
 condition, by Lemma~\ref{lem:1-qubit-2},
\begin{align}
 &D(\sigma_1||\rho) =D(\sigma_2||\rho)
 \Leftrightarrow D(\rho||\sigma_1) =D(\rho||\sigma_2)
 \nonumber\\
 &\quad \Leftrightarrow
 x\tilde{x}_1+y\tilde{y}_1+z\tilde{z}_1
 =x\tilde{x}_2+y\tilde{y}_2+z\tilde{z}_2. 
\end{align} 
 On the other hand,
 \begin{align}
  d_\mathrm{E}(\rho,\sigma_1)&=d_\mathrm{E}(\rho,\sigma_1)\notag\\
  &\Leftrightarrow 
  \sqrt{r^2+1-2(x\tilde{x}_1+y\tilde{y}_1+z\tilde{z}_1)}
  =\sqrt{r^2+1-2(x\tilde{x}_2+y\tilde{y}_2+z\tilde{z}_2)}\notag\\
  &\Leftrightarrow
  x\tilde{x}_1+y\tilde{y}_1+z\tilde{z}_1
  =x\tilde{x}_2+y\tilde{y}_2+z\tilde{z}_2.
 \end{align}
Thus, the equivalence of the divergence Voronoi diagrams and the
 Euclidean Voronoi diagram is proved. 

 Now suppose that the sites $\sigma_j$. $(j=1,2)$ are expressed as
 \begin{equation}
 \sigma_j=\ket{\psi_j}\bra{\psi_j}
 \end{equation}
 with
 \begin{equation}
 \psi_j=
  \begin{pmatrix}
   s_j\\
   t_j
  \end{pmatrix},
 \end{equation}
 which means
 \begin{equation}
 \sigma_j=
  \begin{pmatrix}
   s_j \bar{s}_j & s_j \bar{t}_j \\
   t_j \bar{s}_j & t_j \bar{t}_j
  \end{pmatrix}.
 \end{equation}
 Then, since $\sqrt{\sigma_j}=\sigma_j$ the Bures distance for an
 arbitrary quantum state $\rho$ is calculated as
 \begin{align}
 \label{expansion-mixed-bures}
 \lefteqn{d_\mathrm{B}(\sigma_j,\rho)}\nonumber\\
 &=1-\Tr\sqrt{\sigma_j \rho \sigma_j}\nonumber\\
 &=1-\Tr\sqrt{\ket{\psi_j}\bra{\psi_j} \rho
 \ket{\psi_j}\bra{\psi_j}}\nonumber\\
 &=1-\Tr\sqrt{\ket{\psi_j}\bigl[\bra{\psi_j} \rho
 \ket{\psi_j}\bigr]\bra{\psi_j}}\nonumber\\
 &=1-\Tr\sqrt{\left(\frac{1+z}{2}s_j\bar{s}_j +\frac{x-iy}{2}t_j\bar{s}_j
 \frac{x+iy}{2}s_j\bar{t}_j +
 \frac{1-z}{2}t_j\bar{t}_j\right)}\ket{\psi_j}\bra{\psi_j}\nonumber\\
 &=1-\sqrt{\frac{1+z}{2}s_j\bar{s}_j +\frac{x-iy}{2}t_j\bar{s}_j
 \frac{x+iy}{2}s_j\bar{t}_j +
 \frac{1-z}{2}t_j\bar{t}_j}.
 \end{align}
 Here we supposed $\rho$ is parameterized as (\ref{eq:one-qubit-rho}). If
 $\sigma_j$'s are also parameterized in the same way as
 \begin{equation}
 \sigma_j=
  \begin{pmatrix}
   \frac{1+z_j}{2} & \frac{x_j-iy_j}{2}\\
   \frac{x_j+iy_j}{2}& \frac{1-z_j}{2},
  \end{pmatrix}
 \end{equation}
 Formula (\ref{expansion-mixed-bures}) can be more simplified as
 \begin{equation}
 d_\mathrm{B}(\sigma_j,\rho)=
 1-\sqrt{\frac{1}{2}\left(x_j x + y_j y + z_j z\right)}
 \end{equation}
 Thus,
 \begin{align}
 & d_\mathrm{B}(\sigma_1,\rho)-d_\mathrm{B}(\sigma_2,\rho)=0\nonumber\\
 \Longleftrightarrow & 
 \left( x_1 x + y_1 y + z_1 z\right)-
 \left( x_2 x + y_2 y + z_2 z\right)=0.
 \end{align}
 This means the Bures diagram is the same as the Euclidean Voronoi
 diagram in the $xyz$-space.
\end{proof}

Similarly as for pure states, computational complexity can be known for
the diagrams:
\begin{corollary}
 In the space of general one-qubit states, when $n$ sites are given as
 pure states, each of the following Voronoi diagrams can be computed in
 $O(n^2)$-time
 \begin{enumerate}
 \item
 the Voronoi diagram with respect to the Bures distance
 \item
 the Voronoi diagram with respect to the Euclidean distance
 \item
 the Voronoi diagram with respect to the divergences, i.e.~$V_D$ and
     $V_{D^*}$. 
 \end{enumerate}
\end{corollary}
\begin{proof}
 The Euclidean Voronoi diagram can be computed in $O(n^2)$-time (See
 Section~\ref{sec:computation-vd}). Then,
 apply Theorem~\ref{th:coincidence-one-qubit-whole-space}.
\end{proof}

\section{Meaning of the result}
\label{sec:one-qubit-meaning}
The direct application of the fact proved above for the coincidences of
Voronoi diagrams is the algorithm by Hayashi et al.~\cite{hayashi05} and
Oto et al.~\cite{oto04a, oto04b} to
compute the Holevo capacity.

In the source of a channel, they plotted points so that they are
vertices of the mesh which is obtained by dividing the sphere equally
both latitudinally and longitudinally. It is a intuitive way and looks
reasonably well-distributed, but its uniformness of the points is in
terms of Euclidean distance. In the algorithm, the smallest enclosing
ball of the images of the points is computed, and it is in terms of the
``divergence.'' The accuracy depends on how well the points are
distributed with respect to the divergence. However, as we have shown,
the Voronoi diagrams with respect to Euclidean distance and the quantum
divergence are the same, and it guarantee the uniformness of the points.

In the original paper by Hayashi et al.~\cite{hayashi05} and Oto et al.~\cite{oto04a, oto04b}, there is an
implicit assumption that uniformly distributed points in terms of the
Euclidean distance is also uniform in the world of the quantum
divergence. The coincidences of Voronoi diagrams we have shown partly
fill its gap. Since the coincidences we proved is only for the case that
all Voronoi site are pure, we have not showed the uniformness preserve
in general case. However, if the image of a given channel is
sufficiently large (i.e.\ the surface of the image ellipsoid is near to the
unit sphere), the images of the plotted points have a similar property as
pure states because the smoothness of pseudo-distances.

Another expected application of our result is a refinement of
points. Since the algorithm by Hayashi et al.~\cite{hayashi05} approximates a geometry by some
points, it becomes more accurate if the number of plotted points becomes
larger. Then, where should the additional points be located? The
reasonable answer is the Voronoi edges, because the Voronoi
edge can be regarded as the set of the farthest points from the existing
plotted points. In a general case, the problem arises here is difficulty
of non-Euclidean distance, but thanks to the theorems we proved, the
Voronoi edges with respect to the divergence is the same as those of
Euclidean Voronoi diagram. You can easily refine the point set by adding
points on the Voronoi edges, not worrying about the distortion of the
quantum divergence.

From the viewpoint of computational geometry, we showed Voronoi
diagrams can be a tool to compare some measures defined for the same
set. The important point is that the coincidence of Voronoi diagrams can be
a hint for uniformness of a point set in different measures, and we
showed an example in three dimensional space, which can be visualized
and observed intuitively.

\section{Summary of this chapter}
We proved the coincidences of some Voronoi diagrams for one-qubit
states. Some part of the result is just a special case of the general
result explained in Chapter~\ref{chap:higher-level}, but some are
specific for one-qubit system.

More precisely, we showed that when Voronoi sites are given, the
Euclidean Voronoi diagram $V_{d_\mathrm{E}}$, the Bures Voronoi diagram
$V_{d_\mathrm{B}}$ and the divergence Voronoi diagrams $V_{D}, V_{D^*}$ are
the same. Additionally, if it is restricted to the set of pure states,
the diagrams above, the Fubini-Study Voronoi diagram
$V_{d_\mathrm{FS}}$, and the geodesic Voronoi diagram are all the
same. Table~\ref{tab:summary-one-qubit} shows the summary of the proved facts.

\begin{table}[h]
\caption[Coincidences of Voronoi diagrams for one-qubit states and their
 computational complexities]
{Coincidences of Voronoi diagrams for one-qubit states and their
 computational complexities: Note that the Voronoi sites are given as
 pure states in all the cases}
\label{tab:summary-one-qubit}
\begin{center}
  \begin{tabular}{|c|c|c|c|}
 \hline
 space & pseudo-distance & coincidence & complexity \\\hline
 & Fubini-Study & \rightbrace{5}{2cm}[coincide] &\\
 & Bures & & \\
 pure states & geodesic & &  $O(n\log n)$\\
 & Euclidean & & \\
 & divergence & & \\ \hline
 & Bures & \rightbrace{3}{2cm}[coincide] &\\
 mixed states & Euclidean & &  $O(n^2)$\\
 & divergence & & \\ \hline
 \end{tabular}
\end{center}
\end{table}

The known application of it is the algorithm by Hayashi et al.\ to compute the
Holevo capacity. It supports the effectiveness of the algorithm, and we
also suggested the refinement of the point set by add point in its
Voronoi edges. 


\chapter{Voronoi diagrams for 3 or higher level quantum states}
\label{chap:higher-level}

\section{Overview}

Theoretic analysis for Voronoi diagrams for three or higher level
quantum state space is given in this chapter. Our motivation originally
comes from the natural interest about whether the extension of the
theorems shown in Chapter \ref{chap:one-qubit} hold or not.

However, the structure of the quantum state space for three or higher
level system is much more complicated for one-qubit quantum system. One
of the obvious support for its complicatedness is the result by Kimura
\cite{kimura03}, which showed the explicit condition for a complex
matrix to be a density matrix and the inequalities appeared are too much
complicated. Consequently, we could not have done analysis for three or
higher level system about the same condition as in one-qubit case. We shall
only show the case that Voronoi site are given as pure states though the
Voronoi regions may be general. 

Using our result, we can convert a problem about a certain
pseudo-distance into another problem about another pseudo-distance. 

\section{Primal and dual Voronoi diagrams}

The following is the essential property of $V_D$.
\begin{theorem}\label{th:vd-planer-1}
 The boundaries of the Voronoi diagram $V_D$ are linear.
\end{theorem}

Although, only $V_D$ is proved to be linear, the other Voronoi diagram
$V_D^*$ can be obtained by some transformation from a linear Voronoi
diagram. It is stated as a following theorem; this is based on a common
mathematical framework known as {\em Legendre transformation}.
\begin{theorem}\label{th:vd-planer-2}
\label{th:dual-diagram}
Define transformation from $\rho$ to $\hat\rho$ by
\begin{equation}\label{eq:hatrho}
 \hat\rho = \log\rho +\frac{1}{d} \Tr (\log\rho) I,
\end{equation}
and
\begin{equation}
 \hat\xi_i = -\frac{\partial}{\partial \xi_i}\Tr(\rho \log\rho).
\end{equation}
Then, the parameterization of $\hat\rho$ is given as
\begin{align}
\label{eq:multi-level-dual}
\lefteqn{\hat{\rho}(\hat{\xi})=}\nonumber\\
&
\begin{pmatrix}
d\hat{\xi}_1 - \sum_{i=1}^{d-1} \hat{\xi}_i & \hat{\xi}_d - i \hat{\xi}_{d+1} & & \cdots &
 \hat{\xi}_{3d-4} - \hat{\xi}_{3d-3} \\
\hat{\xi}_d + i \hat{\xi}_{d+1} & d\hat{\xi}_2 - \sum_{i=1}^{d-1} \hat{\xi}_i &  & \cdots &
 \hat{\xi}_{5d-8} - i\hat{\xi}_{5d-7} \\
\vdots & & \ddots & & \vdots\\
\hat{\xi}_{5d-8} + i\hat{\xi}_{5d-7} & & & d\hat{\xi}_{d-1} - \sum_{i=1}^{d-1} \hat{\xi}_i &
 \hat{\xi}_{d^2-1} - i \hat{\xi}_{d^2-1}\\
\hat{\xi}_{3d-4} + \hat{\xi}_{3d-3} & & & \cdots &  - \sum_{i=1}^{d-1} \hat{\xi}_i 
\end{pmatrix}
,\nonumber\\
&\quad \hat{\xi}_i \in \mathbb{R},
\end{align}
where $\rho=\rho(\xi)$ is parameterized as in
 Formula~(\ref{multi-level-rho}). 

Moreover, define $\hat{D}$ by
\begin{equation}
\label{eq:hatD-psi}
 \hat{D}(\hat\rho||\hat\sigma)=\psi(\hat{\rho})-\psi(\hat{\sigma})-
\left\langle \hat\xi-\hat\eta , \nabla_\psi(\hat{\eta})\right\rangle,
\end{equation}
where
\begin{equation}
 \label{eq:dual-psi}
 \psi(\hat\rho) = \log \left( \Tr (\exp \hat\rho) \right).
\end{equation}
Then, 
$D(\rho||\sigma)=\hat{D}(\hat\sigma||\hat\rho)$ and the
 Voronoi diagram with respect to $\hat{D}$ with sites as second argument
 of $\hat{D}$ (denoted by $V_{\hat{D}}$) is linear.
\end{theorem}

This theorem is mostly proved by Oto et al.~\cite{oto04a}, but it
contains a logical gap or an ambiguous expression. We give a
self-consistent proof as follows.

\begin{proof}[Proof of Theorem \ref{th:vd-planer-1} and \ref{th:vd-planer-2}] 
Denote $\varphi$ by
\begin{equation}
 \varphi=-\Tr(\rho\log\rho),
\end{equation}
 and suppose that $\rho$ is diagonalized as
\begin{align}
 \rho&=X\Lambda X^*,\\
\Lambda&=
\begin{pmatrix}
 \lambda_1& &&\\
 & \lambda_2& &\\
 & & \ddots &\\
 & & & \lambda_d
\end{pmatrix},
\end{align}
where $X$ is a unitary matrix.
Then,
\begin{align}
 \eta_i&=\frac{\partial \varphi}{\partial \xi_i}\notag\\
 &=-\Tr\left(
\frac{\partial \rho}{\partial \xi_i} \log\rho
+
\rho \frac{\partial \log\rho}{\partial \xi_i}
\right)\notag\\
 &=-\Tr\left(
\frac{\partial \rho}{\partial \xi_i} \log\rho
\right)
-\Tr\left(
\rho \frac{\partial \log\rho}{\partial \xi_i}
\right).
\end{align}
Here,
\begin{align}
&\Tr\left(
\rho \frac{\partial \log\rho}{\partial \xi_i}
\right)\notag\\
&=
\Tr\left(
\rho \frac{\partial }{\partial \xi_i} (X \log\Lambda X^*)
\right)\notag\\
&=
\Tr\left(
X\Lambda X^* \frac{\partial X }{\partial \xi_i} \log\Lambda X^*
+X\Lambda X^* X\frac{\partial \log\Lambda }{\partial \xi_i} X^*
+X\Lambda X^* X \log\Lambda \frac{\partial X^* }{\partial \xi_i}
\right)\notag\\
&=
\Tr\left(
\log\Lambda X^* X\Lambda X^* \frac{\partial X }{\partial \xi_i} 
+X\Lambda X^* X\frac{\partial \log\Lambda }{\partial \xi_i} X^* 
+\Lambda X^* X \log\Lambda \frac{\partial X^* }{\partial \xi_i} X
\right)\notag\\
&=
\Tr\left[
\Lambda \log\Lambda \left( 
X^* \frac{\partial X }{\partial \xi_i} 
+ \frac{\partial X^* }{\partial \xi_i} X\right)
\phantom{
\begin{pmatrix}
 \lambda_1& & \\
 & \ddots & \\
 & & \lambda_d
\end{pmatrix}
}
\right.
\notag\\
&\quad\left.
+X
\begin{pmatrix}
 \lambda_1& & \\
 & \ddots & \\
 & & \lambda_d
\end{pmatrix}
\begin{pmatrix}
 \frac{\partial}{\partial \xi_i} \log\lambda_1& & \\
 & \ddots & \\
 & &\frac{\partial}{\partial \xi_i} \log\lambda_d \\
\end{pmatrix}
X^*
\right]\notag\\
&=
\Tr\left[
\Lambda \log\Lambda \left( 
\frac{\partial X^* X }{\partial \xi_i} 
\right)
+X
\begin{pmatrix}
 \lambda_1& & \\
 & \ddots & \\
 & & \lambda_d
\end{pmatrix}
\begin{pmatrix}
 \frac{\partial\lambda_1}{\partial \xi_i} \frac{1}{\lambda_1}& & \\
 & \ddots & \\
 & &\frac{\partial\lambda_d}{\partial \xi_i} \frac{1}{\lambda_d} \\
\end{pmatrix}
X^*
\right]\notag\\
&=
\Tr\left[
\Lambda \log\Lambda 
 \left(  
\frac{\partial I }{\partial \xi_i} 
\right)
+X
\begin{pmatrix}
 \frac{\partial\lambda_1}{\partial \xi_i} & & \\
 & \ddots & \\
 & &\frac{\partial\lambda_d}{\partial \xi_i}  \\
\end{pmatrix}
X^*\right]
\notag\\
&=
\Tr\left(
\frac{\partial \rho}{\partial \xi_i}
\right)\notag\\
&=
\frac{\partial }{\partial \xi_i} (\Tr \rho)\notag\\
&=
\frac{\partial }{\partial \xi_i} (1)\notag\\
&=0.
\end{align}
Thus,
\begin{equation}\label{eq:hatxi_i}
 \hat\xi_i = \Tr\left(
\frac{\partial \rho}{\partial\xi_i}\log\rho
\right).
\end{equation}
Now compare the each element of Equation~(\ref{eq:hatxi_i}). For $i=1,\ldots,d-1$, 
only $i$-th and $d$-th diagonal elements of $ \frac{\partial
 \rho}{\partial\xi_i}$ are $1/d$ and $-1/d$ respectively and the other
 elements are zero, i.e.,
\begin{equation}
 \frac{\partial \rho}{\partial\xi_i}
=
\bordermatrix{
& & i\text{-th} & & \cr
& & \vdots& &\cr
i\text{-th}& \cdots & 1/d & &\cr
& & & &\cr
& & & &-1/d\cr
}.
\end{equation}
So, if we write $\log\rho$ element-wise as
\begin{equation}
 \log\rho= (\zeta_{ij}),
\end{equation}
then
\begin{equation}
 \hat\xi_i=\frac{1}{d}(\zeta_{ii}-\zeta_{dd}).
\end{equation}
Thus,
\begin{align}
 \sum_{i=1}^{d-1}\hat\xi_i
 &=\frac{1}{d}(\sum_{i=1}^{d-1}\zeta_{ii}-(d-1)\zeta_{dd})\notag\\
 &=\frac{1}{d}\Tr \log\rho - \zeta_{dd},
\end{align}
and 
\begin{equation}
 d\hat\xi_i- \sum_{i=1}^{d-1}\hat\xi_i 
=\zeta_{ii} - \frac{1}{d}\Tr\log\rho.
\end{equation}
Now think of $i=d$ for example. the only non-zero elements of
 $\frac{\partial \rho}{\partial\xi_d}$ are the $(1,2)$ and $(2,1)$-th, and
\begin{align}
 \Tr\left(
\frac{\partial \rho}{\partial\xi_d}\log \rho
\right)
&=
\Tr
\left[
\begin{pmatrix}
 & 1/2 & & &\\
1/2 & & & &\\
& & & &\\
& & & &
\end{pmatrix}
\log\rho
\right]\notag\\
&= \frac{1}{2}\zeta_{21}+ \frac{1}{2}\zeta_{12}\notag\\
&= \mathrm{Re}\; \zeta_{12},
\end{align}
because $\log\rho$ is also Hermitian. Here $\mathrm{Re}$ means a real part.
Similarly for $i=d+1$,
\begin{align}
 \Tr\left(
\frac{\partial \rho}{\partial\xi_{d+1}}\log \rho
\right)
&=
\Tr
\left[
\begin{pmatrix}
 & -i/2 & & &\\
 i/2 & & & &\\
& & & &\\
& & & &
\end{pmatrix}
\log\rho
\right]\notag\\
&= \frac{i}{2}\zeta_{21}- \frac{i}{2}\zeta_{12}\notag\\
 &=\mathrm{Im}\; \zeta_{12},
\end{align}
where $\mathrm{Im}$ means an imaginary part.
Similar observation shows Formula~(\ref {eq:multi-level-dual}).

In the rest of the proof, we assume 
\begin{align}
 \psi(\hat\xi) = \sum_{i=1}^{d^2-1} \hat\xi \xi - \varphi (\xi),
\end{align}
and it is shown in the lemma below. Then,
\begin{align}
 D(\rho(\xi)||\sigma(\eta))
&= \varphi(\xi)-\varphi(\eta)- \left\langle\xi-\eta, \hat\xi\right\rangle\notag\\
&= \left( \sum_i \xi_i \hat\xi_i -\psi(\xi) \right) 
- \left( \sum_i \eta_i \hat\eta_i -\psi(\eta) \right) 
- \left\langle \xi-\eta, \hat\xi \right\rangle\notag\\
&= \psi(\eta)-\psi(\xi)
\left\langle \xi, \hat\xi\right\rangle
-\left\langle  \eta, \hat\eta \right\rangle 
- \left\langle \xi-\eta, \hat\xi \right\rangle\notag\\
&= \psi(\eta)-\psi(\xi)
- \left\langle \eta, \hat\eta - \hat\xi \right\rangle\notag\\
\end{align}
We also show 
\begin{equation}
 \nabla_\psi(\hat\eta)=\eta
\end{equation}
as a lemma, and it completes the proof for $D(\rho||\sigma)=\hat{D}(\hat\sigma||\hat\rho)$

Now we will show both $V_D$ and $V_{\hat{D}}$ are linear.
For $\sigma_1$ and $\sigma_2$, the bisector is given as a set of $\rho$
 which satisfies
\begin{equation}
 D(\rho||\sigma_1)-D(\rho||\sigma_2)=0.
\end{equation}
Since
\begin{equation}\label{eq:div-phi}
  D(\rho(\xi)||\sigma(\eta))
= \varphi(\xi)-\varphi(\eta)- \left\langle\xi-\eta, \nabla_\varphi (\eta) \right\rangle,
\end{equation}
we obtain
\begin{align}
& D(\rho||\sigma_1)-D(\rho||\sigma_2)\notag\\
 &=
\bigl(\varphi(\xi)-\varphi(\eta_1)- \left\langle\xi-\eta_1,
 \nabla_\varphi (\eta_1)\right\rangle
\bigr)
-
\bigl(
\varphi(\xi)-\varphi(\eta_2)- \left\langle\xi-\eta_2,
 \nabla_\varphi (\eta_2)\right\rangle
\bigr)\notag\\
 &=
-\varphi(\eta_1)
+\varphi(\eta_2)
+ \left\langle \eta_1, \nabla_\varphi (\eta_1)\right\rangle
- \left\langle \eta_2, \nabla_\varphi (\eta_2)\right\rangle
- \left\langle \xi, \nabla_\varphi (\eta_1)-\nabla_\varphi (\eta_2)\right\rangle.
\end{align}
This is linear equation in $\rho$. Because of the similarity of
 (\ref{eq:hatD-psi}) and (\ref{eq:div-phi}), $V_{\hat{D}}$ can
 also be proved to be linear in the same way.
\end{proof}

Now we show the following lemma to complete the proof of
Theorem~\ref{th:vd-planer-1} and \ref{th:vd-planer-2}.

\begin{lemma}
The following equations hold.
\begin{align}
  \psi(\hat\xi) &= \sum_{i=1}^{d^2-1} \hat\xi \xi - \varphi (\xi),\\
 \nabla_\psi(\hat\xi)&=\xi.
\end{align}
\end{lemma}
\begin{proof}
\begin{align}
 &\sum_{i=1}^{d^2-1} \hat\xi \xi - \varphi (\xi)\notag\\
&=\sum_i \Tr \left(\xi_i \frac{\partial \rho}{\partial \xi_i}
 \log\rho\right)
-\Tr(\rho\log\rho).
\end{align}
Here,
\begin{align}
 &\sum_i \xi_i \frac{\partial \rho}{\partial \xi_i}
 \log\rho\notag\\
&=
\begin{pmatrix}
 \displaystyle\frac{\xi_1}{d} &
 \!\!\!\displaystyle\frac{\xi_d-i\xi_{d+1}}{2} &
 \cdots & &   
 \displaystyle\frac{\xi_{3d-4}-i\xi_{3d-3}}{2} 
 \smallskip\\
 \displaystyle\frac{\xi_d+i\xi_{d+1}}{2} & 
 \!\!\!\displaystyle\frac{\xi_2}{d} &
 \cdots & & 
 \displaystyle\frac{\xi_{5d-8}-i\xi_{5d-7}}{2} 
 \smallskip\\
 \vdots & & \!\!\!\!\ddots & & \vdots \\
 \displaystyle\frac{\xi_{3d-6}+i\xi_{3d-5}}{2} & \cdots& &
 \!\!\!\!\!\!\!\displaystyle\frac{\xi_{d-1}}{d}& 
 \displaystyle\frac{\xi_{d^2-2}-i\xi_{d^2-1}}{2}\smallskip\\
 \displaystyle\frac{\xi_{3d-4}+i\xi_{3d-3}}{2} & \cdots& &
 \!\!\!\!\!\!\!\displaystyle\frac{\xi_{d^2-2}+i\xi_{d^2-1}}{2} &
 \displaystyle
 \frac{-\sum_{i=1}^{d-1}\xi_i}{d}
\end{pmatrix}\notag\\
&=\rho-\frac{1}{d}I.
\end{align}
Thus,
\begin{align}
 \sum_{i=1}^{d^2-1} \hat\xi \xi - \varphi (\xi)
 &=
\Tr\left[\left(\rho - \frac{1}{d}I\right)\right]-\Tr(\rho\log\rho)\notag\\
&=-\frac{1}{d}\Tr(\log\rho)\notag\\
&= \psi(\hat\xi).
\end{align}

 Moreover, 
\begin{align}
 \frac{\partial \psi}{\partial \hat\xi_i}&=
 \frac{\partial }{\partial \hat\xi_i}\left(
\sum_j \hat\xi_j \xi_j - \varphi(\xi)
\right)\notag\\
&=
\sum_j \left(
\frac{\partial \hat\xi_j}{\partial \hat\xi_i} \xi_j
+
\hat\xi_j \frac{\partial \xi_j}{\partial \hat\xi_i}
\right)
-
\frac{\partial \varphi}{\partial \hat\xi_i}(\xi)\notag\\
&=
\xi_i
+
\sum_j \hat\xi_j \frac{\partial \xi_j}{\partial \hat\xi_i}
-
\sum_j\frac{\partial \varphi}{\partial \xi_j}(\xi)
\frac{\partial \xi_j}{\partial \hat\xi_i}\notag\\
&=\xi_i.
\end{align}
Hence,
\begin{equation}
 \nabla_\psi(\hat\xi)=\xi
\end{equation}
\end{proof}
\begin{lemma}
\end{lemma}

\begin{figure}
\begin{center}
  \includegraphics[scale=0.9]{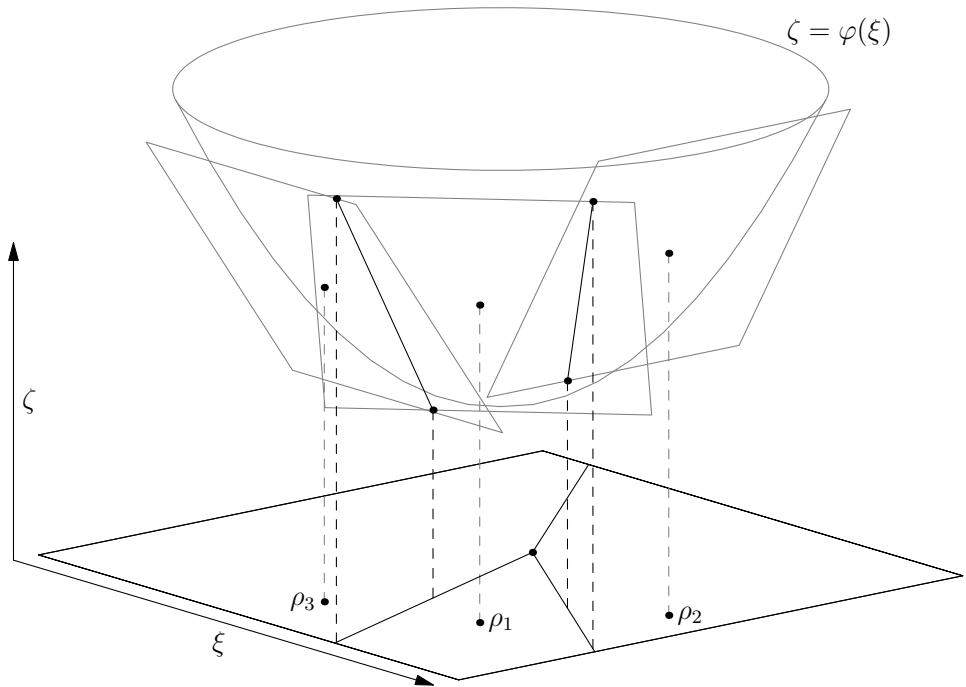}
\end{center}
\caption{An example of a Voronoi diagram obtained from a lower-envelope
 of tangent planes}
\label{fig:qdiv-envelope}
\end{figure}

\begin{figure}
\begin{center}
  \includegraphics{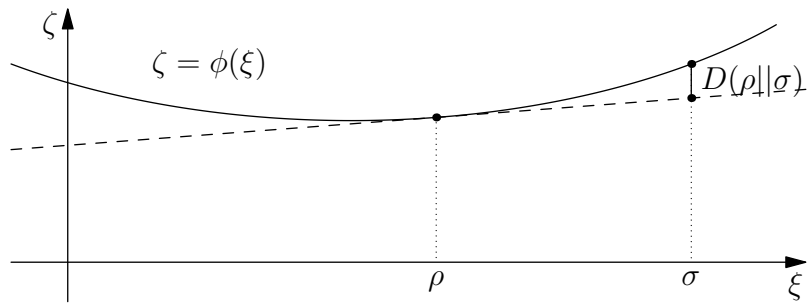}
\end{center} 
\caption{An explanation for a geometric meaning of the divergence}
\label{fig:qdiv-phi}
\end{figure}

The following is another important property of $V_D$.
\begin{theorem}
Consider the surface defined by
\begin{equation}\label{eq:lower-envelope-surface}
 \zeta = \psi(\rho(\xi)).
\end{equation}
 Then, the Voronoi diagram $V_D$ is obtained as a projection of a lower-envelope
 of tangent planes of this surface at the Voronoi sites (Fig.~\ref{fig:qdiv-envelope}).
\end{theorem}
\begin{proof}
 Because of Formula~(\ref{eq:div-phi}), the divergence $D(\rho||\sigma)$ can be
 considered as $\sigma$ minus the value of the tangent surface at
 $\sigma$ (Fig.~\ref{fig:qdiv-phi}).
\end{proof}

 Note that this is another intuitive proof of the fact that $V_D$ is
 linear. Actually, since the Voronoi diagram is a lower envelope of
 planes, its boundaries are linear.

\section{Euclidean Voronoi Diagram and divergence Voronoi diagram}

We show that the coincidence which happens in one-qubit case never
occurs in a higher level case. To show it, it is enough to look at some
section of the diagrams with some subspace. If the diagrams do not
coincide in the section, you can say they are different. It is stated as
follows:
\begin{theorem}
Suppose that $d\geq 3$ and that the space of general quantum states is
expressed as Equation~(\ref{multi-level-rho}). Then, if given Voronoi
sites are in a general position, the Voronoi diagram with respect to the
Euclidean distance and the Voronoi diagram with respect to the quantum
divergence $V_D^*$ are different.
\end{theorem}

Note that the diagramed considered here is $V_D^*$ only. This is
because $V_D$ is not well-defined for pure sites. Actually, define that
\begin{align}
 \sigma_1&=
 \begin{pmatrix}
  1+2\varepsilon & &&&\\
  & \varepsilon & &&\\
  & & \ddots && \\
  & & & \varepsilon & \\
  &&&& -d\varepsilon
 \end{pmatrix},\notag\\
 \sigma_2&=
 \begin{pmatrix}
  1+\varepsilon & &&&\\
  & 2\varepsilon & &&\\
  & & \ddots && \\
  & & & \varepsilon & \\
  &&&& -d\varepsilon
 \end{pmatrix},\notag\\
 \sigma_3&=
 \begin{pmatrix}
  \varepsilon & &&&\\
  & 1+2\varepsilon & &&\\
  & & \ddots && \\
  & & & \varepsilon & \\
  &&&& -d\varepsilon
 \end{pmatrix}.
\end{align}
Then,
\begin{equation}
 \rho=
  \begin{pmatrix}
   1& & & & \\
   & -1 & & & \\
   & & 0 & & \\
   & & & \ddots & \\
   & & & & 0 
  \end{pmatrix}
\end{equation}
is on the bisector of $\sigma_1$ and $\sigma_3$, and is not on the
bisector of $\sigma_2$ and $\sigma_3$. However, 
\begin{equation}
 \lim_{\varepsilon\to0} \sigma_1
= \lim_{\varepsilon\to0} \sigma_2
= \begin{pmatrix}
   1&& &\\
   & 0 &&\\
   & & \ddots &\\
   & & & 0\\
  \end{pmatrix}.
\end{equation}
This shows the Voronoi diagram depends on how Voronoi sites converge.

Think the section of $\rho$ with a $d+1$ dimensional plane:
\begin{equation} \label{rep-section}
 \xi_{d+2} = \xi_{d+3} = \cdots = \xi_{d^2-1}.
\end{equation}
Then the section is expressed as:
\begin{equation}
 \label{section-rho}
\rho=
\begin{pmatrix}
  \frac{\xi_1+1}{d} & \frac{\xi_{d}-i\xi_{d+1}}{2}& & & \smash{\lower1.0ex\hbox{\bg 0}}\\
 \frac{\xi_{d}+i\xi_{d+1}}{2} & \frac{\xi_2+1}{d}& & & \\
 & & \ddots & & \\
 & & & \frac{\xi_{d-1}+1}{d}&\\
 \smash{\hbox{\bg 0}} & & & &
 \frac{-\sum_{i=1}^{d-1}\xi_i+1}{d}
\end{pmatrix}
.
\end{equation}
The elements of this matrix are 0 except diagonal, (0,1), and (1,0)
elements. This matrix is diagonalized with a unitary matrix as:
\begin{align}
 \label{diagonalized}
 \rho=
 \begin{pmatrix}
  X & 0 \\
  0 & I_{d-2}
 \end{pmatrix} 
 \begin{pmatrix}
  \lambda_1 & & & & & \\
  & \lambda_2 & & & & \\
  & &\frac{\xi_3+1}{d} & & & \\
  & & &\ddots & & \\
  & & & &\frac{\xi_{d-1}+1}{d} & \\
  & & & & &\frac{-\sum_{j=1}^{d-1} \xi_j + 1}{d}
 \end{pmatrix} 
 \begin{pmatrix}
  X^* & 0 \\
  0 & I_{d-2}
 \end{pmatrix},
\end{align}
where
\begin{equation}
 r = \sqrt{\frac{(\xi_1-\xi_2)^2}{d^2}+\xi_d^2+\xi_{d+1}^2},
\end{equation}
\begin{align}
 \lambda_1=\frac{\xi_1+\xi_2+2}{2d}+\frac{r}{2}, \\
 \lambda_2=\frac{\xi_1+\xi_2+2}{2d}-\frac{r}{2},
\end{align}
\begin{align}
 X=
 \left(
 {
 \begin{array}{cc}
 \displaystyle\frac{\frac{\xi_d - i\xi_{d+1}}{2}}
 {\sqrt{R_+}}
 &
 \displaystyle\frac{\frac{\xi_d - i\xi_{d+1}}{2}}
 {\sqrt{R_-}}
 \vspace{3mm}
 \\
 \displaystyle \frac{\frac{\xi_2-\xi_1}{2d}+\frac{r}{2}}
 {\sqrt{R_+}}
 &
 \displaystyle \frac{\frac{\xi_2-\xi_1}{2d}-\frac{r}{2}}
 {\sqrt{R_-}}
 \end{array}
 }
 \right),\\
 R_+ =
  \frac{\xi_d^2+\xi_{d+1}^2}{4}+\left(\frac{\xi_2-\xi_1}{2d}+\frac{r}{2}\right)^2,\\
 R_- =
  \frac{\xi_d^2+\xi_{d+1}^2}{4}+\left(\frac{\xi_2-\xi_1}{2d}-\frac{r}{2}\right)^2.
\end{align}

Now we will figure out the necessary and sufficient condition for the diagonal
matrix of Equation~(\ref{diagonalized}) to be rank $1$. For that condition to
hold, the following three cases can be considered:
\paragraph{Case 1}
(only $d$-th raw of the matrix is non-zero)
\begin{equation}
 \xi_1=\xi_2=\cdots=\xi_{d-1}=-1,\, \xi_d=\xi_{d+1}=0.\nonumber
\end{equation}
\paragraph{Case 2}
 (only one $i$-th raw ($3\leq i \leq d-1$) is non-zero)
\begin{equation}
 \xi_1=\xi_2=-1,\xi_d=\xi_{d+1}=0,\nonumber
\end{equation}
all of $\xi_j\,(3\leq j \leq d-1)$ are $-1$ except one (let its index to be $k$) and 
$\xi_k=d-3$.

\paragraph{Case 3}
(only $\lambda_2$ is non-zero)
\begin{equation}
 \label{case3}
 \xi_1+\xi_2=d-2,\,
\frac{\xi_2- \xi_1}{d^2} + (\xi_d^2+\xi_{d+1}^2) = 1,
  \xi_3=\xi_4=\cdots=\xi_{d-1}=-1.
\end{equation}
Note that it is impossible that only $\lambda_1$ is non-zero.
In both Case~1 and Case~2, the set of points that satisfies the condition
is just one point, so our main interest is Case~3. The
set of points that satisfies this condition is a manifold. Actually,
Case~3 satisfies
\begin{equation}\label{ellipsoid-pure}
\frac{(d-2- 2\xi_1)^2}{d^2} + (\xi_d^2+\xi_{d+1}^2) = 1, \,
\end{equation}
and this is an ellipsoid.

Then we prepare for workout of the divergence.
The log of $\rho$
is expressed as:
\begin{align}
  \log \rho=
 \begin{pmatrix}
  X & 0 \\
  0 & I_{d-2}
 \end{pmatrix}  
 \begin{pmatrix}
  \log\lambda_1 & & & & & \\
  & \hspace{-10mm}\log\lambda_2 & & & & \\
  & &\hspace{-10mm}\log\frac{\xi_3+1}{d} & & & \\
  & & &\hspace{-10mm}\ddots & & \\
  & & & &\hspace{-10mm}\log\frac{\xi_{d-1}+1}{d} & \\
  & & & & &\hspace{-10mm}\log\frac{-\sum_{j=1}^{d-1} \xi_j + 1}{d}
 \end{pmatrix}  
 \begin{pmatrix}
   X^* & 0 \\
  0 & I_{d-2}
 \end{pmatrix}.
\end{align}
Thus, we obtain
\begin{align}
   \Tr \sigma & \log \rho =
 \frac{\eta_1+1}{d}\cdot\frac{\xi_d^2+\xi_{d+1}^2}{4}\left[\frac{\log
 \lambda_1}{R_+}+\frac{\log \lambda_2}{R_-}\right]\nonumber\\
 & +\frac{\eta_d\xi_d + \eta_{d+1}\xi_{d+1}}{2}
 \left[
 \frac{\frac{\xi_2-\xi_1}{2d}+\frac{r}{2}}{R_+}\!\log \lambda_1 +
 \frac{\frac{\xi_2-\xi_1}{2d}-\frac{r}{2}}{R_-}\!\log \lambda_2 
 \right]\nonumber\\
 & + \frac{\eta_2+1}{d}\left[
 \frac{\left(\frac{\xi_2-\xi_1}{2d}+\frac{r}{2}\right)^2}{R_+}\log \lambda_1 +
 \frac{\left(\frac{\xi_2-\xi_1}{2d}-\frac{r}{2}\right)^2}{R_-}\log \lambda_2
 \right]+ \frac{1-\xi_1-\xi_2}{d}.
\end{align}

With some workout, we get
\begin{equation}
  R_+ = r\left(\frac{\xi_2-\xi_1}{2d}+\frac{r}{2}\right),
 R_- = -r\left(\frac{\xi_2-\xi_1}{2d}-\frac{r}{2}\right).
\end{equation}
Using these fact and the assumption $\eta_1+\eta_2=\xi_1+\xi_2=d-2$, we get
\begin{equation}
  \label{simplified-rho}
 \Tr\sigma\log\rho=
 \left[
 \frac{\eta_d\xi_d+\eta_{d+1}\xi_{d+1}}{2r}
 +\frac{2\left(\eta_1-\frac{d-2}{2}\right)\left(\xi_1-\frac{d-2}{2}\right)}{d^2r}
 \right]\log \frac{\lambda_1}{\lambda_2}
 +\frac{1}{2}\log \lambda_1 \lambda_2.
\end{equation}

Next we think of a Voronoi diagram with only two regions for
simplicity. It is enough for our objective. Let $\sigma$ and
$\tilde{\sigma}$ be two sites, and suppose that $\rho$ moves along the
boundary of the Voronoi regions.  Suppose that $\sigma$ and
$\tilde{\sigma}$ are parameterized by $\{\eta_j\}$ and
$\{\tilde{\eta}_j\}$ respectively in the same way as $\rho$.

We consider what happens if $r(0\leq r<1)$ is fixed and the following holds:
\begin{equation}
 \label{ellipsoid-shrink}
 \xi_1+\xi_2=d-2,\, \xi_3=\cdots=\xi_{d-1}=-1.
\end{equation}
The condition $0\leq r <1$ means that $\rho$ is semi-positive and not a
pure state while $r=1$ in pure states.  In other words, we regard that
$\rho$ is on the same ellipsoid obtained by shrinking the ellipsoid
expressed by Equation~(\ref{ellipsoid-pure}). These settings are in
order to take a limit of a diagram to get a diagram in the pure
states. Taking the limit $r\to 1$, we can get a condition for pure
states. This procedure is analogous to the method used in
Chapter~\ref{chap:one-qubit}.

Now to think of the shape of boundary, we have to solve the equation
\begin{equation}
 D(\sigma||\rho)=D(\tilde{\sigma}||\rho),
\end{equation}
and this is equivalent to
\begin{equation}
  \Tr (\sigma - \tilde{\sigma})\log\rho = 0.
\end{equation}
Using Equation~(\ref{simplified-rho}), we obtain
\begin{align}
   \label{boundary-result}
 \Tr &(\sigma - \tilde{\sigma})\log\rho =\notag\\
 &\frac{1}{2r}
 \left[
 (\eta_d\!-\!\tilde{\eta}_d)\xi_d + (\eta_{d+1} \!-\! \tilde{\eta}_{d+1})\xi_{d+1} +\frac{4(\eta_1\!-\!\tilde{\eta}_1)\left(\xi_1\!-\!\frac{d-2}{2}\right)}{d^2}
 \right]
 \log \frac{\lambda_1}{\lambda_2}.
\end{align}
Here when $r=0$, this is zero because $\lambda_1/\lambda_2=1$. In that
case, $\rho$ can take only one point, but we do not have to care about
this case because we are going to take the limit $r\to1$. From now on,
we suppose $r>0$ and that means $\lambda_1/\lambda_2\neq 1$.

Hence we get the following equation that holds in the boundary of the
Voronoi diagram:
\begin{equation}
 \label{boundary-divergence}
(\eta_d-\tilde{\eta}_d)\xi_d + (\eta_{d+1} - \tilde{\eta}_{d+1})\xi_{d+1}
+\frac{4(\eta_1-\tilde{\eta}_1)\left(\xi_1-\frac{d-2}{2}\right)}{d^2}
=0.
\end{equation}
Consequently, taking the limit $r\to 1$, we get
Equation~(\ref{boundary-divergence}) as the expression of the boundary in
pure states.

A careful inspection of Equation~(\ref{boundary-divergence}) tells us a
geometric interpretation of this boundary. We obtain the following
theorem:
\begin{theorem}
 On the ellipsoid of the pure states which appears in the section with
 the $(d+1)$-plain defined above, if transferred by a linear transform which
 maps the ellipsoid to a sphere, the Voronoi diagram with respect to the
 divergence coincides with the one with respect to the geodesic
 distance.
\end{theorem}
\begin{proof}
 Think of the affine transform defined by
\begin{equation}
  \left(
\begin{array}{c}
 x\\
 y\\
 z
\end{array}
\right)
=
 \left(
\begin{array}{c}
 \frac{\xi_1-\frac{d-2}{2}}{\frac{d}{2}}\\
 \xi_d\\
 \xi_{d+1}
\end{array}
\right),
\end{equation}
then Equation~(\ref{boundary-divergence}) is expressed as
\begin{equation}
 x'(x-\tilde{x})+y'(y-\tilde{y})+z'(z-\tilde{z})=0,
\end{equation}
while Equation~(\ref{ellipsoid-pure}) becomes
\begin{equation}
  x^2+y^2+z^2=1.
\end{equation}
Thus when $(x,y,z)$ and $(\tilde{x},\tilde{y},\tilde{z})$ are fixed, the
 point $(x',y',z')$ which stand for $\eta$ runs along the geodesic.
\end{proof}

Now we work out the Voronoi diagram with respect to Euclidean
distance. Under the assumption above, the Euclidean distance is
expressed as
\begin{align}
 \lefteqn{ d(\sigma,\rho)}\nonumber\\
 &=(\eta_1-\xi_1)^2+\!(\eta_2-\xi_2)^2+
 (\eta_d-\xi_d)^2 + (\eta_{d+1}-\xi_{d+1})^2\nonumber\\
 &=2(\eta_1-\xi_1)^2+
 (\eta_d-\xi_d)^2 + (\eta_{d+1}-\xi_{d+1})^2,
\end{align}
and we get the equation for boundary as
\begin{multline}
  \label{boundary-euclidean}
 d(\sigma,\rho)-d(\tilde{\sigma},\rho)
 =-4(\eta_1-\tilde{\eta}_1)\xi_1
 -2(\eta_d-\tilde{\eta}_d)\xi_d
 -2(\eta_{d+1}-\tilde{\eta}_{d+1})\xi_{d+1}
 +2(\eta_1^2-\tilde{\eta}_1^2)\\
 +(\eta_d^2-\tilde{\eta}_d^2)
 +(\eta_{d+1}^2-\tilde{\eta}_{d+1}^2)=0.
\end{multline}
By comparing the coefficients of $\xi_1$, $\xi_d$, and $\xi_{d+1}$,
we can tell that the boundaries expressed by
Equation~(\ref{boundary-divergence}) and (\ref{boundary-euclidean}) are
different. To show how different they are, we give some examples in the
rest of this section.

\begin{example}
 Suppose that $(\eta_1,\eta_d,\eta_{d+1})=(d-1,0,0)$ and
$(\tilde{\eta_1},\tilde{\eta}_d,\tilde{\eta}_{d+1})=(-1,0,0)$, then the
boundary is $\xi_1=\frac{d-2}{2}$ for the both diagrams.
\end{example}

\begin{example}
 Suppose that $(\eta_1,\eta_d,\eta_{d+1})=(0,1,0)$ and
$(\tilde{\eta_1},\tilde{\eta}_d,\tilde{\eta}_{d+1})=(0,-1,0)$, then
the boundary is, for both the divergence and Euclidean distance,
expressed by $\xi_{d+1}=0$.
\end{example}

\begin{figure}[!ht]
\begin{center}
 \includegraphics[scale=.3,clip]{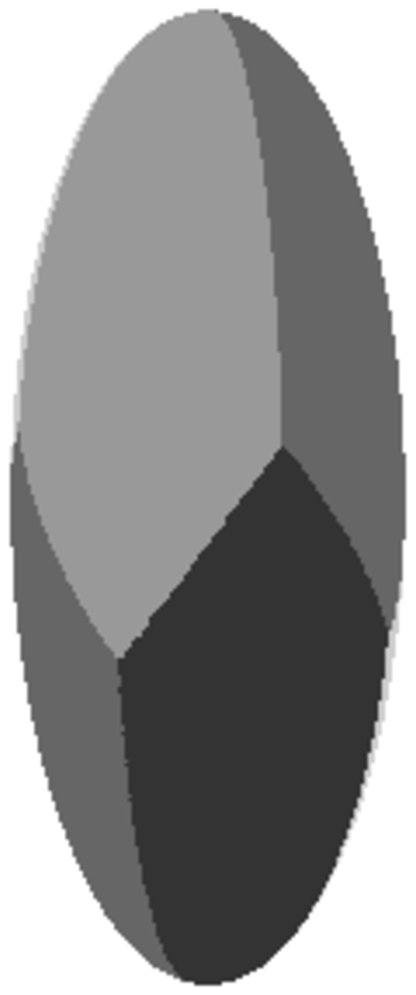}\hspace{-1cm}
 \includegraphics[scale=.3,clip]{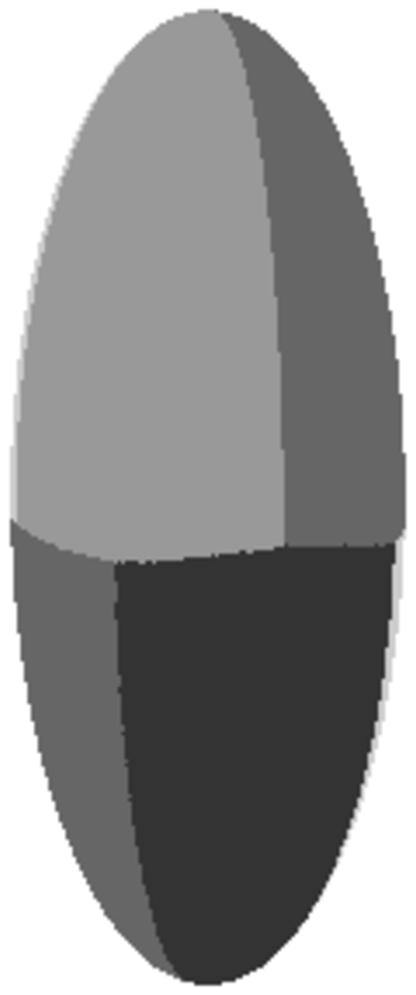} 
\caption[An example of a diagram which appears as a section of a
 Voronoi diagram in three level quantum state space.]  
{An example of a diagram which appears as a section of a Voronoi 
 diagram in three level quantum state space. The left is the diagram 
 by the divergence, and the right is by the Euclidean distance.} 
 \label{fig-example3}
\end{center}
\end{figure}

\begin{example}
Consider the Voronoi diagram with the following eight sites:
\begin{align}
    \left(\frac{d-2}{2}+\frac{d}{2\sqrt{3}},\,\pm\frac{1}{\sqrt{3}},\,\pm\frac{1}{\sqrt{3}}\right),\nonumber\\
  \left(\frac{d-2}{2}-\frac{d}{2\sqrt{3}},\,\pm\sqrt{\frac{2}{3}},\,0\right),\nonumber\\
  \left(\frac{d-2}{2}-\frac{d}{2\sqrt{3}},\,0,\,\pm\sqrt{\frac{2}{3}}\right),
\end{align}
where $\pm$'s mean all the possible combinations. Then the Voronoi diagrams
 look like Fig.\,\ref{fig-example3}. This figure is also for
 $d=5$. Obviously they are different.
\end{example}

\section{Other Parameterization}
We have shown that the Euclidean Voronoi diagram and the divergence
Voronoi diagram are different as far as the regular parameterization of
a quantum state is used. However, the parameterization of the quantum
state is not unique. The condition that a matrix is Hermitian and its
trace is one is expressed in another way. Here, we show an example of
the parameterization with which the section of the diagrams shown in the
previous section coincides. Note that it just means the sections are the
same and it cannot conclude that the diagrams are globally the same.

It is difficult to investigate all the possible parameterization, or
equally all the possible embedding of quantum state space into the Euclidean
space. This section is intended only to show an example to which the proof
of the previous section cannot be applied.

Suppose that a density matrix is parameterized in another way than
Equation (\ref{multi-level-rho}) as
\begin{equation}
\rho=
\left(
 \begin{array}{ccccc}
\displaystyle\frac{\frac{d}{\sqrt{2}}\xi_1+1}{d} &
\!\!\!\displaystyle\frac{\xi_d-i\xi_{d+1}}{2} &
\cdots & &   
\displaystyle\frac{\xi_{3d-4}-i\xi_{3d-3}}{2} 
\smallskip\\
\displaystyle\frac{\xi_d+i\xi_{d+1}}{2} & 
\!\!\!\displaystyle\frac{\frac{d}{\sqrt{2}}\xi_2+1}{d} &
\cdots & & 
\displaystyle\frac{\xi_{5d-8}-i\xi_{5d-7}}{2} 
\smallskip\\
\vdots & & \!\!\!\!\ddots & & \vdots \\
\displaystyle\frac{\xi_{3d-6}+i\xi_{3d-5}}{2} & \cdots& &
\!\!\!\!\!\!\!\displaystyle\frac{\frac{d}{\sqrt{2}}\xi_{d-1}+1}{d}& 
\displaystyle\frac{\xi_{d^2-2}-i\xi_{d^2-1}}{2}\smallskip\\
\displaystyle\frac{\xi_{3d-4}+i\xi_{3d-3}}{2} & \cdots& &
\!\!\!\!\!\!\!\displaystyle\frac{\xi_{d^2-2}+i\xi_{d^2-1}}{2} &
\displaystyle
 \frac{-\frac{d}{\sqrt{2}}\sum_{i=1}^{d-1}\xi_i+1}{d}
\end{array}
\right).
\end{equation}
In other word, we will think what happens if $\xi_i(0\leq i \leq d-2)$
is replaced by $\frac{d}{\sqrt{2}}\xi_i$. 
We will show that Voronoi diagrams with respect to Euclidean distance
and the divergence coincide for pure states in the section expressed by (\ref{rep-section})

Under this parameterization, Equation (\ref{ellipsoid-pure}) is
expressed as
\begin{equation}
\label{ellipsoid-other}
 2\left(\xi_1-\frac{d-2}{\sqrt{2}d}\right)^2+\xi_d^2+\xi_{d+1}=1,
\end{equation}
and Equation (\ref{boundary-divergence}) becomes
\begin{equation}
\label{boundary-other}
 (\eta_d-\tilde{\eta}_d)\xi_d +(\eta_{d+1}-\tilde{\eta}_{d+1})\xi_{d+1}
+2(\eta_1-\tilde{\eta}_1)\left(\xi_1- \frac{d-2}{\sqrt{2}d}\right).
\end{equation}
Assuming that Equation (\ref{ellipsoid-other}) holds, the Euclidean
distance is calculated as
\begin{align}
d(\rho,\sigma)&=2(\eta_1-\xi_1)^2 + (\eta_d-\xi_d)^2 +
 (\eta_{d+1}-\xi_{d+1})^2 \nonumber\\
 &=2\left[\left(\eta_1-\frac{d-2}{\sqrt{2}d}\right)-\left(\xi_1-\frac{d-2}{\sqrt{2}d}\right)\right]^2
 + (\eta_d-\xi_d)^2 + (\eta_{d+1}-\xi_{d+1})^2 \nonumber\\
 &=\left[2\left(\eta_1-\frac{d-2}{\sqrt{2}d}\right)^2 + \eta_d^2 + \eta_{d+1}^2\right]
 +\left[2\left(\xi_1-\frac{d-2}{\sqrt{2}d}\right)^2 + \xi_d^2 + \xi_{d+1}^2\right]\nonumber\\
 &\qquad -4\left(\eta_1-\frac{d-2}{\sqrt{2}d}\right)\left(\xi_1-\frac{d-2}{\sqrt{2}d}\right)
 -2\eta_d \xi_d - 2\eta_{d+1} \xi_{d+1} \nonumber\\ 
&=1+1
 -4\left(\eta_1-\frac{d-2}{\sqrt{2}d}\right)\left(\xi_1-\frac{d-2}{\sqrt{2}d}\right) -2\eta_d
 \xi_d - 2\eta_{d+1} \xi_{d+1}.
\end{align}
Therefore,
\begin{align}
& d(\rho,\sigma)-d(\rho,\tilde{\sigma})=0\nonumber\\ \Longleftrightarrow
& -4(\eta_1-\tilde{\eta}_1)\left(\xi_1-\frac{d-2}{\sqrt{2}d}\right)
-2(\eta_d - \tilde{\eta}_d) \xi_d - 2(\eta_{d+1}-\tilde{\eta}_{d+1})
\xi_{d+1}=0.
\end{align}
This is equivalent to Equation (\ref{boundary-other}), and we have shown
the Voronoi diagrams are the same.

Again, we mention this does not conclude that the diagrams are the same
globally. We conjecture that they are different. More generally, we
conjecture that even with {\em any} parameterization of quantum state
space, the Euclidean Voronoi diagram and the divergence Voronoi diagrams
are different.

\section{Bures distance and Fubini-Study Distance}

In this section, we prove the following theorem:
\begin{theorem}
 In a general level quantum system, for pure states,
 the following diagrams are equivalent:
\begin{itemize}
 \item diagram with respect to the divergence,
       i.e.\ $\mathrm{Closure}(V_D)^*\cap \mathcal{S}^\mathrm{pure}$
\item diagram with respect to Fubini-Study distance
\item diagram with respect to Bures distance
\end{itemize}
\end{theorem}

The equivalence between the Fubini-Study diagram and the Bures diagram
is obvious because
\begin{align}
\label{equivalence_of_equations}
 & d_\mathrm{B}(\rho,\sigma)\leq d_\mathrm{B}(\rho,\tilde{\sigma})\nonumber\\
\Longleftrightarrow\quad & \sqrt{1-\Tr\rho\sigma}\leq\sqrt{1-\Tr\rho\tilde{\sigma}}\nonumber\\
\Longleftrightarrow\quad & \Tr\rho\sigma\geq \Tr\rho\tilde{\sigma}\nonumber\\
\Longleftrightarrow\quad & \cos^{-1}\sqrt{\Tr\rho\sigma}\leq
 \cos^{-1}\sqrt{\Tr\rho\tilde{\sigma}}\nonumber\\
\Longleftrightarrow\quad & 
d_\mathrm{FS}(\rho,\sigma)\leq d_\mathrm{FS}(\rho,\tilde{\sigma})\nonumber\\
\end{align}
Hence we will show the coincidence between the diagram by Bures distance
and the diagram by divergence.

For $\epsilon>0 \in \mathbb{R}$ we define
\begin{equation}
 \rho_\epsilon = 
  X
  \begin{pmatrix}
   1-(d-1)\epsilon & & & \\
    & \epsilon & &  \\
   & & \ddots & \\
   & & & \epsilon 
  \end{pmatrix}
  X^*, 
\end{equation}
where
\begin{equation}
 X \text{ is a unitary complex matrix expressed by } X=(x_{ij}).
\end{equation}
In other words, $\rho_\epsilon$ is parameterized by one
real parameter $\epsilon$ and $d^2$ complex parameters $x_{ij}$. Note that
in this parameterization, there might be more than one representations for one
given density matrix.

This $\rho_\epsilon$ converges to a pure state when you take the limit
$\epsilon \downarrow 0$. Moreover $\rho_\epsilon$ has rank $d$ if
$\epsilon>0$, and all pure states in space of $d$ level system can be
expressed by the limit of $\rho_\epsilon$. Actually, taking the limit,
we obtain
\begin{equation}
 \lim_{\epsilon \downarrow 0} \rho_\epsilon = 
  \begin{pmatrix}
   x_{11} \bar{x}_{11} & x_{11} \bar{x}_{12} & \cdots & x_{11} \bar{x}_{1d}\\
   x_{12} \bar{x}_{11} & x_{12} \bar{x}_{12} & \cdots & x_{12} \bar{x}_{1d}\\
   \vdots & \vdots & \ddots & \vdots\\
   x_{1d} \bar{x}_{11} & x_{1d} \bar{x}_{12} & \cdots & x_{1d} \bar{x}_{1d}
  \end{pmatrix}
  =
  \begin{pmatrix}
   x_{11} \\
   x_{12} \\
   \vdots \\
   x_{1d}  
  \end{pmatrix}
  \begin{pmatrix}
   \bar{x}_{11} &
   \bar{x}_{12} &
   \cdots &
   \bar{x}_{1d}    
  \end{pmatrix}
,
\end{equation}
and any pure state can be written in this way.

About the divergence, we will calculate the boundary of the Voronoi
diagram as a set of $\rho_\epsilon$'s for a fixed $\epsilon$, and take
its limits. Considering the equivalence of the equations stated in
(\ref{equivalence_of_equations}), we can restate the problem:
\begin{claim}
\label{claim:equality_of_sets}
 For given pure states $\sigma_1$ and $\sigma_2$, let
\begin{equation}
 S_\epsilon(\sigma_1,\sigma_2)= 
  \Bigl\{\rho_\epsilon \Bigm|
   D(\sigma_1||\rho_\epsilon)=D(\sigma_2||\rho_\epsilon)\Bigr\}
\end{equation}
and
\begin{equation}
 T(\sigma_1,\sigma_2)=\Bigl\{ \rho \Bigm| \Tr(\sigma_1\rho)=\Tr(\sigma_2\rho) \Bigr\}.
\end{equation}
Then, for any $\sigma_1, \sigma_2$ the following equation holds:
\begin{equation}
 \lim_{\epsilon\to 0} S_\epsilon(\sigma_1,\sigma_2) = T(\sigma_1,\sigma_2)
\end{equation}
\end{claim}
\begin{proof}
 Suppose that $\sigma_i$'s are expressed as 
\begin{equation}
 \sigma_i = 
  \begin{pmatrix}
   y_{i1}\\
   y_{i2}\\
   \vdots\\
   y_{id}
  \end{pmatrix}
  \begin{pmatrix}
   \bar{y}_{i1} & \bar{y}_{i2} & \cdots & \bar{y}_{id}
  \end{pmatrix}
\quad (i=1,2).
\end{equation}
Defining the matrix $Y$ by
\begin{equation}
 Y=
\begin{pmatrix}
 y_{11} & y_{21} & 0 &\cdots & 0\\
 y_{12} & y_{22} & 0 &\cdots & 0\\
 \vdots & \vdots & \vdots &\ddots & \vdots \\
 y_{1d} & y_{2d} & 0 &\cdots &0
\end{pmatrix},
\end{equation}
we can also express
\begin{equation}
  \sigma_1 = Y
\begin{pmatrix}
 1 & & &&\\
 & 0 & &&\\
 & & 0 &&\\
 & & &\ddots & \\
 & & & &0
\end{pmatrix}
Y^*, \quad
 \sigma_2 = Y 
\begin{pmatrix}
 0 & & &&\\
 & 1 & &&\\
 & & 0 &&\\
 & & &\ddots & \\
 & & & &0
\end{pmatrix}
Y^*.
\end{equation}
Now we obtain
\begin{align}
\label{expansion-general_dim_boundary}
  &D(\sigma_1||\rho)-D(\sigma_2||\rho) \nonumber\\
 =&\Tr \left( \sigma_1 \log \sigma_1 - \sigma_1 \log \rho \right)
 - \Tr \left( \sigma_2 \log \sigma_2 - \sigma_2 \log \rho \right)\nonumber\\
 =& \Tr (\sigma_2-\sigma_1)\log \rho\qquad (\text{because } \Tr \sigma_1 \log
 \sigma_1 = \Tr \sigma_2 \log \sigma_2 =0 ) \nonumber\\
 =& \Tr Y
\begin{pmatrix}
 -1 & & &&\\
 & 1 & &&\\
 & & 0 &&\\
 & & &\ddots & \\
 & & & &0
\end{pmatrix}
 Y^* X
 \begin{pmatrix}
  \log (1-(d-1)\epsilon) & & & \\
  & \log \epsilon &&\\
  && \ddots &\\
  &&& \log \epsilon 
 \end{pmatrix}
 X^* \nonumber\\
 =& \Tr  
\begin{pmatrix}
 -1 & & &&\\
 & 1 & &&\\
 & & 0 &&\\
 & & &\ddots & \\
 & & & &0
\end{pmatrix}
 Y^* X
 \begin{pmatrix}
  \log (1-(d-1)\epsilon) & & & \\
  & \log \epsilon &&\\
  && \ddots &\\
  &&& \log \epsilon 
 \end{pmatrix}
 X^* Y
\end{align}
Denoting $X^* Y$ by $Z=(z_{ij})$,
 (\ref{expansion-general_dim_boundary}) can be expanded further as
\begin{align}
 \text{(\ref{expansion-general_dim_boundary})}= & \Tr  
 \begin{pmatrix}
 -1 & & &&\\
 & 1 & &&\\
 & & 0 &&\\
 & & &\ddots & \\
 & & & &0
 \end{pmatrix}
 Z^*
 \begin{pmatrix}
  \log (1-(d-1)\epsilon) & & & \\
  & \log \epsilon &&\\
  && \ddots &\\
  &&& \log \epsilon 
 \end{pmatrix}
 Z\\
 =& \Tr  
 \begin{pmatrix}
 -1 & & &&\\
 & 1 & &&\\
 & & 0 &&\\
 & & &\ddots & \\
 & & & &0
 \end{pmatrix}
 Z^*\nonumber\\
&\qquad\times
 \begin{pmatrix}
  z_{11}\log (1-(d-1)\epsilon) & z_{12}\log (1-(d-1)\epsilon)& \cdots & 0 \\
  z_{21}\log \epsilon & z_{22}\log \epsilon & \cdots & 0 \\
  \vdots & \vdots & \ddots & \vdots\\
  z_{d1}\log \epsilon & z_{d2}\log \epsilon & \cdots & 0
 \end{pmatrix}\nonumber\\
=&
  \Tr
  \begin{pmatrix}
   -1 & & &&\\
   & 1 & &&\\
   & & 0 &&\\
   & & &\ddots & \\
   & & & &0
  \end{pmatrix}
\nonumber\\
&\times
\begin{pmatrix}
    \bar{z}_{11} z_{11}\log (1-(d-1)\epsilon)+\bar{z}_{21} z_{21}\log
  \epsilon+\cdots+ \bar{z}_{d1} z_{d1}\log \epsilon  & ? & & & \mspace{-50mu}\\
  ? &
 \mspace{-400mu}
 \bar{z}_{12} z_{12}\log (1-(d-1)\epsilon)+\bar{z}_{22} z_{22}\log
  \epsilon+\cdots+ \bar{z}_{d2} z_{d2}\log \epsilon  & &  & \mspace{-100mu} \\
   & & 
 \mspace{-400mu}
 0 & &  \mspace{-100mu}\\
   & & & 
 \mspace{-200mu}
 \ddots &  \mspace{-100mu}\\
   & & & & 
 \mspace{-100mu}
 0
\end{pmatrix}
\nonumber\\
 &\text{( ``?''
 stands for a non-zero element that does not}\nonumber\\
 &\quad \text{
 affect the result of the
 calculation)}
 \nonumber\\
=&
  -[\bar{z}_{11} z_{11}\log (1-(d-1)\epsilon)+\bar{z}_{21} z_{21}\log
  \epsilon+\cdots+ \bar{z}_{d1} z_{d1}\log \epsilon]\nonumber\\
 & \quad + [\bar{z}_{12} z_{12}\log (1-(d-1)\epsilon)+\bar{z}_{22} z_{22}\log
  \epsilon+\cdots+ \bar{z}_{d2} z_{d2}\log \epsilon].
\end{align}
Here note that $z_{ij}=0$ for $j\geq 3$ because the elements $X^*$ are
 all zero except the first two columns. Thus, we get
\begin{align}
\label{equation_for_condition_of_divergence}
 &D(\sigma_1||\rho)-D(\sigma_2||\rho)=0 \nonumber\\
\Longleftrightarrow\; & 
  -\Bigl[|z_{11}|^2\log \bigl(1-(d-1)\epsilon\bigr)+|z_{21}|^2\log
  \epsilon+\cdots+ |z_{d1}|^2\log \epsilon\Bigr]\nonumber\\
 &\quad + \Bigl[|z_{12}|^2\log \bigl(1-(d-1)\epsilon\bigr)+|z_{22}|^2\log
  \epsilon+\cdots+ |z_{d2}|^2\log \epsilon\Bigr]=0\nonumber\\
\Longleftrightarrow\; & 
  -\biggl\{|z_{11}|^2\Bigl[\log \bigl(1-(d-1)\epsilon\bigr)-\log
 \epsilon \Bigr]+\log \epsilon\biggr\} \nonumber\\
 &\quad + \biggl\{|z_{12}|^2\Bigl[\log \bigl(1-(d-1)\epsilon\bigr) -\log
 \epsilon\Bigr]+ \log \epsilon\biggr\}=0\nonumber\\
\Longleftrightarrow\; & 
  \Bigl(|z_{12}|^2-|z_{11}|^2\Bigr)\Bigl[\log \bigl(1-(d-1)\epsilon\bigr) 
 -\log \epsilon \Bigr]=0.
\end{align}
Here we used the fact that
\begin{align}
 |z_{11}|^2+|z_{21}|^2+\cdots+|z_{d1}|^2
=
 |z_{12}|^2+|z_{22}|^2+\cdots+|z_{d2}|^2=1,
\end{align}
 because $X$ is unitary and vectors $(y_{11},y_{12},\ldots,y_{1d})$ and
 $(y_{21},y_{22},\ldots,y_{2d})$ have a unit length.  For Equation
 (\ref{equation_for_condition_of_divergence}) to hold independent of
 $\epsilon$, the necessary and sufficient condition is
\begin{equation}
|z_{11}|=|z_{12}|.
\end{equation}
Writing down the elements by $x_{ij}$'s and $y_{ij}$'s, we obtain the
 condition written as
\begin{equation}
\label{bures-divergnece-result}
|\bar{x}_{11} y_{11}+\bar{x}_{12} y_{12}+\cdots +\bar{x}_{1d} y_{1d}|
=|\bar{x}_{11} y_{21}+\bar{x}_{12} y_{22}+\cdots +\bar{x}_{1d}y_{2d}|.
\end{equation}

Now we consider the condition for Bures (or Fubini-Study)
 distance. Similarly we can extend the formula as follows:
\begin{align}
  &\Tr\sigma_1\rho_0-\Tr\sigma_2\rho_0 \nonumber\\
 =& \Tr Y
\begin{pmatrix}
 1 & & &&\\
 & -1 & &&\\
 & & 0 &&\\
 & & &\ddots & \\
 & & & &0
\end{pmatrix}
 Y^* X 
 \begin{pmatrix}
  1 & & & \\
  & 0 &&\\
  && 0 &\\
  &&& 0 
 \end{pmatrix}
 X^*\nonumber\\
=&
\Tr  
\begin{pmatrix}
 1 & & &&\\
 & -1 & &&\\
 & & 0 &&\\
 & & &\ddots & \\
 & & & &0
\end{pmatrix}
 Z^*
 \begin{pmatrix}
  1 & & & \\
  & 0 &&\\
  && 0 &\\
  &&& 0 
 \end{pmatrix}
 Z\nonumber\\
=&
\Tr  
\begin{pmatrix}
 1 & & &&\\
 & -1 & &&\\
 & & 0 &&\\
 & & &\ddots & \\
 & & & &0
\end{pmatrix}
 \begin{pmatrix}
  \bar{z}_{11} z_{11} & \bar{z}_{11} z_{12} & \cdots & \bar{z}_{11} z_{1d} \\
  \bar{z}_{12} z_{11} & \bar{z}_{12} z_{12} & \cdots & \bar{z}_{12} z_{1d} \\
  \vdots & \vdots & \ddots & \vdots \\
  \bar{z}_{1d} z_{11} & \bar{z}_{1d} z_{12} & \cdots & \bar{z}_{1d} z_{1d} 
 \end{pmatrix}\nonumber\\
=& \bar{z}_{11} z_{11}-\bar{z}_{12} z_{12}.
\end{align}
Thus,
\begin{equation}
 \Tr(\sigma_1-\sigma_2)\rho_0=0\;\Longleftrightarrow\;
  |z_{11}|=|z_{12}|,
\end{equation}
and finally we obtain the same condition as (\ref{bures-divergnece-result}).
\end{proof}

\section{Expected applications}
Although the Euclidean distance is an exception and the space is
restricted to pure states, we have shown that the coincidence of the
Voronoi diagrams also happens in three or higher system. This means some
problem about a distance for pure states can be translated into another
problem in another distance. In such a sense, we can say we have
clarified the structure of pure states as a space.

As is explained in Section~\ref{sec:one-qubit-meaning}, effectiveness of
the algorithm by Hayashi et al.\ to compute the Holevo capacity of one-qubit
channel is partially supported by the coincidence of Voronoi
diagrams. It is natural to think the extended coincidence for higher
level system might become useful in the extended algorithm. The extended
algorithm is introduced in Chapter~\ref{chap:numerical}, but we have not
found a concrete application in it. We also have to mention that the
main part of the theorem about the coincidence used to support the
effectiveness of the algorithm for one-qubit system is the one between
Euclidean Voronoi and the divergence Voronoi, and it is proved not to
happen in a higher level system.

The distances in a quantum state space are originally considered as a
tool to distinguish the states by measuring. We expect our result will
be used in the identification of states. Actually, how well a message is
coded in quantum system partially depends on how uniformly distributed
points we can get in a quantum state space. The uniformness of a point
set is preserved between multiple pseudo-distances if their Voronoi
diagrams are the same.

As a real vector space, a $d$-level quantum state space has a quadratic
dimension $d^2-1$. From the viewpoint of computational geometry, our
contribution is an implication about such a large dimensional space
about its actual computation. It provides a new category of research
fields: distortion measure in a high dimensional space.

\section{Summary of this chapter}
We have investigated whether the same thing as the coincidence of
Voronoi diagrams in the one-qubit system also occurs in three or higher
level system.  The Euclidean Voronoi diagram and the divergence Voronoi
diagram are proved to be different in three or higher level system. On
the other hand, for pure states, the coincidence of the Euclidean
distance, the Bures distance, and the Fubini-Study distance occurs even
in three or higher level system.

Some problems about a pseudo-distance for pure quantum states can be
translated into a problem about another pseudo-distance. There is still
no concrete application of it, but by the analogy of the numerical
computation of capacity of one-qubit channel, it is likely to become
useful in the future.

Another point is that we found a connection among measures which have
apparently have no relation. Especially, the connection between a distance
used for quantum database search and a pseudo-distance used in quantum
information theory is meaningful. 

\chapter{Numerical Computation and Experiment}
\label{chap:numerical}

\section{Overview}
We propose a new algorithm to compute the Holevo capacity of a quantum
channel especially for three or higher level system. Our algorithm is
based on a global optimization and never converges to a local
optimum. It is the main merit of our algorithm compared to the preceding
one.

Osawa and Nagaoka \cite{osawa01} proposed an algorithm to compute the
Holevo capacity, which is an extension of the algorithm by Arimoto and
Blahut to compute the capacity of a classical channel. The main deficit
of the algorithm by Osawa and Nagaoka is that it only computes a local
optimum, although Arimoto--Blahut's algorithm is guaranteed to reach
the global optimum.  It is because the objective function in the
optimization is concave for a classical capacity while it is not concave
for a quantum one.

The algorithm we propose is an extension of the one by Hayashi et
al.~\cite{hayashi05} and Oto et al.~\cite{oto04a,oto04b}.  We follow
their idea of using the smallest enclosing ball problem after
approximating the geometric object by a set of points.  However, to
solve the smallest enclosing ball problem, the method based on a
farthest Voronoi diagram explained in \cite{oto04a, oto04b} cannot be
directly applied to a higher level system because of its complexity.
Our main contribution is to show theoretically and practically that
Welzl's algorithm to solve the smallest enclosing ball problem is also
useful for a quantum state space.

\begin{figure}
\begin{center}
 \subfigure[In Osawa--Nagaoka's algorithm, the objective function is not
 concave and it might converge to a local optimum]{
 \includegraphics[angle=-90,scale=0.7,clip]{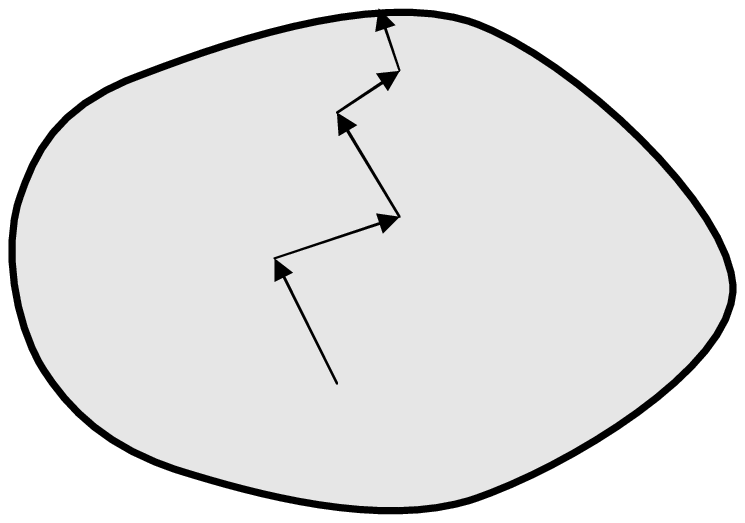} 
 \label{subfig:capacity1}
 }
 \hspace{7mm}
 \subfigure[In our proposed algorithm, the smallest enclosing ball
 problem is computed for plotted points and it is the global optimum]{
 \includegraphics[angle=-90,scale=0.7,clip]{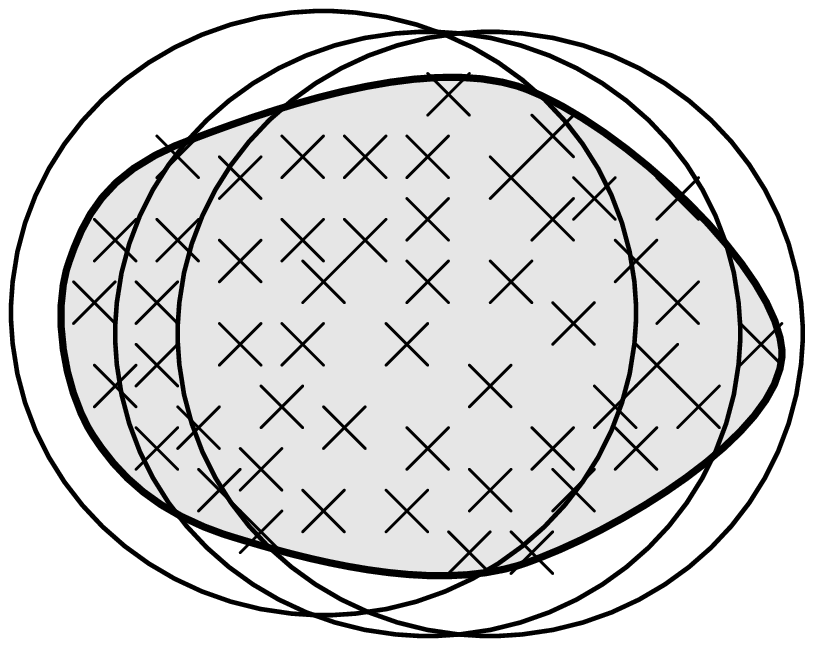} 
 \label{subfig:capacity2}
 }
\end{center}
\caption{Conceptual explanation of Osawa--Nagaoka's algorithm and our algorithm} \label{fig:capacity}
\end{figure}

The main merit of our algorithm is that it can find a global optimum
although it is only an approximate algorithm. The difference of the
ideas between our algorithm and Osawa--Nagaoka's one is conceptually
explained by Fig.~\ref{fig:capacity}.

\section{Smallest enclosing ball problem in a quantum state space}
\label{sec:seb-qdiv}

In this section, we show that Welzl's algorithm is also applicable for
the quantum state space with the quantum divergence as its
pseudo-distance.

Matou\v{s}ek and al.\ \cite{matousek96} showed that the smallest
enclosing problem belongs to the class called linear programming-type
({\em LP-type}) problem which is introduced as an abstraction of some
algorithm for linear programming. The LP-type problem is defined as
follows \cite{matousek96}. Think about the class of the optimization
problems defined by pairs $(H,w)$, where $H$ is a finite set and $w:2^H
\to W$ is a function with a values in a ordered set $W$. The goal is to
find the optimal set $B_H$ which satisfies
\begin{quote}
 $w(B_H)=w(H)$ and for any $G\subset B_H$, $w(B)<w(B_H)$.
\end{quote}
This problem is {\em LP-type} when the following axioms are satisfied.
\begin{axiom}[Monotonicity]
For any $F$ and $G$ with $F\subset G \subset H$, we have
\begin{equation}
 w(F)\leq w(G)
\end{equation}
\end{axiom}
\begin{axiom}(Locality)
For any $F\subset G \subset H$ with $(-\infty <)w(F)=w(G)$, and any $h\in H$,
\begin{equation}
 w(F)<w(F\cup \{h\}) \ \text{implies}\  w(G)<w(G\cup \{h\}).
\end{equation} 
\end{axiom}

The smallest enclosing ball problem in Euclidean space is obviously
belongs to this class. For the smallest enclosing ball problem with
respect to a given arbitrary distance in a given space, whether Welzl's
algorithm is also applicable, or in other words, has the same complexity
order as in Euclidean space is certified by checking if the axioms above
are satisfied.

Nielsen and Nock \cite{nielsen07a} showed Bregman divergence
satisfies the axioms above. We will show that the quantum divergence
also satisfies the axioms in the rest of this section, but it is by
mostly the same idea of Nielsen and Nock.

The latter axiom, locality, can be proved by showing the uniqueness of
the optimal ball. The uniqueness is follows from the linearity of the
bisectors (Theorem~\ref{th:vd-planer-1}).

Note that there are two types of the smallest enclosing ball problem can
be considered as
\begin{align}
 \mathrm{SEB}_{D}(\mathcal{P})
 = \min_{\rho\in \mathcal{S}^\mathrm{faithful} }
 \max_{\sigma\in \mathcal{P}} D(\rho||\sigma),\\
 \mathrm{SEB}_{D}^*(\mathcal{P})
 = \min_{\rho\in \mathcal{S}^\mathrm{faithful} }
 \max_{\sigma\in \mathcal{P}} D(\sigma||\rho),
\end{align} 
where $\mathcal{P}$ means a given set of quantum states. Our main
interest is the second type because it is used to compute the Holevo
capacity. The locality can be proved for both. First it is directly
proved for the first type. For the second type, we take dual of the
problem as in Theorem~\ref{th:dual-diagram}, and similarly in the dual
space, we can determine the smallest enclosing ball uniquely.

We have shown that for the smallest enclosing ball problem with respect
to the quantum divergence, we can reach the optimum following the same
process as Welzl's algorithm. However, note that it is only under the
assumption that the most primitive part of the algorithm works. In other
words, it can come to no conclusion about the complexity. Actually, it
needs a complicated non-linear optimization to find a ball in terms of
the divergence which passes through given points. Not only the complexity
for it is not known, but there is no guarantee that the solution can be
found.

\section{Algorithm for numerical computation of the Holevo capacity}
\label{sec:algo-holevo}

For the fixed level $d$ of quantum system, the top level procedure of
the algorithm we propose is as follows:
\begin{algorithm}\mbox{}
 \begin{algorithmic}[1]
  \Procedure{main}{$\Gamma$ : quantum channel}\Comment{Compute the
  Holevo capacity of a given channel} 
  \State $P \gets $ {\sc dist\_points}
  \State $Q \gets \Gamma(P)$
  \State $B\gets${\sc qminball}($Q$)
  \State \Return radius of $B$
  \EndProcedure
 \end{algorithmic}
\end{algorithm}

The function {\sc dist\_points} computes the reasonably distributed
points to some extent, {\sc qminball} computes the smallest enclosing
ball with respect to the quantum divergence. As an algorithm for {\sc
dist\_points}, we employed the following:
\begin{algorithm}\mbox{}\label{alg:dist-points}
 \begin{algorithmic}[1]
  \Function{dist\_points}{}\Comment{Compute the reasonably distributed
  point in the quantum state space}
  \State $\varphi =(\varphi_1,\ldots,\varphi_{2(d-1)}) \gets (0,\ldots,0)$
  \Repeat
  \State $\psi \gets 1-\sum_i^{2(d-1)} \varphi$\label{line:psi}
  \State $\Phi \gets (\varphi_1+i\varphi_2,\; \varphi_3+i\varphi_4,\; \ldots
  ,\;\varphi_{2d-3}+i\varphi_{2d-2},\; \psi)^T$\label{line:Phi}
  \State $R=R\cup \left\{ \ket{\Phi}\bra{\Phi}\right\}$
  \State $\varphi \gets $ {\sc next\_state}$(\varphi,1)$
  \Until{$\varphi$ is null}
  \State \Return $R$
  \EndFunction
 \end{algorithmic}
The function {\sc next\_state} called from this is as follows.
Here, $\Delta$ is a constant given as a parameter. 
 \begin{algorithmic}[1]
  \Function{next\_state}{$\varphi$,$i$}
  \If{$i>2(d-1)$}
  \State \Return null
  \EndIf
  \State $\varphi_i\gets \varphi_i + \Delta$
  \If{$\varphi_i >1-(\varphi_1 + \cdots + \varphi_{i-1)}$}
  \State $\varphi_i\gets 0$
  \State \Return {\sc next\_state}($\varphi$, $i+1$)
  \Else
  \State \Return $\varphi$
  \EndIf
  \EndFunction
 \end{algorithmic}
\end{algorithm}

This algorithm is equivalent to run through all the possible tuples of
integers $(n_1,\ldots,n_{2(d-1)})$ which satisfies
$(n_1\Delta)^2+\cdots+(n_{2(d-1)}\Delta)^2\leq 1$. The similar mechanism
as in a carry-up in the computation of sum of two integer is used in
{\sc next\_state}. It tries to add $\Delta$ to the $i$-th value
$\phi_i$, but if the resulting value does not satisfy
$(\varphi_1,\ldots,\varphi_{2(d-1)}) \leq 1$, $\phi_i$ is set to $0$ and
it tries to carry up to the next value $\phi_{i+1}$.

Since $\ket{\Phi}\bra{\Phi}$ has the same value up to the multiplication
of a complex number $z$ to $\Phi$ (of course, under the condition that
$|z|=1$ because $\ket{z\Phi}$ must be a pure quantum state), $d$-th
value of vector $\Phi$ can be restricted to $\mathbb{R}$. So, lines
\ref{line:psi} and \ref{line:Phi} of {\sc dist\_points} do not lose
generality.

We are not convinced that this is the best algorithm for {dist\_points},
but we take a trade-off between the simpleness of the algorithm and the
goodness of its distribution. The algorithm is easy to understand, and
looks reasonable to generate uniformly distributed points. Nevertheless,
there is no theoretical guarantee that it really generates uniformly
distributed points.

The main part of our algorithm is the procedure to solve the smallest
enclosing ball problem. It is described as follows:
\begin{algorithm}\mbox{}
  \begin{algorithmic}[1]
  \Procedure{qminball}{$P$ : set of points}
  \State {\sc b\_qminball($P$, $\emptyset$)}
  \EndProcedure
  \Procedure{b\_qminball}{$P$,$R$}
  \If{$P=\emptyset$ or $R=d^2$}
   \State \Return {\sc b\_qmb}($R$)
  \Else
   \State Choose $p\in P$
   \State $B \gets \text{\sc b\_qminball}(P-\left\{p\right\}, R)$
   \If{$p\not\in B$}
    \State $B \gets \text{\sc b\_qminball}{P-\left\{p\right\}, R \cup
  \{p\}}$
   \EndIf
   \State \Return $B$
  \EndIf
  \EndProcedure
 \end{algorithmic}
\end{algorithm}

The hardest part is the algorithm for {\sc b\_qmb}. {\sc b\_qmb}
computes the sphere which passes through all the given points. To
implement {\sc b\_qmb}, the non-linear (convex) optimization is
necessary, and we have not found the algorithm to compute it
certainly. However, we believe that for a small $d$, we can use some
general-purpose optimization libraries. We show in the next section that
it is a really practical option.

Although there is no 100\% reliable way to compute {\sc b\_qmb}, the
characteristic of the algorithm that it runs through points globally is
a merit to cover that problem. To approximate the final result, it is
not necessary to compute the exact value for each call of {\sc b\_qmb},
but it is enough with mostly correct value for {\em mostly} all the call
of {\sc b\_qmb}. Since it runs through all the plotted points globally,
it is enough if the correct value is computed in one of the neighbor of
the real optimal point. Of course, here we assume the error is
one-sided, i.e.\ when it returns a wrong ball, the ball is always
bigger than the real smallest ball.

The computational error must be considered after divided into two
phases. About the smallest enclosing ball problem of points, the error
is one-sided. In that phase, the error is because of the optimization
process, and if the optimization does not converges to the real optimum,
the computed value is always bigger then the real optimum. However, our
final objective to compute the Holevo capacity and it is equivalent to
solve the smallest enclosing ball of a continuous geometric
object. Since a geometric object is approximated by plotting points,
one-sidedness of the error is not guaranteed at all.

Of course, we can say we can obtain a better result by plotting more
points. Then, how many points are needed to achieve a given upper bound
of the error? It is left to be open. It is essential to make the
proposed algorithm really practical.

\section{Numerical experiment}
Now, we show the algorithm we proposed is really practical. We will
check whether the samples computed by Osawa and Nagaoka \cite{osawa01}
is correct.

The most primitive part to compute the sphere with respect to the
divergence which passes through given points ({\sc b\_qmb} in
Section~\ref{sec:algo-holevo}) is implemented using GNU Scientific Library
(GSL) \cite{gsl}. We used the function {\tt gsl\_multimin\_function\_fdf}
in GSL to compute the minimum value of a given function with its domain
in $\mathrm{R}^n$. The optimization problem we have to solve is
formulated as follows:

{\parindent=0mm
 \begin{quote}
  Input: $n\leq d^2$ and $\sigma_1,\ldots,\sigma_n$\\
  \qquad Minimize $D(\sigma_1||\rho)$\\
  \qquad Subject to $D(\sigma_1||\rho)=\cdots=D(\sigma_n||\rho)$
 \end{quote}
}

However, GSL has no non-linear optimization function which process the
constrains like this. To fit to the specification of {\tt
gsl\_multimin\_function\_fdf} of GSL we rewrite the problem with a
sufficiently large number $A$ as follows:
{\parindent=0mm
 \begin{quote}
  Input: $n\leq d^2$ and $\sigma_1,\ldots,\sigma_n$
  \begin{align}
   &\text{Minimize } D(\sigma_1||\rho)\notag\\
   &\quad-A\left[\left(D(\sigma_1||\rho)-D(\sigma_2||\rho)\right)^2
 +\cdots
 +\left(D(\sigma_{n-1}||\rho)-D(\sigma_n||\rho)\right)^2
 \right]
  \end{align}  
 \end{quote}
}

Note that the only exception is the case $n=d^2$. In that case, the
computation of the center of the ball does not include optimization. The
center is uniquely determined by equations. Actually, the equation
\begin{equation}\label{eq:maximal-boudary-points}
D(\sigma_1||\rho)=D(\sigma_2||\rho)=\cdots=D(\sigma_{d^2}||\rho)
\end{equation}
includes $d^1-1$ equal marks and $d^2-1$ unknown variables, and it has a
unique solution in general.

We employed a restricted version of the algorithm which ignore the ball
with $d^2$ points on its boundary. In other words, we solved the problem
under the assumption that the optimal ball is determined by less than
$d^2$ points. This restriction is because there is no easy way to
compute the solution of Equation (\ref{eq:maximal-boudary-points}). Although
this restriction seems unnatural, we expect that it makes no difference
for the result, or at worst, the difference is small because of the
characteristic of the problem.

We compute the capacity for the channel $\Gamma_5$ in
\cite{osawa01}. $\Gamma_5$ is given by
\begin{equation}
 \Gamma(\rho)=V_1 \rho V_1^* + V_2 \rho V_2^* + V_3 \rho V_3^*.
\end{equation}
where $V_1, V_2, V_3$ are given as
\begin{align}
 V_1&=
 \begin{pmatrix}
  0.2&0.3&0.4\\
  0&0.5i&0\\
  0.1i&0.4i&0.5i
 \end{pmatrix},\\
 V_2&=
\begin{pmatrix}
 0.1-0.3i&0&0\\
 0&-0.3i&0.1-0.2i\\
 0.3-0.3i&0.2+0.1i&0
\end{pmatrix},\\
 V_3&-\sqrt{I-V_1^*V_1-V_2^*V_2}.
\end{align}

The step value ($\Delta$ in Algorithm~\ref{alg:dist-points}) is set to
$0.1$, and the number of plotted points are $49486$. This value is
determined so that the computation ends in a reasonable time, and returns
a reasonable result.

The result of our experiment was $0.672\cdots$, while the optimal value
written in \cite{osawa01} is $0.677\cdots$. The arguments to achieve
the optimum is shown in Table~\ref{tab:numerical-result}, where the
optimal value for $\rho$ is expressed by
\begin{equation}
 \rho^* = \sum_i w_i \sigma_i,
\end{equation}
and the approximated capacity is given by
\begin{equation}
 C = D(\sigma_1||\rho^*)=D(\sigma_2||\rho^*)=\cdots.
\end{equation}
The machine used in experiment is Intel Xeon 2.3GHz (64bit) with Linux
x86\_64 installed. The time for computation was 6 hours and 55 minutes.
 
\begin{table}[h]
 \caption{Result of numerical computation}
 \label{tab:numerical-result}

Result: 0.6729054

Time for computation: 6h55m

Variables to achieve the optimum:\\
 {\small
 \begin{tabular}{c|c}
 $w_i$& $\sigma_i$\\\hline
 0.082689 
 &
 $
 \begin{pmatrix}
 0.65000 + 0.00000i & -0.37000 - 0.29000i &  0.08000 + 0.01000i\\
 -0.37000 + 0.29000i &  0.34000 + 0.00000i & -0.05000 + 0.03000i\\
 0.08000 - 0.01000i & -0.05000 - 0.03000i &  0.01000 + 0.00000i\\
 \end{pmatrix}
 $\\
 0.15770 
 &
 $
 \begin{pmatrix}
 0.16000 + 0.00000i &  0.28000 + 0.04000i &  0.23324 - 0.00000i\\
 0.28000 - 0.04000i &  0.50000 + 0.00000i &  0.40817 - 0.05831i\\
 0.23324 + 0.00000i &  0.40817 + 0.05831i &  0.34000 + 0.00000i\\
 \end{pmatrix}
 $\\
 0.13717 
 &
 $
 \begin{pmatrix}
 0.17000 + 0.00000i &  0.29000 - 0.03000i &  0.22978 - 0.05745i\\
 0.29000 + 0.03000i &  0.50000 + 0.00000i &  0.40212 - 0.05745i\\
 0.22978 + 0.05745i &  0.40212 + 0.05745i &  0.33000 + 0.00000i\\
 \end{pmatrix}
 $\\
 0.17875
 &
 $
 \begin{pmatrix}
 0.01000 + 0.00000i &  0.06000 + 0.00000i & -0.07937 - 0.00000i\\
 0.06000 - 0.00000i &  0.36000 + 0.00000i & -0.47624 - 0.00000i\\
 -0.07937 + 0.00000i & -0.47624 + 0.00000i &  0.63000 + 0.00000i\\
 \end{pmatrix}
 $\\
 0.21126 
 &
 $
 \begin{pmatrix}
 0.05000 + 0.00000i &  0.10000 + 0.05000i & -0.16733 - 0.08367i\\
 0.10000 - 0.05000i &  0.25000 + 0.00000i & -0.41833 - 0.00000i\\
 -0.16733 + 0.08367i & -0.41833 + 0.00000i &  0.70000 + 0.00000i\\
 \end{pmatrix}
 $\\
 0.23243
 &
 $
 \begin{pmatrix}
 0.73000 + 0.00000i & -0.36000 - 0.23000i &  0.11314 - 0.04243i\\
 -0.36000 + 0.23000i &  0.25000 + 0.00000i & -0.04243 + 0.05657i\\
 0.11314 + 0.04243i & -0.04243 - 0.05657i &  0.02000 + 0.00000i\\
 \end{pmatrix}
 $
 \end{tabular}
 }
\end{table}

\begin{figure}
\begin{center}
  \includegraphics{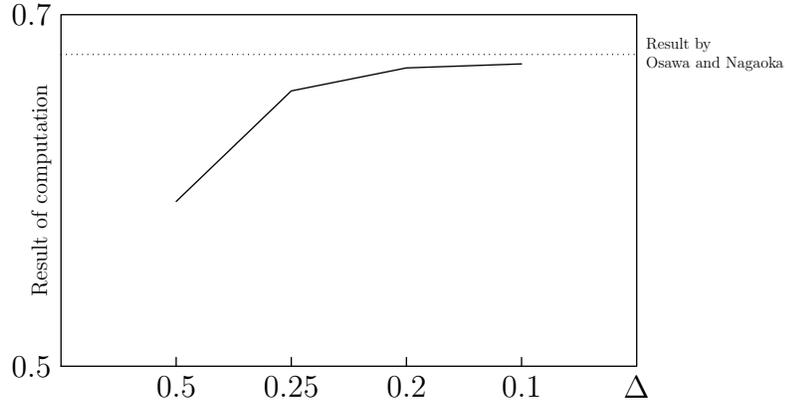}
 \caption{Result of numerical computation for various $\Delta$}
 \label{fig:res-comp}
\end{center}
\end{figure}

Fig.~\ref{fig:res-comp} shows the result of the computation for various
$\Delta$. Since we can approach to the real solution with a smaller
$\Delta$, it seems our result is approaching to the result by Osawa and
Nagaoka. However, we cannot conclude because we have not estimated the
computational error and the error for our algorithm is not one-sided.
 
\section{Discussion}
We obtained the similar result as that of Osawa and Nagaoka. Since we
have not done any error estimation, we can not know how correct the
result is. However, it is still meaningful because we achieved a good
approximation for the result of Osawa and Nagaoka, and showed their
result is likely to be the global optimum.

In our experiment, the number of quantum states which appears on the
surface of the optimum ball is 6, which is bigger than the number 3
shown in the paper by Osawa and Nagaoka. Although it is probably because
of an error of the approximation, another possibility is that for
the smaller number of points, the optimum is not correctly
calculated. Actually we observed that smaller number of points as a
constraint makes the optimization more unstable. It is because with
fewer constraints, the dimension of the space where the variables can go
becomes higher. Anyway, further observation is needed to know the reason
for the difference.

To make the approximation better, we have to make the step value
$\Delta$ smaller. The problem is the computational complexity which
rapidly increases as $\Delta$ gets smaller. For three level system, when
$\Delta$ is $1/n$, the complexity becomes $O(n^4)$.

To avoid the computationally difficult case, our experiment was under
the assumption that the optimum ball is determined by less than $d^2$
points. To remove the restriction is naturally considered as a next
work. Although the direct computation of the ball which pass through
$d^2$ is to solve a $d^2-1$ dimensional non-linear equation, the dual
problem by Legendre transform becomes a linear equation. Using the dual
problem is a probable option.

Another problem is how much likely this method can be used for an arbitrary
$d$-level system. For a $d$-level system, the complexity becomes
$O(n^{2d-2})$, where $\Delta=1/n$. Because of lack of error estimation,
we cannot say any concrete criterion for $n$, but it is likely that the
computation for $d=4$ is far more difficult than the case $d=3$.

As is mentioned by Welzl \cite{welzl91}, the possible way of improvement
is to care for the order of the points. The original algorithm to
compute the smallest enclosing ball is a Las Vegas algorithm and its
computational complexity is based on the randomness of the order of
points. Some heuristics might improve the actual performance of the
computation, even if it does not improve the complexity expressed as an
order. In particular, since the optimization process for fixed bounding
points is very heavy in the proposed algorithm, the heuristics in the
order of the points is more likely to contribute to the improvement
than the usual Euclidean smallest enclosing ball problem.

\section{Summary of this chapter}
We proposed a new algorithm to compute the Holevo capacity of a quantum
channel which uses arranged version of Welzl's algorithm to compute the
smallest enclosing ball problem. The proposed algorithm is for an
arbitrary $d$-level system, and it is natural extension of the
existing algorithm for one-qubit system.

Although the proposed algorithm includes a non-linear optimization in
its process which is likely to be very unstable, we have shown it is
actually works for some examples of channels in 3-level system. We
calculated the same examples as the ones by Osawa and Nagaoka
\cite{osawa01}, and certified the results by them are real optima though
there had been a concern that those might be only local optima. The main
merit of our algorithm is that it searches an optimum globally.

Although proposed algorithm can practically approximate the result,
analysis for its error is not done yet, and the error is not
one-sided. Estimating the error and computing the lower-bound for the
real value by the computed value are to be done.


\chapter{Conclusion}
\label{chap:conclusion}

In this dissertation, we have given another geometric interpretation to
a quantum state space. It indicates a connection among some
quantum-related research fields. 

More precisely, what we have shown in this dissertation mainly consists of
the following two parts:
\begin{itemize}
 \item some Voronoi diagrams in quantum state space coincides
 \item Welzl's algorithm for the smallest enclosing ball problem is also
       applicable for a quantum state space.
\end{itemize}

In this chapter, we summarize again the main results and discuss the
meaning of them and future potential of development of related
researches. 

\section{Coincidences of Voronoi diagrams}
We proved that in one-qubit system, the Voronoi diagrams with respect to
the divergence and Euclidean distance are the same.  We also proved that
in an $n$-level system for $n \geq 3$, that coincidence does not occur.

In three or higher level system, restricting on a certain subspace,
we have investigated some distances. We showed that with some
unnatural parameterization, Voronoi diagrams with respect to the divergence
and Euclidean distance coincide on the supspace. We also showed that
Voronoi diagrams with respect to the divergence and the Fubini-Study
distance coincide.

The existing only application of these fact is the algorithm by Hayashi et
al.~\cite{hayashi05} to compute the Holevo capacity of a quantum
channel. Nevertheless, we believe those result have clarified some
aspect of the structure of a quantum state space. 

Our result shows thatthere is a connection among different distances
which are used in different cotexts. Especially a significant thing is a
connection between the distance used in quantum computation and the
pseudo-distance which is used for quantum information theory. We regard
we indicated a bridge between those research fields.

From the viewpoint of computational geometry, what we have done can be
a methodological hint. Generally in a situation that some measures are
associated to a set, analyzing those Voronoi diagrams will some
indication about the relation of those measures. Especially for a
computational purpose, since a Voronoi diagram is a popular tool to
approximate a continuous geometric object in a computer, comparing
Voronoi diagrams is a reasonable start point for the discussion about
how we can deal with a given continuous object with some associated
metrics.

\section{Numerical computation of Holevo capacity}
We have shown that Welzl's algorithm for the smallest enclosing problem
is also applicable to a quantum state space. Using Welzl's algorithm, we
proposed a new algorithm to compute the Holevo capacity of a quantum
channel. Although it leads to no conclusion about its complexity, we
showed it is practical by an experiment.

Although the experiment is for only few samples and under some
restrictions, we showed the proposed algorithm really works for a real
computation. We have not compared its practical performance to the
existing algorithm by Osawa and Nagaoka \cite{osawa01}, but it has at
least one merit: there is no worry for local optima.

Since our algorithm is an approximation by plotting points on a continuous
geometric object, its performance is trade-off with its
error. Consequently, for the future research, the improvement of
``uniformness'' of plotted points and the error estimation should be
considered as a set. The proposed method to generated a ``reasonably''
uniform points as pure states is only intended to be implemented easily,
and no mathematical analysis is given for its uniformness. To improve
the algorithm of that part, consideration about what error bound can be
attained will be necessary. 

Welzl's algorithm is classified as a Las Vegas algorithm, and its
expected time of computation is linear. As is mention by Welzl himself
\cite{welzl91}, the expected time is based on the randomness of the
order of points, but some heuristics about ordering of points might
improve the performance. To seek for heuristics specific for a quantum
state space is one of the possible extensions of this research.


\bibliography{kkato2006}
\bibliographystyle{habbrv}

\end{document}